\documentclass[
  ,draft            
  ,numberedheadings 
  ]
  {aipproc}

\layoutstyle{6x9}

\begin{document}

\newcommand{\be}{\begin{eqnarray}}
\newcommand{\ee}{\end{eqnarray}}
\newcommand\del{\partial}

\newcommand{\mat}{\left ( \begin{array}{cc}}
\newcommand{\emat}{\end{array} \right )}
\newcommand{\matt}{\left ( \begin{array}{ccc}}
\newcommand{\ematt}{\end{array} \right )}
\newcommand{\matf}{\left ( \begin{array}{cccc}}
\newcommand{\ematf}{\end{array} \right )}
\newcommand{\vect}{\left ( \begin{array}{c}}
\newcommand{\evect}{\end{array} \right )}
\newcommand{\Tr} {\rm Tr}
\newcommand{\nn} {\nonumber}


\begin{flushright}
SUNY-NTG-04/06
\end{flushright}

\title{The  Supersymmetric Method in Random Matrix Theory and Applications to QCD
}

\author{Jacobus Verbaarschot}{  
address={Department of Physics and Astronomy,\\SUNY at Stony Brook,
NY 11794, US,\\
jacobus.verbaarschot@stonybrook.edu}
}

\begin{abstract}

The supersymmetric method is a powerful method for the nonperturbative
evaluation of quenched averages in disordered systems. Among others,
this method has been applied to the statistical theory of S-matrix
fluctuations, the theory of universal conductance fluctuations and the
microscopic spectral density of the QCD Dirac operator.

We start this series of lectures with a general review of Random Matrix
Theory and the statistical theory of spectra.
An elementary introduction of
the supersymmetric method in Random Matrix Theory is given
in the second and third lecture. We will show that
a Random Matrix Theory can be rewritten as an integral over a
supermanifold. This integral will be worked out in detail for the
Gaussian Unitary Ensemble that describes level correlations in systems
with broken time reversal invariance. We especially emphasize the role
of symmetries.

As a second example of the application of the supersymemtric method
we discuss the calculation of the microscopic spectral density of
the QCD Dirac operator. This is the eigenvalue density
 near zero on the scale of the average level spacing
which is known to be given by chiral Random Matrix Theory.
Also in this case we use symmetry considerations to rewrite the
generating function for the resolvent as an integral over a
supermanifold.

The main topic of the second last lecture is the  recent developments 
on the relation
between the supersymmetric partition function and integrable hierarchies 
(in our case the Toda lattice hierarchy).
We will show that this relation is an efficient way 
to calculate superintegrals.
Several examples that were given in  previous lectures will be
worked out by means of this new method. Finally, we will 
discuss the quenched
QCD Dirac spectrum at nonzero chemical potential. Because of the 
nonhermiticity of the Dirac operator the usual supersymmetric method
has not been successful in this case. However, we will show that the
supersymmetric partition function can be evaluated by means
of the replica limit of the Toda lattice equation.

\end{abstract}

\maketitle
\newpage
\tableofcontents

\newpage
\section{Introduction}
\label{intro}

Random Matrix Theories are ensembles of matrices with random numbers
as matrix elements. They were first introduced in physics in 1955 by Wigner
\cite{Wigner} to describe the spacing distribution of nuclear levels.
The mathematical problem of Random Matrix Theory is to derive 
eigenvalue correlations in the limit of large matrices. This problem
was solved by Dyson \cite{Dyson} and Mehta and Gaudin \cite{mehta-np,Mehta} 
who introduced  
orthogonal polynomials to perform the matrix integrals.
About the same time the Anderson model \cite{Anderson} was introduced to 
describe the localization of wave functions in one-dimensional 
systems which led to the theory of disordered systems. 
Random Matrix Theory, with its main applications to nuclear physics, and
the theory of disordered systems, with its main applications to solid
state physics, developed independently until the mid eighties when
it was realized that Random Matrix Theory is the zero dimensional
limit of a disordered system \cite{Efetov,EfetovBook,VWZ-an,VWZ}. 
This was made possible
in particular by the introduction of a field theoretical formulation
of the theory of the disordered systems \cite{Wegner}, and by 
the introduction of the 
supersymmetric method which made it possible to perform nonperturbative
calculations. After these developments a rapid
growth of Random Matrix Theory followed 
in the late eighties  after  the discovery of universal
conductance fluctuations \cite{Wash86}
and applications of Random Matrix Theory to quantum gravity
\cite{david,kazakov-migdal}. Both applications
greatly enhanced the intellectual breath of the subject. 

Random Matrix Theory has many applications. We already mentioned
the correlations of nuclear levels \cite{Haq}. It was realized
in 1985 \cite{Bohigas} that the necessary ingredient
for levels to be correlated according to
Random Matrix Theory is not the complexity of the system
but rather that the system is classically
chaotic. This led to many applications in the field of Quantum Chaos.
A second application of Random Matrix Theory in nuclear physics is
the description of $S$-matrix fluctuations . 
This problem was solved
by means of the supersymmetric method \cite{VWZ} and the same
approach was later applied to the theory of universal conductance
fluctuations \cite{iida}. 

A new class of Random Matrix Ensembles was introduced in the early
nineties \cite{SV,V} to describe the eigenvalue correlation of
the Euclidean Dirac operator in QCD. This made it possible
to derive exact analytical results for Dirac spectra that have
been generated by means of lattice QCD Monte-Carlo simulations.
Random Matrix Theory in this field made it possible to get
a much better grip on finite size corrections.

In Mathematics, Random Matrix Theory was applied to the correlations
of the nontrivial zeros of the Riemann $\zeta$-function
\cite{montgomery}. After the impressive
numerical work by Odlyzko \cite{Odly87} this has been understood in
great detail by Berry and Keating (see \cite{Keating} for a review).
More recently, Random Matrix Theory has been applied to growth 
processes \cite{PS01,wiegmann},
the longest increasing subsequence of random permutations 
\cite{baik}
and the theory
of random partitions \cite{okoun,Jo02}.

There are several books and reviews on Random Matrix Theory that should
be mentioned. The classic
text is the book of Mehta \cite{Mehta} which emphasizes the orthogonal
polynomial method.
The supersymmetric method
is discussed in detail in the book by 
Efetov \cite{EfetovBook}. 
In the book by Forester \cite{ForresterBook} the emphasis is
on the relation with solvable models and mathematical physics.
The historical developments of Random Matrix Theory 
are discussed in the comprehensive review by Guhr, M\"uller-Groeling,
and Weidenm\"ller. \cite{Guhr98}. A short historical overview of
Random Matrix Theory can be found in the special edition of Journal of
Physics A which contains a broad selection of recent papers on
Random Matrix Theory as well as several reviews \cite{special}.   
Applications to mesoscopic physics are reviewed in \cite{Beenreview},
applications to QCD in \cite{anrev} and applications to Quantum Gravity
in \cite{zinn}. Among the pedagogical reviews we mention 
a general review of Random Matrix Theory \cite{ieee}, 
an introduction to the supersymmetric method \cite{fyodorov,zirnency} and
applications of random matrix theory to quantum gravity \cite{francesco}.

\section{Random Matrix Theory}
\label{rando}

In this section we give an elementary introduction to Random Matrix
Theory. Some of the material in the subsections below is based on
the encyclopedia article \cite{ieee}.

\subsection{The Wigner-Dyson ensembles}
\label{rando1}

The Wigner-Dyson ensembles are ensembles of hermitian matrices with
matrix elements distributed according to
\be
P(H)DH = {\cal N}e^{-\frac {N\beta}{4} {\rm Tr} H^\dagger H} DH.
\label{probwd}   
\ee
Here, $H$ is a Hermitian  $N\times N$ matrix with real, complex, or
quaternion real matrix elements, and  $\beta$ is the Dyson index of the
ensemble. Its value is one for real matrix elements, two for complex
matrix elements and four  for 
quaternion real matrix elements. Real quaternions can be
expressed in terms of the Pauli spin matrices as
\be
q =a_0 + i a_k \sigma_k,
\ee
where the $a_\mu$ are real variables. The measure $DH$ is
the product over independent differentials.
The normalization constant of the probability distribution is denoted by 
${\cal N}$.
The probability distribution is invariant under the  transformations
\be
H \rightarrow U H U^{-1},
\label{sim}
\ee
where $U$ is an orthogonal transformation for $\beta =1$, a unitary 
transformation for $\beta = 2$, and a symplectic transformations for
$\beta =4$. This is the reason why these ensembles are known as the
 Gaussian Orthogonal Ensemble (GOE), the  Gaussian Unitary 
  Ensemble (GUE), and the  Gaussian Symplectic Ensemble (GSE), 
respectively. Since both the eigenvalues of $H$ and the Haar measure $DH$ are
invariant with respect to (\ref{sim}), the eigenvectors and the
eigenvalues are independent with the distribution of the eigenvectors
given by the invariant measure of the corresponding orthogonal,
unitary, or symplectic group. Notice that eigenvalues of a quaternion
real matrix are quaternion scalars and are doubly degenerate.

\vspace*{0.2cm}
\noindent {\it Exercise}. Show that the symplectic ensemble is invariant
under symplectic similarity transformations. A symplectic transformation
is a unitary transformation that leaves the matrix $Z$ invariant where,
in quaternion notation, $Z$ is the direct product of 
$i\sigma_2$ and the identity.

\vspace*{0.2cm}
\noindent {\it Exercise}. Show that the measure, $dH$, is invariant with
respect to the similarity transformations (\ref{sim}). Prove this
for each value of $\beta$.
\vspace*{0.2cm}

One way of deriving the probability distribution (\ref{probwd}) is to
look for a distribution that minimizes the information entropy under the
constraints that the average, variance and normalization of the 
probability distribution are fixed. The proof goes as follows \cite{balian}. 
The sum of the information
entropy and the constraints is given by
\be
S = \int DH[ -P(H) \log P(H) + aP(H)  + b P(H){\rm Tr} H + 
c P(H){\rm Tr} H^2] .
\ee
Minimizing this with respect to $P(H)$ we obtain
\be
-1 -\log P(H) + a + b{\rm Tr} H +c {\rm Tr} H^2=0,
\ee
resulting in a Gaussian distribution for $P(H)$.
The maximum entropy approach has also been successfully applied to the
joint probability distribution of $S$-matrix elements \cite{Seligman}
and the transport properties of quantum dots (see \cite{mello,Beenreview}).

The joint probability distribution of the Wigner-Dyson ensembles is
given by
\be 
P(\{\lambda_k\}) \prod_kd \lambda_k = {\cal N} |\Delta(\{\lambda\})|^\beta
\prod_k e^{-N\beta \lambda_k^2/4} d\lambda_k,
\label{jointwd}
\ee
where the Vandermonde determinant is defined by 
\be
\Delta(\{\lambda\}) =\prod_{k>l}( \lambda_k -\lambda_l). 
\label{vandermond}
\ee
 
The Vandermonde determinant originates from the Jacobian of
the transformation from the matrix elements as integration variables to
the eigenvalues and eigenvectors as integration variables. The distribution
of the eigenvectors completely factorizes from the probability distribution.
As an example, let us calculate the Jacobian for $\beta =2$.
One way to do this is to consider the invariant distance
\be
{\rm Tr}\, d H d H^\dagger.
\ee
By differentiating (\ref{sim}) we find 
\be
U^{-1} dH U = \delta U \Lambda - \Lambda \delta U + d \Lambda,
\ee
where $\delta U = U^{-1} dU$. For the invariant distance we then obtain
\be
{\Tr}\, d H d H^\dagger &=& 2 {\rm Tr} \delta U \Lambda \delta U \Lambda
-2 {\rm Tr} \delta U \Lambda ^2\delta U +{\rm Tr }d \Lambda^2\nonumber\\
 &=&2\sum_{k<l} \delta U_{kl}\delta U_{kl}^*(\Lambda_k -\Lambda_l)^2
+ \sum_k (d\Lambda_k)^2.
\ee
We can immediately read off the Jacobian from the Lam\'e coefficients
and the result is given by the square of the Vandermonde determinant.

\vspace*{0.2cm}
\noindent {\it Exercise}. Calculate the Jacobian for $\beta =1 $ and
$\beta = 4$.

\subsection{ The chiral ensembles}
\label{rando2}

A second class of ensembles are the chiral ensembles with the chiral
symmetry of the QCD Dirac operator \cite{SV,V}.
 They are defined as the ensembles of $N\times N$
Hermitian matrices with block structure
\be\label{hc}
D= \left ( \begin{array}{cc} 0 & C \\ C^\dagger & 0 \end{array} \right )
\label{chRMT}
\ee
and probability distribution given by
\be
P(C)DC = {\cal N}{\det}^{N_f}
\left ( \begin{array}{cc} 0 & C \\ C^\dagger & 0 \end{array} \right )
e^{-\frac {N\beta}{4} {\rm Tr} C^\dagger C}
DC.
\label{chRMTprob}
\ee
Again, $DC$ is the product of differentials of the independent
parts of the matrix elements of $C$, and $N_f$ is a real parameter
(corresponding to the number of quark flavors in QCD).
The matrix $C$ is a rectangular $n \times (n+\nu)$ matrix.
The nonzero eigenvalues of the matrix $D$ occur in pairs $\pm \lambda_k$.
This can be seen as follows. If 
\be
\left ( \begin{array}{cc} 0 & C \\ C^\dagger & 0 \end{array} \right )
\left ( \begin{array}{c} a \\ b \end{array} \right )
= \lambda \left ( \begin{array}{c} a \\ b \end{array} \right ),
\ee
then
\be
\left ( \begin{array}{cc} 0 & C \\ C^\dagger & 0 \end{array} \right )
\left ( \begin{array}{c} a \\ -b \end{array} \right )
= -\lambda \left ( \begin{array}{c} a \\ -b \end{array} \right ).
\ee
Generically, the matrix $D$ in (\ref{hc})
 has exactly $|\nu|$ zero eigenvalues. Also generically,
the QCD Dirac operator corresponding to a field configuration
with the { topological charge} $\nu$ has exactly $|\nu|$ zero eigenvalues,
in accordance with the { Atiyah-Singer index theorem}.
For this reason, $\nu$ is identified as the topological quantum
number. The normalization constant of the probability distribution is
denoted by ${\cal N}$.  Also in this case one can distinguish 
ensembles with real, complex, or quaternion real matrix elements. They
are denoted by $\beta= 1$, $\beta = 2$, and $\beta = 4$, respectively.
The unitary invariance of the chiral ensembles is given by
\be
C \rightarrow U C V^{-1},
\ee
where $U$ and $V$ are orthogonal, unitary, or symplectic matrices,
respectively. For this reason, the corresponding ensembles are known
as the  chiral Gaussian Orthogonal Ensemble (chGOE), the    
  chiral Gaussian Unitary Ensemble (chGUE), and the  chiral
  Gaussian Symplectic Ensemble (chGSE), in this order. 
   
\vspace*{0.2cm}
\noindent{\it Exercise.} Show that the eigenvalues of the matrix
with structure (\ref{chRMT}) occur in pairs $\pm \lambda$ or are zero. If
$C$ is an $n\times (n+ \nu)$ block matrix 
show that there are exactly $\nu$ zero
eigenvalues.
\vspace*{0.2cm}

It is always possible to decompose the matrix $C$ in (\ref{chRMTprob}) as
\be
C = U \Lambda V^{-1},
\ee
where $\Lambda$ is a diagonal matrix  with $\Lambda_{kk}\ge 0$,
and $U$ and $V$ are orthogonal matrices
for $\beta = 1$, unitary matrices for $\beta = 2$ and symplectic matrices
for $\beta = 4$. 
The joint probability distribution for the eigenvalues
is obtained by transforming to $\Lambda$, $U$ and $V$ 
as new integration variables. 
 The Jacobian can be obtained in the same
way as in the previous section and is given by
\be
J \sim \prod_k \lambda_k^{\beta \nu+\beta -1}
\prod_{k<l} |(\lambda^2_k -\lambda^2_l)|^\beta 
\label{jchiral}
\ee
resulting in the joint eigenvalue distribution
\be
P(\{\lambda\}) d \{\lambda\} = {\cal N} |\Delta(\{\lambda^2\})|^\beta
\prod_k \lambda_k^\alpha e^{-N\beta \lambda_k^2/4} d\lambda_k,
\label{pchjoint}
\ee
where  $\alpha= \beta -1 +\beta \nu + 2 N_f$.
As in the case the Wigner-Dyson ensembles the distribution of the eigenvectors
and the distribution of the eigenvalues factorize.

\vspace*{0.2cm}
\noindent {\it Exercise}. Find the Jacobian (\ref{jchiral}) using
only symmetry arguments and dimensional arguments.
\vspace*{0.2cm}
 
\subsection{The superconducting ensembles}
\label{rando3}

A third class of random matrix ensembles occurs in the description of
disordered superconductors \cite{Oppermann,AZ}. 
Such ensembles with the symmetries
of the Bogoliubov-de Gennes Hamiltonian have the block structure
\be
H =\left ( \begin{array}{cc}  A & B \\ B^\dagger & -A^T \end{array} \right),
\ee
where $A$ is Hermitian and, depending on the underlying symmetries, 
the matrix $B$ is symmetric or anti-symmetric. The probability
distribution is given by
\be
P(H)DH = {\cal N}\exp
\left (-\frac {N\beta}{4} {\rm Tr} H^\dagger H\right ) DH,
\label{prob}
\ee
where $DH$ is the Haar measure and ${\cal N}$ is a normalization
constant.  For a symmetric matrix $B$ the matrix elements of $H$ can be either
complex ($\beta =2$) or real ($\beta =1$). 
One can easily verify that
in this case $H$
satisfies the relation
\be
H^T = \Gamma_A H \Gamma_A,
\ee
where
\be
\Gamma_A = \left ( 
\begin{array}{cc} 0 & 1_N \\ -1_N & 0 \end{array} \right ).
\ee
One can then show that if                $\phi$ is an eigenvector
of $H$ with eigenvalue $\lambda$, then $\Gamma_A \phi$ is an eigenvector of
$H^T$ with eigenvalue $-\lambda$. Because $H$ and $H^T$ have the same
eigenvalues, all eigenvalues occur in pairs $\pm \lambda$.

For an anti-symmetric matrix $B$ the matrix elements
of $H$ can be either complex ($\beta =2$) or quaternion real ($\beta = 4$).
In that case $H$ satisfies the relation
\be
H^T = -\Gamma_S H \Gamma_S,
\ee
where
\be
\Gamma_S = \left ( \begin{array}{cc} 0 & 1_N \\ 
1_N & 0 \end{array} \right ).
\ee
From this relation it follows that if $\phi $ is an eigenvector of $H$ with
eigenvalue $\lambda$, then $\Gamma_S \phi$ is an eigenvector of $H^T$ with
eigenvalue $-\lambda$. Also in this case, all eigenvalues occur in pairs
$\pm \lambda$.

The joint eigenvalue probability distribution 
of these so called Oppermann-Altland-Zirnbauer ensembles
has the same general form  as the joint probability
distribution chiral ensembles given in (\ref{pchjoint}).
The value of 
$\alpha = 1$ for $ \beta = 1$ and $\beta = 4$ whereas for $\beta = 2$ the
value of $\alpha$ is 0 or 2. 

\vspace*{0.2cm}
\noindent{\it Exercise}.  Derive the joint eigenvalue distribution for
the Oppermann-Altland-Zirnbauer ensembles.
\vspace*{0.2cm}

\subsection{Anti-Unitary symmetries and the Dyson index}
\label{rando4}

The value of the Dyson index is determined by the anti-unitary symmetries
of the system. If there are no anti-unitary symmetries the Hamiltonian
is Hermitian and the value of $\beta = 2$. 

An anti-unitary symmetry operator, which can always be written as $A= UK$ with
$U$ unitary and $K$ the complex conjugation operator,  commutes
with the Hamiltonian of the system
\be
[H,UK] = 0.
\ee 
We can distinguish
two possibilities
\be
(UK)^2 = 1 \quad {\rm or} \quad (UK)^2 =-1.
\ee
The argument goes as follows. The 
symmetry operator $A^2=(UK)^2=UU^*$ is unitary, and in an irreducible
subspace of the unitary symmetries, 
it is necessarily a multiple of the identity, $UU^* = \lambda
{\bf 1}$.  Because of this relation, $U$ and $U^*$ commute so that
$\lambda$ is real. By unitarity we have $|\lambda| = 1$ which yields
$\lambda=\pm1$.

The case $(UK)^2 =1$ corresponds to $\beta =1$ and the case $(UK)^2= -1$
to $\beta = 4$.
In the first case it is always possible to find a basis in which the
Hamiltonian is real. Starting  with basis vector $\phi_1$ we
construct $\psi_1 =\phi_1 + UK \phi_1$. Then choose $\phi_2$ perpendicular
to $\psi_1$ and define $\psi_2 =\phi_2 + UK \phi_2$.
Then
\be
&&(\phi_2+ UK \phi_2, \psi_1) \nonumber\\
&=& (UK \phi_2,\psi_1) \nonumber\\
&=& ((UK)^2 \phi_2,UK\psi_1)^* \nonumber\\
&=& (\phi_2,\psi_1)^* = 0.
\ee
The next basis vector is found by choosing $\phi_3$ perpendicular
to $\psi_1$ and $\psi_2$, etc.\,.
In this basis the Hamiltonian is real
\be
H_{kl} &=& (\psi_k, H \psi_l) \nonumber\\
&=& (UK\psi_k, UKH \psi_l)^* \nonumber\\
&=& (\psi_k, HUK \psi_l)^* \nonumber\\
&=& (\psi_k,H \psi_l)^* =H_{kl}^*.
\ee

The best known anti-unitary operator in this class is the time-reversal
operator for which  $U$ is the identity matrix.

In the case $(UK)^2 = -1$ all eigenvalues of the Hamiltonian are
doubly degenerate. This can be shown as follows. If $\phi_k$ is and
eigenvector with eigenvalue $\lambda_k$, then it follows from the
commutation relations that also $UK\phi_k$ is an eigenvector of
the Hamiltonian with the same eigenvalue. The important thing is
that this eigenvector is perpendicular to $\phi_k$,
\be
(\phi_k, UK\phi_k) = (UK \phi_k , (UK)^2 \phi_k)^*= -(\phi_k, UK\phi_k).
\label{independent}
\ee
One can prove that in this case it is possible to chose a basis for
which the Hamiltonian matrix can be organized into real quaternions
\cite{Porter}. The eigenvalues of a Hermitian quaternion real matrix
are quaternion scalars, and the eigenvalues of the original matrix 
are thus doubly degenerate in agreement with (\ref{independent}).
The best known example in this class is the Kramers degeneracy  for 
time reversal invariant systems of half-integral spin but no
rotational invariance. Then the time reversal operator is
given by $\sigma_2 K$ with $(\sigma_2K)^2 = -1$.

Later in these lectures we will discuss other anti-unitary symmetries
that enter in the Dirac operator for non-abelian Yang-Mills fields.
They determine the Dyson index of the chiral ensembles.

\subsection{Classification of the Random Matrix Ensembles}
\label{rando5}

The Random Matrix Ensembles discussed above can be classified according to
the Cartan classification of symmetric spaces 
\cite{dyson-class,class,clency}. A symmetric space is a 
manifold such that every point is a fixed point of an involutive isometry
(i.e. $x_i \rightarrow - x_i$).
The Riemann curvature tensor is covariantly constant
in a symmetric space. 
A symmetric space is best characterized via the notion of symmetric pair.
A symmetric pair $(G, H)$ is defined as a pair of a connected Lie group
$G$ and a closed subgroup $H$ such that an involutive analytic
automorphism $\sigma$
of $G$ exists with $H \in H_\sigma$, where $H_\sigma$ is the set of fixed
points of $\sigma$.  Then, with some additional conditions (see the
book by Helgason \cite{Helgason} for more details) the coset $G/H$ is a
symmetric space.

As an example, consider the group $U(p+q)$ and define $\Gamma_{pq}$ as the
diagonal matrix with the first $p$ diagonal matrix elements equal to 1 and
the remaining $q$ diagonal matrix elements equal to $-1$.  Consider the
transformation
\be
\sigma(g) = \Gamma_{pq} g \Gamma_{pq},
\ee
which is an involution because obviously $\sigma(\sigma(g)) = g$.
One can also easily show that it is an analytic automorphism.
The set of fixed points of $\sigma$ is the group $U(p) \times U(q)$. Therefore,
$U(p+q)/U(p)\times U(q)$ is a Riemannian symmetric space.

A symmetric space can be of the compact, noncompact
or the Euclidean type with positive, negative or zero curvature, respectively.
Each of the random matrix ensembles discussed in this section is tangent
to one of the large classes of symmetric spaces. The complete classification
is given in table I \cite{class} where  the corresponding symmetric
space of the compact type is also given.
\begin{table}[ht!]
\vspace*{0.5cm}
\label{tableI}
\centering
\begin{tabular}{cccc}
\hline
  RMT & symmetric space  & group structure & $\beta$ \\
\hline
 GOE  & AI &$U(N)/O(N)$ &1  \\
 GUE  & A & $U(N)$ & 2  \\
 GSE  & AII &$U(2N)/Sp(N)$& 4  \\
chGOE  & BDI & $SO(p+q)/SO(p) \times SO(q)$ &1   \\
chGUE  & AIII & $U(p+q)/U(p) \times U(q)$& 2  \\
chGSE  & CII & $Sp(p+q)/Sp(p) \times Sp(q)$& 4\\
AZ-CI  & CI &$Sp(N)/U(N)$ &1 \\
AZ-D    & D &$ SO(N)$&2  \\
AZ-C    & C & $Sp(N) $& 2 \\
AZ-DIII    &DIII &$ SO(2N)/U(N)$ &4 \\
\hline
\end{tabular}
\caption{Random Matrix Ensemble, Corresponding symmetric space and
the value of the Dyson index $\beta$.}
\end{table}

For example, for special unitary matrices we can write $U = 1 + iH + \cdots$. 
Therefore, the Hermitian matrices are tangent to the space $A$ 
(after dividing out a $U(1)$ factor).
As another example, the generators $U(p+q)/U(p) \times U(q)$
(class AII) are given by
matrices with the structure (\ref{chRMT}).
More examples can be found in the recent review by
Caselle and Magnea \cite{caselle}.

\section{ Spectra of complex systems}
\label{spect}

In this section we discuss quantum spectra and several frequently used
statistics
for the statistical analysis of eigenvalues.

\subsection{General Remarks}
\label{spect1}

Typically, when we talk about spectra we think about quantum mechanics. Among
the best
known examples are the spectrum of a harmonic oscillator with eigenvalues
\be
E_n = (n + \frac 12) \hbar \omega,
\ee
and the spectrum of an angular momentum operator with eigenvalues of $J^2$
given by
\be
\hbar^2 j(j+1).
\ee
However, it is the exception that we are able to write down analytical
formulas for the complete set of eigenvalues of an operator or even
a single eigenvalue. 
One of the questions
we wish to address is whether we are able to find any generic features
in spectra even if we do not have any detailed analytical knowledge of the 
eigenvalues.

Let me first give examples of different systems for which spectra 
play an important role.
One of the most important examples is the Schr\"odinger
equation 
\be
(-\nabla^2 + V(x)) \psi = \lambda \psi,
\ee 
of course with boundary conditions on the wave functions. 
One particularly simple class of potentials are billiards where $V(x)$
is zero inside a connected domain and infinite outside this domain. In 
particular, billiards in the two-dimensional plane have been studied 
extensively in the literature. Although it is trivial to solve the 
Schr\"odinger equation, the complication arises because of the boundary
conditions which can only be satisfied for a discrete set of eigenvalues.

A second example is the Helmholtz equation for the electric and magnetic
field inside a resonance cavity
\be
(\nabla^2 + k^2)\vec E = 0, \qquad (\nabla^2 + k^2)\vec B = 0.
\ee
In this case one can distinguish different types of boundary conditions,
transverse electromagnetic, transverse magnetic and transverse electric. 
One case of special interest are transverse magnetic modes
in  a  two-dimensional cavity. For  $E_z$ we find the wave
equation
\be
(\nabla^2 + k^2)E_z = 0,
\ee
with the boundary condition that $E_z$ vanishes on the boundary. This is
exactly the Schr\"odinger equation for a two-dimensional billiard. 
This equivalence has been exploited for the experimental analysis of 
quantum billiards \cite{richter-koch}.

A third example is that of sound waves in materials. In that case the
wave equation is given by
\be
(\nabla^2 + k^2)\vec u = 0,
\ee
where $\vec u $ can be decomposed in the longitudinal and the transverse
displacement. This equation is more complicated than either the Maxwell or 
Schr\"odinger equation because the two types of modes are coupled through
reflections at the boundary. Also this system has been a valuable 
tool for the experimental analysis of spectra of complex systems
\cite{ellegaard}.

A fourth example is the Dirac operator in QCD. In this case the wave
equation is given by
\be
\gamma_\mu (\partial_\mu + iA_\mu) \psi =i\lambda \psi.
\ee
Here, the $\gamma_\mu$ are the Euclidean gamma matrices and the $A_\mu $
are $SU(N_c)$ valued  gauge fields. We will discuss
this example in much more detail later in these lectures.

As a fifth example I mention the zeros of the Riemann $\zeta$ function. 
This function
is defined as
\be
\zeta(s) = \sum_n \frac 1{n^s}.
\ee
Empirically, it is know that all zeros are on the line 
${\rm Re} \,s = \frac 12$ (also known as the Riemann conjecture). An
intriguing question is what these zeros
have in common with the eigenvalues of a complex dynamical 
system (for answers we refer to \cite{Keating}).

All differential operators we have discussed up to now are Hermitian operators.
They can be diagonalized by a unitary transformation and have real eigenvalues,
i.e.
\be
H = U \Lambda U^{-1}.
\ee
Later we will see applications of non-hermitian differential operators. 
They can be diagonalized by a similarity transformation with eigenvalues
scattered in the complex plane.  

In many cases spectra can be observed experimentally. The best-known 
examples are atomic and nuclear spectra. However, a great deal of spectroscopy 
has been performed on  molecular spectra as well. As a more recent
example I mention  quantum dots in which electrons are enclosed inside a 
heterostructure and collide elastically against the boundary. 
Other examples are the spectra of small metallic particles, 
which  are determined by the boundary conditions on the surface. 
In this case the specific heat can be related to the spectral properties of
the particles. For references and an extensive discussion of these
and other examples we refer the \cite{Guhr98}.

\subsection{Statistical analysis of spectra}
\label{spect3}
 
Although it is the exception that quantum spectra can be obtained analytically,
it is often possible to obtain a large number of
different eigenvalues either experimentally or by numerical calculations. 
This cries out for a reduction of information,
and it is  natural the perform a statistical
analysis of the spectra. The average spectral density can be written as
\be
\rho(\lambda) = \langle \sum_k \delta(\lambda- \lambda_k) \rangle
\ee
with normalization given by (for a finite Hilbert space)
\be
\int d\lambda \rho(\lambda) = N.
\ee
The average can be either over an ensemble of different systems 
(ensemble average) or
over different parts of a spectrum (spectral average). 
Of course it is also possible that
the average is a mixture of both types of averages.

One  frequently employs the integrated spectral density
\be
N(E) = \int_{-\infty}^E \rho(\lambda) d\lambda.
\ee
Since $N(E)$ jumps by one at the position of each eigenvalue,  it is also
know as the staircase function.

Typically, the spectral density varies little over the  scale
of the  average level spacing. Therefore the average spectral density
can be obtained by locally averaging over an interval much larger than 
the average level spacing. Since $\langle\rho(\lambda) \rangle$ depends on
the specific properties of the system one would like to eliminate
this dependence. 
This is achieved by the so-called
unfolding procedure. The unfolded spectrum is given by
\be
\lambda_k^{\rm unf} = \int_0^{\lambda_k}  \langle\rho(\lambda) \rangle
d \lambda.
\ee
One can easily verify that the average spacing of the unfolded sequence
is equal to unity. 

The statistical analysis of the unfolded eigenvalues can be performed by 
a variety of statistics.
The simplest statistic is the so called nearest neighbor
spacing distribution. It is denoted by $P(S)$ and is 
just a histogram of neighboring unfolded 
levels. A second class of statistics is obtained from counting the number
of levels in  $q$ intervals along the spectrum
that contains $n$ levels on average.
If the number of levels in consecutive intervals is
given by $n_k$, we can define the moments
\be
M_p(n) = \frac 1q\sum_{k=1}^q n_k^p.
\ee
Of course, we have that $M_1(n) \rightarrow  n$ for a large sample.
The number variance is defined by 
\be
\Sigma^2(n) = M_2(n) - n^2.
\ee
Similarly, one can define higher order cumulants. 
 
Another frequently used statistic is the $\Delta_3$ statistic which is
defined as
\be
\Delta_3(L) = \frac 2{L^4} \int_0^L (L^3 - 2L^2r + r^3) \Sigma^2(r) dr.
\ee
It has the property that quadratic functions are projected to zero
by the kernel of this
integral operator. Generally, this statistic is much smoother than
the number variance 
and for that reason it has been used widely in the
literature.

\vspace*{0.2cm}
\noindent
{\it Exercise.} Calculate the number variance and the $\Delta_3$ statistic
for a picket fence spectrum, i.e. a spectrum of equally spaced eigenvalues.
\vspace*{0.2cm}

\subsection{Statistics for uncorrelated eigenvalues}

Let us calculate $P(S)$ and $\Sigma_2(n)$ for independently distributed
eigenvalues with average level spacing equal to unity.
Then $P(S) dS$ is the probability that there are no eigenvalues inside
the interval $[0,S]$ and one eigenvalue in $[S, S+dS]$. Dividing the
first interval in $n$ equal pieces we find
\be
P(S) dS = \left (1- \frac Sn \right)^n DS \rightarrow e^{-S} d S
\quad {\rm for} \quad n \to \infty.
\ee

The number variance can be expressed as
\be
\Sigma^2(n) &=& \langle \int_0^n \rho(\lambda) d\lambda 
\int_0^n \rho(\lambda') d\lambda' \rangle - n^2,\nonumber \\
&=&  \int_0^n \int_0^n d\lambda d\lambda'
\langle \sum_{k,l} \delta(\lambda-\lambda_k)
 \delta(\lambda'-\lambda_l)\rangle -n^2,\nn \\
\ee
where the $\lambda_k$ are the unfolded eigenvalues.
The average factorizes for different eigenvalues. For a finite
number of levels we find
\be
\Sigma^2(n) 
&=&  \int_0^n \int_0^n d\lambda d\lambda'[
\delta(\lambda - \lambda') + \frac{N(N-1)}{N^2}
 \sum_{k,l} \langle \delta(\lambda-\lambda_k)\rangle
\langle \delta(\lambda'-\lambda_l)\rangle]  -n^2.\nn \\
\ee
The first term in the brackets is the diagonal contribution with $k =l$, and
in the second term we have corrected the absence of the diagonal
terms in the sum by a factor $N(N-1)/N^2$. 
After integration over $\lambda$ and $\lambda'$ we obtain
\be
\Sigma_2(n) = n -\frac{n^2}N,
\ee
where $N$ is the total number of eigenvalues.

The $\Delta_3$-statistic is obtained by a simple integration resulting in
\be
\Delta_3(L) = \frac{L}{15} + O(1/N).
\ee

However, if one calculates $P(S)$ and $\Sigma_2(n)$ for  spectra
of interacting systems
one finds different results. In particular, one 
notices a suppression of small spacings and a strongly reduced number
variance for large $n$. These properties are characteristic for spectra
of many complex systems and are known as level repulsion and spectral
rigidity, respectively.

\subsection{Correlation functions}

The number variance can be expressed in terms of the two-point correlation
function of the unfolded eigenvalues. The two-point correlation function
of the not necessarily unfolded eigenvalues is defined by
\be 
\rho_2(\lambda, \lambda') = \langle \rho(\lambda) \rho(\lambda')\rangle
             -  \langle  \rho(\lambda) \rangle
               \langle  \rho(\lambda') \rangle.
\ee
The nearest neighbor spacing distribution cannot be expressed in terms
a two-point
correlator only. 
The reason is that it measures the probability that none of the
other eigenvalues are inside the interval $[0,S]$.

The correlation function $\rho_2(\lambda,\lambda')$ includes a term
in which the eigenvalues are equal, and can thus be decomposed as
\be
\rho_2(\lambda, \lambda') = \delta(\lambda - \lambda') 
\langle\rho(\lambda) \rangle + R_2(\lambda, \lambda').
\ee
The two-point correlation function satisfies the sum rule
\be
\int d\lambda \rho_2(\lambda, \lambda') = 0,
\ee
where the integral is over the complete spectrum.

In the literature one also frequently uses the quantity $Y_2(\lambda, 
\lambda')$ for the two-point correlation function of the 
unfolded eigenvalues \cite{Mehta}
\be
Y_2(\lambda,\lambda') = - R_2(\lambda, \lambda'),
\ee
where the minus sign is conventional.
If a smoothened average spectral density exists, it is natural
to expect that $Y_2$ is translational invariant, i.e.,
\be
Y_2(\lambda,\lambda') = Y_2(\lambda-\lambda').
\ee
In that case, the number variance can be expressed as
\be
\Sigma^2(L) = L - \int_0^L (L-r) Y_2(r) dr.
\ee

\section{The Supersymmetric Method in Random Matrix Theory}
\label{super}

The supersymmetric has been introduced in Random Matrix Theory
and the theory of disordered systems to calculate quenched
averages according to \cite{Efetov,EfetovBook,brezin-sup}
\be
\left .\left \langle {\rm Tr} \frac 1{z+D} \right \rangle
=\del_z \left \langle \frac {\det(D+z)}{\det(D+z')} \right \rangle
\right |_{z'=z}.
\ee
The determinant can be expressed as a fermionic integral and the 
inverse determinant as a bosonic integral. For $z'=z$ the generating function
has an exact supersymmetry. In this lecture we will discuss in detail
the supersymmetric method. We start with the introduction of 
Grassmann variables and Grassmann integration. Our main objects are
graded vectors and graded matrices that consist of both bosonic and
fermionic variables. As application we will discuss the one-point function
and the two-point function of the Gaussian Unitary Ensemble. For more details
we refer to the book by Efetov \cite{EfetovBook} and \cite{VWZ}
where the class $\beta=1$ is worked out in detail.

\subsection{Definitions}
\label{super1}

Grassmann variables are anti-commuting variables
\be
\{ \chi_k, \chi_l\}= 0,
\ee
which are necessarily nilpotent, i.e. $\chi^2_k = 0$. 
One can introduce two types of conjugation, namely, conjugation of the
first kind with the property that
\be
\chi^{**} = \chi,
\ee
and conjugation of the second kind with the property
\be
\chi^{**} = -\chi.
\ee
Below it will become clear that it is often advantageous to use conjugation
of the second kind.

The integration
over Grassmann variables is defined by
\be
\int d\chi = 0, \qquad \int \chi d\chi = 1,
\ee
 and thus more resembles  a differentiation than an integration.  
As an example, consider the integral
\be
\int d\chi d \chi^* e^{-\chi a \chi^*} = \int d\chi d \chi^* (1 - 
{\chi a \chi^*}) = a.
\ee
Notice that also the differentials $d\chi$ are anti-commuting elements.

Next let us make a change of integration variables, $\xi = a \chi$.
Then
\be
1= \int \xi d \xi = \int a \chi d(a \chi),
\ee
so that we have the relation
\be
d(a \chi) = \frac 1a d\chi.
\ee
For a general transformation given by $\xi_k = A_{kl} \chi_l$ we have that
\be
d\xi_1 \cdots d\xi_n = \frac 1{\det A} d\chi_1 \cdots \chi_n.
\ee

\vspace*{0.2cm}
\noindent {\it Exercise.} Prove this relation.
\vspace*{0.2cm}

An important class of integrals are Gaussian Grassmannian integrals. 
Let us consider
\be
I\equiv \int d\chi_1 \cdots d \chi_n d\chi_1^* \cdots d \chi_n^* 
e^{-\chi_k A_{kl}\chi_l^*}. 
\ee
Because of the nil-potency and the definition of the Grassmann integration
we obtain
\be
I\equiv \int d\chi_1 \cdots d \chi_n d\chi_1^* \cdots d \chi_n^* 
\frac 1{n!}(-{\chi_k A_{kl}\chi_l^*})^n.
\ee
By renaming the integration variables one easily finds that 
$n!$ of the terms contributing to the integral are the same. We thus find
\be
I = \int  d\chi_1 \cdots d \chi_n d\chi_1^* \cdots d \chi_n^* 
(-1)^n\sum_{\pi} 
\chi_1 A_{1\pi(1)}\chi_{\pi(1)}^* \cdots\chi_n A_{n\pi(n)}\chi_{\pi(n)}^*  .
\ee
where the sum is over all permutations of $1, \cdots, n$. By rearranging 
the differential it easily follows that the result of
Grassmann integrations is exactly the sign of the permutation, and
\be
I = \det A.
\ee
This should be compared to a bosonic integral 
\be
\frac 1{(-2\pi i)^n}\int d\phi_1 
\cdots d\phi_n d\phi_1^* \cdots d\phi_n^* e^{-\phi_k 
A_{kl} \phi_l^*} = \frac 1{\det A} .
\ee

\vspace*{0.2cm}
\noindent {\it Exercise.} Prove this result.
\vspace*{0.2cm}

An essential difference between fermionic and bosonic integrals is that 
fermionic integrals are always convergent, whereas bosonic integrals
only exist for specific integration contours.

\subsection{Graded vectors and Graded Matrices}
\label{super2}

An $(m|n)$ graded vector is defined by
\be
\phi_k \equiv \left ( \begin{array} {c} S_1 \\ \vdots\\ S_m \\ \chi_1 \\
\vdots \\ \chi_n \end{array} \right ).
\ee
The matrices acting on these vectors have the structure
\be
M = \left ( \begin{array}{cc} a &\sigma \\ \rho & b \end{array} \right ),
\label{typmat}
\ee
where $a$ and $b$ are matrices with commuting entries and $\rho$ and
$\sigma$ are matrices with anti-commuting entries. The block structure
is such that $a$ is an $m\times m$ matrix and $b$ is an $n\times n$ matrix.

The trace of a graded matrix is defined by
\be
{\rm Trg} M = {\rm Tr } a - {\rm Tr} b.
\label{trg}
\ee
A graded trace of a product of graded matrices is invariant under cyclic
permutations.  

The graded determinant is defined by the relation
\be
{\rm detg} M = \exp {\rm Trg} \log M=\exp {\rm Trg} 
\log \left [ \left ( \begin{array}{cc} a& 0 \\ 0 & b \end{array} \right )
[1 +\left ( \begin{array}{cc} 0& a^{-1} \sigma \\  b^{-1} \rho & 0
\end{array} \right )]\right ].
\ee
The logarithm can be expanded in a power series and only the even
powers contribute to the graded trace. Resumming the power series again
into a logarithm, one finds
\be
{\rm detg} M = \frac{ \det (a - \sigma b^{-1} \rho )}{\det b}.
\label{detg}
\ee 

\vspace*{0.2cm}
\noindent{\it Exercise.} Show that
${\rm detg} (AB) = {\rm detg}\, A {\rm detg} B$.
\vspace*{0.2cm}

The transpose of the matrix $M$ is defined by
\be
M^T = \left ( \begin{array}{cc} a^T & \rho^T \\ -\sigma^T & B^T
\end{array} \right ).
\ee
Because of the additional minus sign we have the relation
\be
(M_1 M_2)^T = M_2^T M_1^T.
\ee
The Hermitian conjugate is defined in the usual way
\be
M^\dagger = M^{*T}.
\ee
A super-matrix is Hermitian if $ M^\dagger = M$. 
For a scalar product the Hermitian conjugate satisfies the usual relation
\be
(\phi_1, M \phi_2) = (M^\dagger \phi_1, \phi_2),
\label{hconj}
\ee
where we have used the conjugation of the second kind.

The eigenvalues of a supermatrix are defined by
\be
M \phi_k = \lambda_k \phi_k.
\ee
For two different eigenvectors of a super-Hermitian matrix $M$ we have
\be
(\phi^k, M \phi^l) = (M \phi^k, \phi^l).
\ee
This can be rewritten as
\be
(\lambda_k - \lambda_l) (\phi_k, \phi_l) = 0,
\ee
so that the eigenvectors corresponding to different eigenvalues are
orthogonal. As a consequence, a super-Hermitian matrix can be diagonalized
by a super-unitary matrix, where a super-unitary matrix is defined in the
usual way
\be
U^\dagger U = 1.
\ee

The eigenvalues of a graded matrix with block structure (\ref{Qmat})
are given by the zeros and the
poles of the secular equation
\be
{\rm detg} (Q- \lambda)  = \frac{\det(a-\lambda -\sigma (b-\lambda)^{-1} \rho)}
{\det(b -\lambda)}= 0.
\label{zerosec}
\ee
The poles of
the secular equation can be obtained from the equation
\be
{\rm detg} \frac 1{(Q- \lambda)}  =\frac{\det(b-\lambda -
\rho (a-\lambda)^{-1} \sigma)}
{\det(a -\lambda)}= 0.
\label{polesec}
\ee
As an illustration, it is suggested to the reader to
calculate the eigenvalues of a diagonal matrix. 

The ordinary parts of the eigenvalues are obtained by putting $\sigma$
and $\rho$ equal to zero and are thus given by the eigenvalues of
$a$ and $b$. The eigenvalues with ordinary part equal to one of the
eigenvalues of $a$ are given by the solutions of eq. (\ref{zerosec}), and the
eigenvalues with ordinary part equal to one of the eigenvalues of $b$
are given by the solutions of eq. (\ref{polesec}). Notice that 
eq. (\ref{zerosec}) is singular for $\lambda$ equal to one of the eigenvalues
of $b$. 

\vspace*{0.2cm}
\noindent{\it Exercise.} Show that the eigenvalues of
 a $(1|1) $ super-matrix 
\be
\mat a& \sigma \\ \rho & b \emat
\ee
 are given by
\be
\lambda_1 &=& a + \frac {\sigma \rho}{a-b}, \nn \\
\lambda_2 &=& b + \frac {\sigma \rho}{a-b}.
\ee
\vspace*{0.2cm}

Next, let us calculate the following Gaussian graded integral
\be
I \equiv \frac 1{(-2\pi i)^m} \int d\phi d\phi^* e^{- \phi_k^* M_{kl} \phi_l},
\ee
where, for reasons of convergence, the matrix $M$ is a super-Hermitian
matrix of the form (\ref{typmat}). 
In the block notation introduced earlier in this chapter, the
exponent can be written as
\be
\phi_k^* M_{kl} \phi_l &=& S^\dagger a S + S^\dagger \sigma \chi
+ \chi^\dagger \rho S + \chi^\dagger b \chi\nonumber\\
&=& (S^\dagger + \chi^\dagger \rho a^{-1})a(S + a^{-1}\sigma \chi)
+\chi^\dagger (b-\rho a^{-1} \sigma ) \chi.
\ee
After shifting the bosonic integration variables, the integral factorizes
into a bosonic and fermionic piece which we already know how to do. 
The result is
\be
I = \frac{\det (b-\rho a^{-1} \sigma  )}{\det a} = \frac {\det b}
{\det (a - \sigma b^{-1} \rho)} = \frac 1{{\rm detg } M}.
\ee
The second identity follows by writing the determinant as an exponent of the
trace of a logarithm, and perform a cyclic permutation of the graded trace. 
The minus sign resulting in the inverse determinant comes from the 
anti-commutation of the Grassmann variables.

It is now clear why the graded trace was introduced as in (\ref{trg}):
A graded Gaussian integral  is given by the inverse graded
determinant. 

\vspace*{0.2cm}\noindent{\it Exercise}.
Calculate the superintegrals for the case that the Grassmannian blocks
of $M$ are equal to zero.
\vspace*{0.2cm}

Finally, a work of caution about notation. In the physics literature ``graded''
and ``super'' are used as synonyms. For example, graded vectors and
supervectors, graded matrices and supermatrices, etc. are synonyms. The same 
is true for abbreviations. For example, ${\rm Str} 
\equiv {\rm Trg}$, ${\rm Sdet} \equiv {\rm detg}$.
We will use both notations interchangeably.
In quantum field theory, supersymmetry is used for a Grassmannian extension
of the Poincar\'e group, and is different from the use of supersymmetry in
these lectures.

\section{Integration Theorems}
\label{integ}

If an integrand is invariant under super-unitary transformations, there
exist a number of powerful integration theorems. In character they
are comparable to a complex contour integration.

\subsection{The Parisi-Sourlas reduction}
\label{integ1}

Let us first consider the simplest case involving the graded vectors 

\be
p = \left ( \begin{array}{c} a \\ \theta \end{array} \right ),
\qquad  p^\dagger = \left (  a^*,  \theta^* \right  ).
\ee
Let $F(p, p^\dagger)$ be an invariant function, i.e.,
\be
F(Up, U^\dagger p^\dagger) =  F(p, p^\dagger) ,
\ee
for an arbitrary super-unitary transformation $U$. In this case
the integration theorem states that
\be
\frac 1{2\pi} \int da da^* d \theta d \theta^* F(p, p^\dagger) = 2iF(0,0),
\label{intps}
\ee
which is also known as Parisi-Sourlas dimensional 
reduction \cite{parisi}.
To prove this theorem, we expand $F$ in powers of the Grassmann variables
(see also Appendix L of \cite{VWZ})
\be
 F(p, p^\dagger) = F_{00} + F_{01} \theta^* + F_{10} \theta + F_{11} \theta^*
\theta.
\ee
If we choose 
\be
u = \left ( \begin{array}{cc} 1 & 0 \\0 & e^{i\phi} \end{array} \right ),
\ee
then the r.h.s. is only invariant if $F_{01}= F_{10} = 0$.
By performing an infinitesimal transformation 
\be
u = \left ( \begin{array}{cc} 1 &\alpha \\ \alpha^* & 1 \end{array} \right ),
 \ee
we find that
\be
F_{11} = \frac 1a \partial_{a^*} F_{00}.
\ee

\vspace*{0.2cm}
\noindent{\it Exercise.} Show that $F_{00}$ is a function of $a a^*$.
\vspace*{0.2cm}

After performing the Grassmann integration, the integral (\ref{intps})
is given by
\be
\frac 1{2\pi} \int da da^*  \frac 1a \partial_{a^*} F_{00}.
\ee
In polar coordinates the integral can be rewritten as
\be
\frac {-2i}{2\pi} \int \rho d \rho d\phi\frac 1\rho \partial_\rho F_{00}.
\ee
For a function that is well-behaved  
at infinity we then find that the integral is
given by $2iF(0,0)$.
 
This theorem can be extended to supervectors with an arbitrary, but equal
number of commuting and anticommuting components. We leave it up to the
reader to formulate this theorem.

\subsection{The Wegner Theorem}
\label{integ2}

Next we consider an integration theorem for integrals over 
invariant functions of a supermatrix. This theorem was formulated
and proved in lectures by Wegner 
\cite{Wegnerpr} but appeared in published form six years later
by different authors \cite{Groote}.
In this case the invariant function satisfies the relation
\be
F(Q) = F(U Q U^{-1}),
\ee
where $U$ is a super-unitary matrix. The general form of an invariant function
is thus given by
\be
F(Q) = \sum_k a_k {\rm Trg} Q^k.
\label{fseries}
\ee
Therefore, automatically, $F$ is also invariant under general super-linear
transformations.
The theorem states that
\be
\int DQ F(Q) = cF({\lambda\bf 1}),
\ee
where $\lambda$ is an arbitrary constant 
Notice that this constant is not 
determined for an invariant function.
The constant $c$ depends on the 
size of the matrix $Q$ and the choice of the integration contour. In the
example to be discussed below its value is equal to $i$.

We first prove this theorem 
for the simplest case of a $(1|1)$ matrix.  In this case, any function
of the graded  matrix
\be
Q = \left ( \begin{array}{cc}a &\sigma \\ \rho& b \end{array} \right )
\label{Qmat}
\ee
can be expanded as
\be
F(Q) = F_{00} + \sigma F_{01} + \rho F_{10} + \sigma \rho F_{11}.
\label{fexpand}
 \ee
If we choose 
\be
U= \left ( \begin{array}{cc}e^{i\phi} &0 \\ 0& e^{i\theta} \end{array} 
\right ),
\ee
Then,
\be
U Q U^{-1} =  \left ( \begin{array}{cc}a &\sigma e^{i\phi-i\theta}\\
\rho e^{i\theta-i\phi} & b\end{array} \right ).
\ee
 Invariance of $F(Q)$ then requires that $F_{10} =F_{01} = 0$ in the
expansion (\ref{fexpand}). Next consider the invariance under
\be
U =  \left ( \begin{array}{cc}1 &\omega \\ \zeta& 1 \end{array} 
\right ).
\ee
Then to first order in $\omega$ and $\zeta$
\be
U Q U^{-1} = \left ( \begin{array}{cc}a + \omega \rho + \zeta\sigma 
&\omega(b-a) + \sigma \\ \zeta(a-b) + \rho& b+\omega \rho + \zeta \sigma 
\end{array} \right ).
\label{u11}
\ee
From the invariance of $F$ we then find
\be
 (\omega \rho + \zeta\sigma)(\partial_a F_{00} + \partial_b F_{00})
+(b-a)(\omega\rho +\zeta \sigma)F_{11} =0.
\ee
This results in
\be
F_{11} = \frac {\partial_a F_{00} + \partial_b F_{00}}{a-b}.
\ee

\vspace*{0.2cm}
\noindent {\it Example.} Consider the invariant function $F(Q) = {\rm Trg}
Q^2$. In components we obtain $F(Q) = a^2 - b^2 + 2\sigma \rho$, which
is in agreement with the above general result.
\vspace*{0.2cm}

 After performing the Grassmann integrals we find
\be
\int DQ F(Q) = \frac 1 {2\pi }
\int da db  \frac {\partial_a F_{00} + \partial_b F_{00}}{a-b}.
\ee
This integral is not well defined if $a$ and $b$ are along the same path
in the complex plane. For definiteness let us take $a$ along the real
axis and $b$ along the imaginary axis. In polar coordinates 
\be
a= \rho \cos \phi,\nn \\
b = i \rho\sin \phi.
\ee
Then  
\be
\partial_a F_{00} + \partial_b F_{00} =
(\frac {a-b}\rho \partial_\rho + \frac {a-b}{i\rho^2}\partial_\phi) F_{00}.
\ee
The integral over $\phi$ of the second term gives zero provided that 
the point $\rho = 0$ is suitably regularized. The  integral over $\rho$ of the
first term is the integral of a total derivative. 
For well-behaved functions $F$ we 
thus find
\be
\int DQ F(Q) = iF(0) = i F(\lambda \bf 1),
\label{theorem}
\ee
where the second equality follows from the expansion (\ref{fseries}).

An inductive proof for the general case of this theorem was also given by
Wegner \cite{Wegnerpr}. 
He considered the case of an arbitrary $(m|n)$ matrix $Q$. In that
case an integral over an $(m-n)\times (m-n) $ ordinary matrix remains
in the final result
(for $m>n$). However, for $m = n$, the theorem is exactly as in (\ref{theorem})
but with $Q$ equal to an $(m|n)$ super-matrix. 

The proof of the theorem starts from the observation that an arbitrary $(m|n)$
matrix can be decomposed into the $(1|1)$ super-matrix
\be
\bar Q = \left ( \begin{array}{cc} Q_{11} & Q_{1,n+m}\\Q_{n+m,1} & 
Q_{n+m,n+m} \end{array} \right ),
\ee
the $(1|1)$ super-vectors
\be
P_\alpha = \left ( \begin{array}{c} Q_{1 \alpha} \\ Q_{n+m \alpha}\end{array}
\right ), \qquad \alpha = 2, \cdots n+m-1,
\ee
and
\be
\bar P_\alpha = \left ( Q_{\alpha 1},  Q_{\alpha n+m }
\right ), \qquad \alpha = 2, \cdots n+m-1.
\ee
The remaining matrix elements will be  denoted by $Q_R$. We can thus write
\be
F(Q) = \bar F( \bar Q, P, \bar P, Q_R).
\ee
We now consider the following invariance
\be
\bar F( U\bar QU^{-1},U P,\, \bar PU^{-1}, Q_R) =  
\bar F( \bar Q, P, \bar P, Q_R).
\ee
and define
\be
I(P, \bar P, Q_R) = \int D\bar Q \bar F(\bar Q, P,\bar P, Q_R).
\ee
By using the invariance of the measure one easily obtains that
\be
I(U P, \bar PU^{-1}, Q_R) = I(P, \bar P, Q_R).
\ee
We thus can apply the Parisi-Sourlas theorem to the $P$ and $\bar P$
integrations. Next we can apply Wegner's theorem for $(1|1)$ matrices
to the $\bar Q$ integrations. What remains is an  integral over the
smaller super-matrix $Q_R$, thereby proving the general form of the
theorem.

\subsection{ The Efetov-Zirnbauer theorem}
\label{integ3}

In general, one is interested in super-integrals of a non-supersymmetric
integrand. As an example we work out  
the $(1|1)$ case given by
\be
\int DQ ( a_{00}(Q) + a_{10}(Q) \sigma + a_{01}(Q)\rho + a_{11}(Q) 
\sigma \rho),
\label{ezint}
\ee
where $Q$ is parametrized as in (\ref{Qmat}) and $a_{ij}(Q)$ are 
invariant functions. Because of the general decomposition of
invariant functions, we immediately find that the integrals of the
second and third terms are zero. However, as was observed by Zirnbauer,
there is another way of deriving this result.

The proof by Zirnbauer \cite{Zirnbauer} is based on the 
invariance of the measure. If we apply the transformation
(\ref{u11}) to the $Q$ variables in  the integral
we obtain
\be
\int DQ a_{10}(Q) \sigma \rho = 
\int DQ  a_{10}(Q) (\sigma + \omega (b-a))
(\rho + \zeta(a-b)).
\ee
Since this equation is valid for arbitrary $\omega$,  
we find that the
second term in (\ref{ezint}) integrates to zero. In the same way we find
that the third term in (\ref{Qmat}) integrates to zero.

To better appreciate this theorem let us
consider a more complicated integral involving an invariant function
of $Q^\dagger Q$ with $Q$ the $(1|1)$ supermatrix given in eq. (\ref{Qmat}). 
The matrix $Q^\dagger Q$ is then given by
\be
Q^\dagger Q = \mat a^* a + \rho^* \rho & a^* \sigma + b \rho^* \\
                  -a\sigma^* + b^* \rho & b^*b -\sigma^*\sigma \emat.
\ee
The eigenvalues of this matrix are given by
\be
\lambda_1 &=& a^* a + \rho^* \rho -
\frac{a^*a \sigma^*\sigma + b^*b\rho^*\rho + a^*b^*\sigma\rho 
+ab\sigma^*\rho^*}{a^*a-b^*b} + \frac{(a^*a+b^*b)\sigma^*\sigma+\rho^*\rho}
{(a^*a -b^*b)^2},\nn\\
\lambda_2 &=& \lambda_1 - a^*a +b^*b -\sigma^*\sigma-\rho^*\rho.
\ee
A general invariant function of $Q^\dagger Q$ can therefore be decomposed as
\be
F(Q^\dagger Q) = F_{00} + F_{11} \sigma \rho + F_{11}^* \sigma^* \rho^*
+ F_{20} \sigma^* \sigma + F_{02} \rho^* \rho + 
F_{22} \sigma^*\sigma \rho^*\rho,
\label{QdQ}
\ee
where the $F_{ij}$ are functions of the commuting variables only.
The invariance is given by
\be
Q \rightarrow U Q V,
\ee
where $U$ and $V$ are super-unitary matrices.

Because of this decomposition the super-invariant integral over the
product of an invariant function and an odd number of Grassmann 
variables is trivially zero. According to a theorem due to Zirnbauer 
\cite{Zirnbauer} any
super-invariant integral over the product of an invariant function and
two Grassmann variables also vanishes. The proof proceeds as in our
previous example and in this case we consider the infinitesimal transformation
\be
\mat a& \sigma \\ \rho & b \emat \rightarrow
\mat 1& \alpha \\ \beta & 1 \emat 
\mat a& \sigma \\ \rho & b \emat 
\mat 1& \zeta \\ \omega & 1 \emat.
\ee
and
\be
\mat a& \sigma \\ \rho & b \emat^\dagger \rightarrow
\mat 1& -\zeta \\ -\omega & 1 \emat
\mat a& \sigma \\ \rho & b \emat^\dagger 
\mat 1& -\alpha \\ -\beta & 1 \emat .
\ee
If we apply this transformation to the integrand
\be
\int d Q F(Q^\dagger Q) \sigma\sigma^* \rho \rho^*,
\label{start}
\ee
and use the invariance of $F(Q^\dagger Q) $ and the invariance of the
measure we find that the integral over the product of an invariant function
and two Grassmann variables vanishes. Since this result was also used
in Efetov's calculation \cite{Efetov}, this theorem is known as
the Efetov-Zirnbauer theorem. 

One might think that the same argument can be applied to the integral of
an invariant function. Obviously, because of Wegner's theorem this is
not true.  The reason is that the integral (\ref{start}) diverges in
this case. Typically, the invariant measure diverges if two eigenvalues
coincide. In the next subsection we will analyze this singularity in detail.

\subsection{Efetov-Wegner terms}
\label{integ4}

Let us study the following $(1|1)$ super-invariant integral
\be
I =\frac 1 {\pi} \int da db d\sigma d \rho e^{-{\rm Trg } Q^\dagger Q} ,
\label{gint}
\ee
where 
\be
Q = \mat a & \sigma \\ \rho & ib \emat.
\ee
The additional factor $i$ has been inserted so that the integral 
converges for the $b$-integration along the real axis.
By expanding the Grassmann variables and performing the Gaussian integral
one immediately finds that $I = 1$. Next we choose the eigenvalues
of $Q$ and $\sigma$ and $\rho$ as new integration variables. The Jacobian 
of the transformation from $(\lambda_1, \lambda_2)$ to $(a,b)$ is given by
\be
J = \det \mat 1 + \frac{\sigma\rho}{(a-ib)^2} & \frac {-i\sigma\rho}{(a-ib)^2}
\\ \frac{\sigma\rho}{(a-ib)^2} & i - \frac{i\sigma\rho}{(a-ib)^2} \emat = i.
\ee
We thus find that our integral is given by
\be
I = \int \frac {d\lambda_1 d \lambda_2 d\sigma d\rho}{\pi i}
e^{-\lambda_1^2 + \lambda_2^2}=0,
\ee
where the integral over $\lambda_2$ is along the imaginary axis.
However, because of the Grassmann integration the result of the integral
is zero. Obviously, something went wrong! The only place where we could
have made an error  is at
the point $a=ib$, where the variable transformation becomes singular. 

Let us do the integral by excluding an infinitesimal sphere around
this point from the integration domain. This approach was followed
by Haldane and Zirnbauer \cite{Zirnhal}
and by Gossiaux, Pluhar and Weidenm\"uller \cite{GPW}.
In terms of the original variables this changes the integral
by only an infinitesimal amount. Our integral is thus
given by 
\be
I = \lim_{\epsilon \rightarrow 0} I_\epsilon
\ee
with
\be
I_\epsilon = \int d Q e^{-{\rm Trg } Q^2 } \theta (\sqrt{a^2+b^2} - \epsilon).
\ee 
In terms of our new integration variables we have that
\be
\sqrt {a^2 + b^2} = \sqrt{\lambda_1^2 - \lambda_2^2} -\frac{\sigma\rho}
{\sqrt{\lambda_1^2 -\lambda_2^2}}, 
\ee
and our $\theta$ function is given by
\be
 \theta (\sqrt{a^2+b^2} - \epsilon) = \theta(\sqrt{\lambda_1^2 -\lambda_2^2} 
-\epsilon) -\sigma\rho \frac 1{\sqrt{\lambda_1^2 -\lambda_2^2}}
\delta (\sqrt{\lambda_1^2 -\lambda_2^2}-\epsilon).
\ee 
Additional terms due to the $\delta$-function are known as Efetov-Wegner terms.
If we use polar coordinates for $\lambda_1$ and $\lambda_2/i$, the 
anomalous contribution to our integral can be written as
\be
\frac 1{\pi} \int r dr d\phi d\sigma d\rho
\frac 1r \delta(r - \epsilon) e^{-r^2}\rho\sigma =1,
\ee
which is the correct answer for the Gaussian integral (\ref{gint}. 
In this case the non-anomalous piece is zero. 

In general one is interested in non-superinvariant integrals of the
following         type 
\be
I=\int dQ (f_0(Q) + \rho\sigma  f_1(Q)). 
\ee
with $f_0$ and $f_1$ depending only of the eigenvalues of $Q$.
In terms the new integration variables the integral is given by
\be
I = f_0(0,0) + \frac 1{2\pi i}\int d \lambda_1 d\lambda_2 
f_1(\lambda_1,\lambda_2),  
\ee
This result can also be interpreted in terms of the integration measure
which is now given by
\be
\frac 1{2\pi i} d\lambda_1 d\lambda_2 d\sigma d\rho + 
\frac 1{2\pi i}\frac{\delta(\sqrt{\lambda_1^2-\lambda_2^2})}
{\sqrt{\lambda_1^2-\lambda_2^2}} d\lambda_1 d\lambda_2.
\ee
\section{The Supersymmetric Method of Random Matrix Theory: The One-Point
Function}
\label{sumat}

In this chapter we calculate the one-point function of the GUE by means of the 
supersymmetric method. It is evaluated
starting from the RMT formulation of the problem.

\subsection{ The One-Point Function}
\label{sumat1}

In this subsection we calculate the average resolvent for the Gaussian Unitary
Ensemble by means of the supersymmetric method. This section closely follows
the corresponding section in the article \cite{ieee}. 
The starting point is
\be
\langle{ G(z)}\rangle = \frac 1N\left\langle {\rm Tr }\frac 1{z-H} 
\right\rangle 
= \frac 1N\left . \frac {\partial}{\partial J}\right |_{J=0}
Z(J),
\ee
where the generating function is defined by
\be
Z(J) = \int DH P(H)\frac {\det (z-H+J)}{\det(z-H)},
\ee
and the integral is over the probability distribution of one of the
random matrix ensembles defined in the first chapter.  The
determinants can be expressed in terms of Gaussian integrals,
\be
\frac {\det (z-H+J)}{\det(z-H)} &=& \int d\psi
\exp\biggl( -\sum_{kl}[i\phi_k^* (z-H)_{kl} \phi_l 
+i\chi_k^* (z+J-H)_{kl} \chi_l]\biggr)\:, \nn \\
\ee
where the measure is defined by
\be
d\psi = \prod_{j=1}^N {d\phi_j d\phi^*_j d\chi_j
d\chi^*_j\over {-2\pi i}}\ .
\ee
For convergence the imaginary part of $z$ has to be positive. 

\vspace*{0.2cm}\noindent{\it Exercise.}
Show that that $Z(0) = 1$.
\vspace*{0.2cm}

For the GUE (with Dyson index $\beta = 2$) the probability 
distribution is given by
\be
P(H)DH = {\cal N}e^{-\frac {N}{2} {\rm Tr} H^\dagger H} DH. 
\label{probwd2}
\ee
The Gaussian integrals
over $H$ can be performed trivially, resulting in the generating function 
\be 
Z(J)&=& \int d \psi \exp\Biggl[ -\frac 1{2N} {\rm Trg}  \left (
\begin{array}{cc} \sum_j \phi_j^* \phi_j & \sum_j \chi_j^* \phi_j\\
\sum_j \chi_j \phi_j^*  & \sum_j \chi_j^* \chi_j \end{array} \right )^2
 \nonumber \\ &&\hspace*{18mm}
- i\sum_j (\phi_j^* z \phi_j + \chi_j^*(z+J)\chi_j) \Biggr],
\ee 
where the sums over $j$ run from $1$ to $N$.  
The quartic terms in $\phi$ and
$\chi$ can be expressed as Gaussian integrals by means of a 
  Hubbard-Stratonovitch transformation. This results in
\be
Z(J)&=& \int d \psi d\sigma
\exp\Biggl[ -\frac N2{\rm Trg\:} \sigma^2 - 
i\sum_j\left (\begin{array}{c} \phi_j^* \\ 
\chi_j^* \end{array} \right )^T
(\sigma + \zeta)
 \left (\begin{array}{c} \phi_j \\ \chi_j \end{array} \right )\Biggr],
\ee 
where 
\be
\sigma =
\left (\begin{array}{cc} \sigma_{BB} & \sigma_{BF} \\ \sigma_{FB} & 
i\sigma_{FF} \end{array} \right )
\label{sigma}
\ee
and
\be
\zeta =
\left (\begin{array}{cc} z & 0 \\ 0 & 
z + J \end{array} \right ).
\ee
The variables $\sigma_{BB}$ and $\sigma_{FF}$ are commuting (bosonic) 
variables that range over the real axis. Both $\sigma_{BF}$ and
$\sigma_{FB}$ are Grassmann (fermionic) variables. One easily verifies
that by completing the square and carrying out the $\sigma$-integrations
one recovers the original quartic interaction.

The integrals over the $\phi$ and the $\chi$ variables are now
Gaussian and can be performed trivially. This results in the
$\sigma$-model
\be
Z(J)= \int d\sigma
\exp\left[ -\frac N2{\rm Trg} \sigma^2 - N{\rm Trg } \log (\sigma +
\zeta)\right].
\label{1psigma}
\ee 
For $J=0$ we deal with a super-invariant integral, and, according to
Wegner's theorem, we have $Z(0) = 1$.
By shifting the integration variables according to $\sigma \rightarrow
\sigma -\zeta$ and carrying out the differentiation with respect to
$J$ one easily finds that
\be
\langle{G(z)}\rangle = \langle z-i\sigma_{FF} \rangle.
\ee
In the large $N$ limit, the expectation value of $\sigma_{FF} $ 
follows from a saddle-point 
analysis.
The saddle point equation for $\sigma_{FF}$ is given by
\be
\sigma_{FF} + i z = 1/\sigma_{FF} 
\ee
resulting in the resolvent 
\be
\langle G(z)\rangle = \frac z2 - \frac i2 \sqrt{4 -z^2}.
\label{result}
\ee
Using the relation between the resolvent and the spectral density
we find that the average spectral
density is a semi-circle.

\subsection{Exact calculation of the One-Point Function}
\label{sumat2}

In fact, the integrals in (\ref{1psigma}) can be evaluated exactly
\cite{guhr} instead of performing a saddle point approximation. 
If we parameterize $\sigma $ as
\be
\sigma= \mat a & \alpha \\ \beta & ib\emat ,
\ee
the partition function  can be rewritten as
\be
Z(J) = \frac 1{2\pi} \int  da db d\alpha d \beta 
\frac{ (ib+J+z-\frac{\alpha\beta}{a+z})^N}{(a+z)^N} 
e^{-\frac N2 (a^2 +b^2 + 2\alpha\beta)}.
\ee
By shifting the integration variables and expanding the Grassmann variables,
the integral can be rewritten as
\be
Z(J) =  \frac 1{2\pi }\int da dbd \alpha d\beta \frac{ (ib)^N}{a^N} 
(1 - \frac{N\alpha\beta}{aib}) (1 -N\alpha \beta)
e^{-\frac N2 ((a-z)^2 +(b+iz+iJ)^2)}.
\ee
After differentiating with respect to $J$ and performing the Grassmann 
integrations, the resolvent is given by
\be
G(z) = \frac 1{2\pi } \int da db(z-ib)(\frac{ (ib)^N}{a^N} +
\frac{ {(ib)}^{N-1}}{a^{N+1}}) 
e^{-\frac N2 ((a-z)^2 +(b+iz)^2)}. 
\ee
The integrals factorize into known integrals. In particular
\be
\int_{-\infty}^\infty dx (ix)^n e^{-(x+iz)^2} = \frac {\sqrt \pi}{2^n}
H_n(z).
\ee
We thus find that the result of the $b$ integral is real.
The integral over $a$ is singular. Its calculation was discussed in a  
paper   by Guhr \cite{guhr}. The first observation is that 
 $a$ contains an infinitesimal
increment $i\epsilon$. By successive partial integrations we obtain
\be
\int da \frac 1{a^n} e^{-\frac N2(a-z)^2} = \frac 1{(n-1)!}\int da \frac 1a 
\partial_a^{n-1} e^{-\frac N2(a-z)^2}.
\ee
Using the definition of the Hermite polynomials
\be
\partial_a^{n-1} e^{-\frac N2(a-z)^2} = 
\left ( \frac N2 \right )^{(n-1)/2} 
e^{-\frac N2 (z-a)^2} H_{n-1} ((z-a)\sqrt{N/2})
\ee
and the relation
\be
{\rm Im} \frac 1{a-i\epsilon} = \pi \delta(a),
\ee
we find that
\be
{\rm Im} \int da \frac 1{a^n} e^{-\frac N2(a-z)^2} = \frac \pi{(n-1)!}
\left ( \frac N2 \right )^{(n-1)/2} e^{-\frac N2 z^2} H_{n-1} (z\sqrt{N/2}).
\ee
To obtain the spectral density we use that
\be
\rho(\lambda) = \frac {iN}{2\pi} 
(G(\lambda + i\epsilon) - G(\lambda -i\epsilon)).
\ee

\vspace*{0.2cm}
\noindent{\it Exercise.} Use these results to derive an exact result
for the spectral density in terms of Hermite polynomials.

\section{The Supersymmetric Method of Random Matrix Theory: The Two-Point
Function}
\label{twopo}

In this chapter we discuss the two-point correlation function for the
GUE using the supersymmetric method. As a warm-up exercise we first consider
the replica trick with fermionic replicas and bosonic replicas.  
Contrary to the one-point function  we will obtain the effective partition
function  by using symmetry arguments. 
Both the replica calculations and the supersymmetric calculation
can be found in \cite{critique}. 
Other useful sources are the book by Efetov \cite{EfetovBook},
the papers by Zirnbauer \cite{Zirnbauer,zirnency}, 
the detailed review 
by Zuk \cite{zuk} and the lecture
notes by Fyodorov \cite{fyodorov}.

\subsection{The Replica Trick}
\label{twopo1}

Before embarking on the supersymmetric method, we first study the calculation
of the two-point function by means of the replica trick. In this case 
we have two possibilities for the generating function: a fermionic
generating function or a bosonic generating function.
The fermionic
generating function for the two-point function is defined
by
\be
Z_n(J_1,J_2) = \int dH P(H) {\det}^n (x+J_1+i\epsilon +H)
{\det}^n (y+J_2-i\epsilon +H),
\label{zfer}
\ee
so that 
\be
\langle G(x+i\epsilon) G(y-i\epsilon)\rangle =\lim_{n\to 0} \frac 1{n^2}
\left . 
\partial_{J_1} \right |_{J_1= 0} \left . \partial_{J_2} \right |_{J_2 =0} 
\log Z_n(J_1,J_2)
\label{twopoint}
\ee
and can be represented as a Grassmann integral.

The bosonic
generating function  is defined
by
\be
Z_{-n}(J_1,J_2) = \int dH P(H) {\det}^{-n} (x+J_1+i\epsilon +H)
{\det}^{-n} (y+J_2-i\epsilon +H)
\ee
and can be represented as a bosonic integral. The two-point function
is also given by (\ref{twopoint}).

\subsubsection{The Fermionic Partition Function}
\label{twopo11}

The determinants in the fermionic partition function can be written
as
\be
 {\det}^n (x+J_1+i\epsilon +H)
{\det}^n (y+J_2-i\epsilon +H) =
\int d\chi d\bar\chi \exp [-\bar \chi^k_{\mu} D_{\mu\nu} \chi^k_\nu],
\ee
where
$\chi_k =( \chi_1^1, \cdots, \chi_n^1,\chi_1^2,\cdots,\chi_n^2)$
and $D$ is the block matrix
\be
D = \mat x+J_1+i\epsilon +H & 0 \\   0&y+J_2-i\epsilon +H \emat
\equiv M  + \mat i\epsilon +H & 0 \\  0& -i\epsilon +H  \emat
\ee
For $x = y $, $J_1=J_2$ and $\epsilon = 0$
the symmetry of the generating function is $U(2n)$. One easily verifies
that the resolvent changes
sign when $\epsilon$ crosses the real axis so that 
for $ x=y$ the expectation value
\be
\langle  \chi_1^{k\,*} \chi_1^k  \rangle = - 
\langle  \chi_2^{k\,*} \chi_2^k  \rangle.
\ee
This implies that the $U(2n)$ symmetry is broken spontaneously
to $U(n) \times U(n)$. In the large $N$ limit the massive 
modes can be integrated out and we are left with a theory of
Goldstone modes. The Goldstone fields are in
the coset $Q \in U(2n) /U(n) \times U(n)$ and are coupled to
the microscopic variables as $\bar\chi_k Q_{kl} \chi_l$. Therefore,
they transform as
\be
Q \to U Q U^{-1}
\ee
 under a transformation of the microscopic theory as
\be
\chi_k \to U_{kl} \chi_l\quad {\rm with} \quad U \in U(2n).
\ee
The $U(2n)$ invariance is broken by the mass term. However, invariance
can be recovered if  we  transform 
the mass matrix according to
\be
M \to U M U^{-1}.
\ee
The low energy-effective theory should have the same transformation
properties. To first order in $M$ we can  write down only one invariant term,
\be
{\rm Tr } M Q,
\ee
resulting in the effective partition function
\be
Z_n = \int_{Q \in U(2n)/U(n) \times U(n)} dQ\, e^{iN\Sigma{\rm Tr}MQ}.
\label{partferm}
\ee
Here, $dQ$ is the $U(2n)$ invariant measure to be discussed below.
We will consider only the case that $x$ and $y$ are near the center of the 
spectrum. In particular, we take
$x = \omega/2 $ and $ y = - \omega/2$ so that 
the mass matrix for $J_1=J_2=0$ is equal to
\be 
 M = \frac \omega 2\Sigma_3 \quad {\rm with} \quad \Sigma_3 =
\mat {\bf 1}_n & 0 \\0 & -{\bf 1}_n \emat.
\ee
The constant $\Sigma$ is determined by the requirement
\be
\lim_{n\to 0}
\del_{J_1}|_{J_1=J_2 = 0}\frac 1{nN }\log Z_n  = \langle G(x)\rangle,
\ee
which is equal to $\pi \rho(0)/N$ for $x = 0$.
The partition function (\ref{partferm}) is the exact low energy limit
of (\ref{zfer}) in the thermodynamic
limit for fixed $\omega N$.

We can parameterize $Q$ as,
\be
Q = U \Sigma_3 U^{-1}\quad {\rm with} \quad U = \exp\left[ i \mat
0 & A \\ A^\dagger & 0 \emat \right],
\label{Qpar}
\ee
where $A$ is an arbitrary complex matrix. In a polar decomposition of
$A= v_1^{-1} \theta v_2$  with $v_1 \in U(n)$ and $v_2\in U(n)/U^n(1)$ 
and $\theta$ a
positive definite diagonal matrix, the matrix $U$ is given by
\be
U = \mat v_1^{-1} & 0 \\ 0 & v_2^{-1} \emat
    \mat \lambda & i\mu \\i\mu & \lambda \emat 
    \mat v_1 & 0 \\ 0 & v_2 \emat \quad {\rm with }\quad
    \mu = \sin\theta, \,\,
    \lambda =\cos\theta.
\ee
Therefore, $Q$ can be written as
\be
Q =\mat v_1^{-1} & 0 \\ 0 & v_2^{-1} \emat
    \mat \lambda^2 -\mu^2 & 2i\mu \lambda \\-2i\mu \lambda& \lambda^2-\mu^2 
\emat     \mat v_1 & 0 \\ 0 & v_2\emat
\ee

The integration measure in (\ref{partferm}) is given by
\be
\prod_{k,l} [\delta U_{12}]_{kl}[\delta U_{21}]_{kl}\,, \quad {\rm with}\quad
\delta U = U^{-1} dU.
\ee 
The Jacobian from $[\delta U_{12}]_{kl}$ and
$[\delta U_{21}]_{kl}$ to the variables $[\delta v_1]_{kl}$ and
$[\delta v_2]_{kl}$  is given by the determinant (for $k\ne l$)
\be
\det\mat -i\lambda_k \mu_l & i \mu_k\lambda_l \\ i\mu_k\lambda_l &
-i\lambda_k \mu_l \emat = \lambda_l^2 - \lambda_k^2.
\label{gue-measure-lam}
\ee
For $k =l$ the transformation from $[\delta_{12}]_{kk}$ and
$[\delta_{21}]_{kk}$ to the new variables  $[\delta v_1]_{kk}$
  and $\lambda_k$ so that the Jacobian is given by
\be
\det \mat -i\mu_k \lambda_k &-i/\mu_k \\ i\mu_k\lambda_k & -i /\mu_k
\emat = 2\lambda_k.
\ee
The integration measure for the new variables is therefore given by
\be
\prod_{k,l} [\delta U_{12}]_{kl}[\delta U_{21}]_{kl}
= \prod_k 2\lambda_k \prod_{k<l} (\lambda_k^2 - \lambda_l^2)^2.
\label{gue-measure}
\ee

Using $u_k \equiv \lambda_k^2 -\mu_k^2 $ as new integration variables
we finally obtain
\be
Z_n(x) = \int_{-1}^1 \prod_kd u_k \prod_{k<l}(u_k-u_l)^2  
e^{iN \omega \Sigma \sum_k u_k},
\label{partferm2}
\ee
with the thermodynamic limit taken at fixed
\be
x = N \omega \Sigma.
\ee

The partition (\ref{partferm2}) can be written as a $\tau$ function. The first
step is to expand the Vandermonde determinant
\be
Z_n(x) =\int_{-1}^1 \prod_kd u_k \sum_{\sigma\pi}{\rm sg}(\sigma\pi)
u_1^{\sigma(1) +\pi(1)} \cdots u_n^{\sigma(n) +\pi(n)}   
e^{ix \sum_k u_k},
\ee
where $\sigma$ and $\pi$ are permutations of $\{1,\cdots, n\}$, and sg
denotes the sign of the permutation.
Next we use that
\be
\int_{-1}^1 \prod_k d u_k u_k^a e^{ix u_k} = (\del_{ix})^a Z_1(x),
\ee
which results in 
\be
Z_n(x) = n! [\det (\del_{ix})^{i+j} Z_1(x)]_{0 \le i,j \le n-1}.
\ee

\subsubsection{The Bosonic Partition Function}
\label{twopo12}

The treatment of the bosonic partition function is quite similar to the
fermionic partition function. Now
the determinants in the  partition function can be written
as
\be
 {\det}^{-n} (x+J_1+i\epsilon +H)
{\det}^{-n} (y+J_2-i\epsilon +H) =\frac 1{(2\pi)^{2Nn}}
\int d\phi_1 d\phi_2 \exp[ i\phi^{*\, k}_{\mu} D_{\mu\nu} \phi^k_\nu],
\nn \\
\ee 
where
$\phi_k =( \phi_1^1, \cdots, \phi_1^n,\phi_2^1,\cdots,\phi_2^n)$
and $D$ is the block matrix
\be
D = \mat x+J_1+i\epsilon + H & 0 \\ 0&  -y-J_2+i\epsilon -H  \emat
\equiv M  + \mat i\epsilon +H & 0 \\  0& i\epsilon -H  \emat.
\ee
The essential difference with the fermionic case is that 
we have to make sure that the integrals are convergent. This is the
reason that the sign  of the lower right block of $D$ has been
reversed.
For $x = y $, $J_1=J_2$ and $\epsilon = 0$
the symmetry of the generating function is not  $U(2n)$ but rather
$U(n,n)$. As in the fermionic case, this symmetry is broken
spontaneously 
to $U(n) \times U(n)$. In this case, we have that for $x= y$
\be
\langle\sum_k \phi_1^{k\,*} \phi_1^{k} \rangle= +
\langle\sum_k \phi_2^{k\,*} \phi_2^{k} \rangle .
\ee
In the large $N$ limit the massive modes can be 
integrated out by a leading order saddle point approximation 
and we are left with a theory of
Goldstone modes. The Goldstone fields are in
the coset $Q \in U(n,n) /U(n) \times U(n)$ and transform
as
\be
Q \to U Q U^{-1}
\ee
 under a transformation of the microscopic theory as
\be
\phi_k \to U_{kl} \phi_l,
\ee
where $U\in U(n,n)$ satisfies
\be
U^\dagger \Sigma_3 U = \Sigma_3.
\ee
In this case the Goldstone modes couple to the microscopic variables
as $\phi^*_k \Sigma_3 Q_{kl} \phi_l$.
The $U(n,n)$ invariance of 
the microscopic theory is broken by the mass term. However, this invariance
is recovered if we transform the mass matrix by
\be
M \to U M U^{-1}.
\ee
The low energy-effective theory should have the same transformation
properties. To first order in $M$ we can only write down the term
\be
{\rm Tr } M Q,
\ee
resulting in the effective partition function
\be
Z_{-n} = \int_{Q \in U(n,n)/U(n) \times U(n)} dQ e^{iN\Sigma{\rm Tr}MQ}.
\label{partferm3}
\ee
The mass matrix is defined as in the fermionic case and the constant
$\Sigma$ is also determined by the requirement
\be
\lim_{n\to 0}
-\frac 1{Nn}  \del_{J_1}|_{J_1=J_2 = 0}\log Z_{-n}(x=0) = \langle G(x=0)\rangle
= \pi \rho(0)/N .
\ee
In the second identity we have used that ${\rm Im}(x)>0$.
We can parameterize $Q$ as
\be
Q = U \Sigma_3 U^{-1}\quad {\rm with} \quad U = \exp  \mat
0 & A \\ A^\dagger & 0 \emat,
\label{Qpar2}
\ee
where $A$ is an arbitrary complex matrix.
Notice that there is no factor $i$ in the exponent in the definition
of $U$.
 In a polar decomposition of
$A= v_1^{-1} \theta v_2$  with $v_1 \in U(n)$ and $v_2\in U(n)/U^n(1)$ 
and $\mu$ a
positive definite diagonal matrix, the matrix $U$ is given by
\be
U = \mat v_1^{-1} & 0 \\ 0 & v_2^{-1} \emat
    \mat \lambda & \mu \\\mu & \lambda \emat 
    \mat v_1 & 0 \\ 0 & v_2 \emat \quad {\rm with } \quad 
\mu = \sinh\theta \quad {\rm and } \quad
    \lambda =\cosh\theta\nn .\\
\ee
Therefore, $Q$ can be written as
\be
Q =\mat v_1^{-1} & 0 \\ 0 & v_2^{-1} \emat
    \mat \lambda^2 +\mu^2 & 2\mu \lambda \\-2\mu \lambda& -\lambda^2-\mu^2 
\emat     \mat v_1 & 0 \\ 0 & v_2 \emat.
\ee

Also in this case the integration measure is given by
(\ref{gue-measure}).
The Jacobian from $[\delta U_{12}]_{kl}$ and
$[\delta U_{21}]_{kl}$ to the variables $[\delta v_1]_{kl}$ and
$[\delta v_1]_{kl}$  is given by the determinant (for $k\ne l$)
\be
\det\mat -\lambda_k \mu_l &  \mu_k\lambda_l \\ \mu_k\lambda_l &
-\lambda_k \mu_l \emat = \lambda_l^2 - \lambda_k^2.
\ee
For $k =l$ the new variables are $[\delta v_1]_{kk}$
  and $\lambda_k$ so that the Jacobian is given by
\be
\det \mat \mu_k \lambda_k &1/\mu_k \\ -\mu_k\lambda_k & 1 /\mu_k
\emat = 2\lambda_k.
\ee
The integration measure is therefore also given by (\ref{gue-measure}).

Using $u_k \equiv \lambda_k^2 +\mu_k^2 $ as new integration variables
we finally obtain
\be
Z_{-n}(x) = \int_{1}^\infty \prod_kd u_k \prod_{k<l}(u_k-u_l)^2  
e^{ix \sum_k u_k}.
\label{zbos}
\ee

The bosonic partition function can also be written as a $\tau$-function.
By expanding the Vandermonde determinant we can express
this generating function as a determinant of derivatives
\be
Z_{-n}(x) = n! [\det (\del_{ix})^{i+j} Z_{-1}]_{0 \le i,j \le n-1}.
\ee

\subsection{The Supersymmetric Generating Function}
\label{twopo2}

Next we discuss the supersymmetric generation function for 
the two-point function defined by 
\be
Z(J_1,J_2) = \int dH P(H)
\frac{\det(x +J_1+i\epsilon +H)}{\det(x -J_1+i\epsilon +H)}
\frac{\det(y +J_2-i\epsilon +H)}{\det(y-J_2- i\epsilon +H)}.
\label{gen0}
\ee
Here, $\epsilon$ is an infinitesimal increment, and $P(H)$ is the Gaussian 
probability distribution for the unitary ensemble. The correlation
function of two resolvents is given by
\be
\langle G(x+i\epsilon) G(y-i\epsilon)\rangle =    \frac 1{4N^2}
 \left . 
\partial_{J_1} \right |_{J_1= 0} \left . \partial_{J_2} \right |_{J_2 =0} 
Z(J_1,J_2).
\ee
The generating function can be written in terms of Gaussian integrals
\be
Z(J_1, J_2) &=& \int dH P(H) 
\int d\psi d\psi^* e^{i\phi_1^* (x+J_1+i\epsilon +H) \phi_1
+i\chi_1^* (x-J_1+i\epsilon +H)\chi_1}
\nn \\ && \times  e^{-i\phi_2^*(y+J_2-i\epsilon +H) \phi_2
+i\chi_2^*(y-J_2 -i\epsilon +H) \chi_2},
\label{gen1}
\ee
where the vector $\psi$ is defined by
\be
\psi = \vect \phi_1 \\ \chi_1 \\ \phi_2 \\ \chi_2 \evect,
\ee
and
\be
d\psi = \prod_{\mu=1}^N \frac{d\chi_{1\,\mu}d\chi_{2\,\mu}
d\phi_{1\,\mu}d\phi_{2\,\mu}}{\pi}.
\ee
The matrix $D$ is defined by
\be
 D &=&\matf x+J_1+i\epsilon +H & 0 & 0& 0 \\
0 & x-J_1+i\epsilon +H & 0 & 0 \\
0 & 0 & -[y+J_2-i\epsilon +H] & 0\\
0 & 0 & 0 & y-J_2-i\epsilon +H \ematf\nn \\
&\equiv& LM + \matf H & 0 & 0& 0 \\ 0 & H & 0 & 0 \\
                 0 & 0 & -H &0\\ 0 & 0 & 0  & H \ematf
\ee

Two widely used matrices that enter in the generating function are 
the  matrix $L$, which breaks the symmetry between
the 1-space and the 2-space, and the matrix 
$k$ which breaks the supersymmetry.
They are defined by
\be
L = \matf 1 & 0 &0 & 0 \\ 0 & 1 & 0 & 0 \\ 0 & 0 & -1 & 0 \\ 0 & 0 & 0 & 1 
\ematf, \qquad
k = \matf 1 & 0 &0 & 0 \\ 0 & -1 & 0 & 0 \\ 0 & 0 & 1 & 0 \\ 0 & 0 & 0 & -1
\ematf
\equiv \mat k_2 & 0 \\ 0 & k_2 \emat .
\label{kL}
\ee
As in the previous sections we introduce the energy difference
\be
\omega = x-y.
\ee

For $\omega = 0$ and $J_1 =J_2 = 0$ the
 partition function (\ref{gen0}) is invariant under the transformations
\be
\psi \rightarrow U \psi, \qquad \psi^* \rightarrow \psi^*  U^\dagger
\label{symgl}
\ee
with
\be
U^\dagger L U = L .
\label{sudefine}
\ee
This group of superunitary transformations is known as
 $U(1,1|2 )$.  This symmetry is broken explicitly by 
the mass term $\psi^\dagger M \psi$
 and spontaneously by the expectation value of
\be
\langle \psi_i^*  L_{ii}\psi_i \rangle =  N\Sigma {T_3}_{ii}.
\label{condensate}
\ee
Here, $T_3$ is defined by
\be
T_3 = 
\matf 1 & 0 &0 & 0 \\ 0 & 1 & 0 & 0 \\ 0 & 0 & -1 & 0 \\ 0 & 0 & 0 & -1 
\ematf .
\ee
This result follows from the same analysis as in the case of 
bosonic or fermionic replicas.
The symmetry of the partition function is thus broken spontaneously
to $U(1|1)\times U(1|1)$.
The Goldstone manifold is given $U(1,1/2)/U(1|1) \times U(1|1)$
and can be parameterized as 
\be
Q=U T_3 U^{-1},
\label{supergoldstone}
\ee
with $U \in U(1,1|2)$. The symmetry transformation (\ref{symgl}) becomes
a symmetry of the microscopic partition function if we also transform the
mass term by
\be
M \to U M U^{-1}.
\ee 

\vspace*{0.2cm} \noindent {\it Example.} The mass term is given by $\psi^* L M
\psi$. Under $\psi \to U \psi$ it transforms as
\be
\psi^* U^\dagger L M U\psi = \psi^* U^\dagger L U U^{-1} M U\psi.
\ee 
Since $U^\dagger LU = L$ we recover invariance if $M \to U M U^{-1}$.
\vspace*{0.2cm}

Because the microscopic fields and the Goldstone fields are coupled
as $\psi^*L Q \psi$, 
the Goldstone fields transforms as
\be
Q \to U Q U^{-1}.
\ee
To lowest order in M we thus can write down only one nontrivial term
 with the same transformation 
properties and the microscopic generating function:
\be 
{\rm Trg} M Q.
\ee
Notice that  $Q^2 =1$.

As before we will study the two-point function in the center of the
spectrum. For zero sources the mass matrix is given by
\be
M = \frac \omega 2 T_3.
\ee
We are interested in the thermodynamic limit at fixed $\omega N$. 
 To leading order in $1/N$ the partition function
is given by the integral over the saddle-point manifold,
\be
Z(J_1,J_2) = \int_{Q \in U(1,1|2)/U(1|1)\times U(1|1)} 
dQ e^{i\frac{ \omega N}2 \Sigma {\rm Trg} (T_3 Q) 
+J_1 iN\Sigma {\rm Trg} k_2 Q_{11} + J_2 iN\Sigma {\rm Trg} k_2 Q_{22}} ,
\ee
with the Goldstone manifold can be parameterized as
as in (\ref{supergoldstone}).
The supersymmetry breaking matrix $k_2$ was introduced in (\ref{kL}) and
is given by
by
\be
k_2 = \mat 1 & 0 \\ 0 & -1 \emat .
\ee 
The subscripts $11$ and $22$ refer to the  projection onto 
the respective blocks of
the $Q$ matrices. By differentiation with respect to the source terms we 
thus obtain
\be
\langle G(x+i\epsilon) G(y-i\epsilon) \rangle = -\frac {\Sigma^2}4
\int_{Q \in U(1,1|2)/U(1|1)\times U(1|1)}  
dQ {\rm Trg} (k Q_{11}){\rm Trg} (k Q_{22})
  e^{\frac{ i\omega N}2 \Sigma {\rm Trg} (T_3 Q)}.\nn \\
\ee
We have seen that the boson-boson block of $Q$ is noncompact. This results
in a mass with a positive real part 
when $Q$ is expanded in the generators of $U$. Because
of the supertrace we have an additional minus sign for the fermion-fermion
term which would result in a negative real part. To compensate this minus
sign we have to include an extra factor $i$ in the generators of $U$ so
that the fermion-fermion block becomes  compact. 
The integral is therefore over the maximum Riemannian 
submanifold of $Gl(2|2)/Gl(1|1)
\times Gl(1|1)$ and the measure is the invariant measure of the unbroken
group \cite{class,foundations}.

If $x$ and $y$ would be on the same side of the real axis, the matrix $T_3$
would not appear in the partition function, and the symmetry between the
1-variable and the 2-variables would not be spontaneously broken. In that
case there is no Goldstone manifold and, to leading order in $1/N$ the 
integrals over the 1-variables and 2-variables factorize. Therefore, they
only contribute to the disconnected piece of the correlation function.

\subsection{The Integration Manifold}
\label{twopo3}

In order to identify the integration manifold,  we will be guided by the
principle of convergence. Grassmann integrals are always convergent, and
we will ignore them for the moment. Then fermion-fermion
blocks and the boson-boson blocks decouple and can be treated separately.
This corresponds to the case of one fermionic replica or one bosonic
replica discussed earlier in this chapter. 

In the case of one bosonic replica for $x$ and one for $y$, the
symmetry transformation satisfies
\be
U^\dagger \sigma_3 U = \sigma_3
\ee
with $\sigma_3$ the third Pauli matrix.
If we parameterize $U$ as
\be
U = \mat a & b \\ c & d \emat
\label{umat}
\ee
we obtain the equations
\be
a^*a - b^* b = 1, \nn \\
c^*c - d^* d = -1,\nn \\
a c^* = b d^*, \nn \\
c a^* = d b^*.\nn\\
\ee
Combining these equations we find that $|a| = |d|$ and $|b| = |c|$.
In our partition function, the matrix $U$ only enters in the combination
of the coset $ U \Sigma_3 U^{-1}$. Without loss
of generality we can omit the phase factors of $a$ and $d$. Then we obtain
$b= c^*$. Our result for the parameterization of $U$ is thus given by
\be
U = \mat \sqrt{1 + a^* a}  & a \\ a^*  & \sqrt{1+a^*a} \emat.
\ee

Next we consider the fermion-fermion block which  enters in the 
effective partition function as
\be
e^{i\frac {N\omega}2[{\rm Tr}  Q^{BB} -{\rm Tr}  Q^{FF}]}.
\label{expbbff}
\ee
With some foresight we have chosen the signs of the Hamiltonian
the same in the two fermionic integrals in the generating function.
The symmetry of the fermionic variables is therefore $U(2)$. The Goldstone
manifold is still given by $Q_{FF} = U \sigma_3 U^{-1}$. Because $U$ is
now compact the expansion of $Q_{FF}$ to second order in the generators
of $U$ results in a convergent integral. If we would have started with a 
noncompact formulation also for the fermionic variables the integral over
$Q_{FF}$ would not have been convergent because of the additional minus
sign due to the supertrace. In fact, we would have been forced to 
rotate the integration contour of $Q_{FF}$ so that we again would have
arrived at a compact parameterization of $Q_{FF}$. 

Convergence of the integrals thus forces us  to parameterize $U$ according 
to the compact subgroup of
$Gl(2)$ which is $U(2)$. Therefore, 
$Q_{FF} \in U(2)/U(1)\times U(1)$, which can be parameterized as
\be
\mat \cos \theta & ie^{i\phi} \sin \theta \\ ie^{-i\phi} \sin \theta 
&\cos \theta \emat.
\label{cosu}
\ee
More suggestively this can be written as
\be
U = \mat \sqrt{1-b^* b} & ib \\ ib^* & \sqrt{1-b^* b} \emat, 
\ee
We observe that the parameterization of the fermion-fermion variable is 
structurally the same as the parameterization of the boson-boson variables.
 
Next we consider the Grassmann variables. A matrix $U \in U(1,1|2)$ contains
4 Grassmannian entries. However, in the coset $U$ four of them can be
absorbed by the $U(1|1) \times U(1|1)$ subgroup. Our final result for
the parameterization of $U$ is then given by
\be
U = \mat \sqrt{1 + t_{12} t_{21}} & t_{12} \\ t_{21} & \sqrt{1 + t_{21}t_{12}}
\emat,
\ee
with
\be
t_{12} = \mat  a & \alpha \\ \beta & ib \emat, \qquad
t_{21} = \mat  a^* & \gamma \\ \delta & ib^* \emat,
\ee
and $ \alpha, \, \beta, \, \gamma$ and $\delta $  Grassmann variables.
From (\ref{sudefine}) we find a relation between the Grassmann variables
by conjugation. However, because a Grassmann variable and its
conjugate are independent integration variables, the conjugation relation
can be ignored as has been done in the above parameterization.

\subsection{The Invariant Measure} 
\label{twopo4}

Now that we have found a parameterization of the integration manifold we are 
able to calculate the invariant measure. From measure theory we know that
the invariant measure is given by
\be
{\rm detg}\left ( \frac{\delta U_{12} \delta U_{21}}
{\delta t_{12} \delta t_{21}} \right ) d t_{12} d t_{21}
\ee
with $\delta U = U^{-1} dU$. This measure is invariant under
$U(1,1|2)$ and left invariant under $U(1|1) \times U(1|1)$ 
\cite{foundations}.

Instead of calculating the graded determinant for the general case we hope to 
convince the reader that it is equal to unity by evaluating it for two special
cases. The general case is left as an exercise.

\vspace*{0.2cm}
\noindent  {\it Example}.
We consider the bosonic matrix
\be
U_1 = \mat \sqrt{1 + a^* a}  & a \\ a^*  & \sqrt{1+a^*a} \emat.
\ee
In order to calculate the invariant measure on the space of $U$ matrices
we have to evaluate
\be
U_1^{-1} dU_1 = \mat \frac 12(a^* da - a da^*) &\sqrt{ 1+ a^* a} da - 
\frac {a (a da^* + a^* da)}{2 \sqrt{ 1+ a^* a}}   \\
\sqrt{ 1+ a^* a} da^* - \frac {a^* (ada^* +a^*da)}{2 \sqrt{ 1+ a^* a}}  &
\frac 12(a da^* - a^* da) \emat.\nn \\
\ee
The Jacobian for the transformation from the variables $(U_1^{-1} dU_1)_{12}$
and $(U_1^{-1} dU_1)_{21}$ to the variables $a$ and $a^*$ is given by
the
determinant of the matrix
\be
\mat \sqrt{1+a^* a} - \frac{a^* a}{2 \sqrt{1+a^*a}} & -\frac {a^2}
{2 \sqrt{1+ a^* a}}  \\-\frac {a^{*2}}
{2 \sqrt{1+ a^* a}} &  \sqrt{1+a^*a} - \frac{a^* a}{2 \sqrt{1+a^*a}} \emat,
\ee
which is easily seen to be equal to unity.
The invariant measure  is thus simply given by $da da^*$.

\vspace*{0.2cm}
\noindent {\it Example}. Next we consider a more complicated example
with $U_2$ given by
\be
U_2 = \matf 1 + \frac {\alpha \delta}2 & 0 & 0 & \alpha \\
0 & 1 + \frac{ \beta\gamma}2 & \beta & 0 \\
0 &\gamma & 1+ \frac{ \gamma\beta}2 & 0 \\
\delta & 0 & 0 & 1 + \frac{\delta \alpha}2 \ematf.
\ee
For $U_2^{-1} dU_2 $ we obtain
\be
\matf -\frac 12 (\alpha d \delta + \delta d\alpha) & 0 & 0 & d\alpha\\
0 & -\frac 12 (\gamma d\beta + \beta d\gamma) &d\beta & 0 \\
0 & d\gamma & -\frac 12(\gamma d\beta + \beta d \gamma) & 0\\
d \delta & 0 & 0 & -\frac 12 (\alpha d \delta + \delta d\alpha) \ematf.  
\nn\\
\ee
One easily sees that the Jacobian for the transformation from
the 12 and 21 blocks of $U_2^{-1} dU_2 $  to $\alpha$, $\beta$, $\gamma$ and 
$\delta$ is equal to unity. 
The invariant measure is therefore given by 
$d \alpha d\beta d\gamma d\delta$.

\vspace*{0.2cm}
\noindent {\it Exercise}. Calculate the Berezinian for the general
case and show that its value is equal to unity as well . The invariant measure
is thus given by $d t_{12} d t_{21}$.
\vspace*{0.2cm}
 
In terms of the $t_{12} $ variables the measure
 is flat with  invariant line element  given by
\be
ds^2 = {\rm Trg} d t_{12} d t_{21}.
\ee
This  measure is invariant under the transformations
\be
t_{12} \rightarrow u t_{12} v^{-1} , \qquad t_{21}\rightarrow v t_{21} u^{-1}.
\ee

\subsection{Transformation to Polar Variables.}
\label{twopo5}

Although an explicit calculation of the Grassmann integrations using 
the parameterization of the previous section is possible, the algebra 
of such calculation is quite laborious. For that reason we
introduce new integration variables defined by the polar decomposition
\be
t_{12} = u \mat \mu_1 & 0 \\ 0 & i \mu_2 \emat v^{-1},
\ee
and 
\be
t_{21} = v \mat \mu_1^* & 0 \\ 0 & i \mu_2^* \emat u^{-1},
\ee
with $u$ and $v$ given by
\be
u = \mat 1 + \frac {\alpha \beta} 2 & \alpha \\ \beta & 1 - \frac {\alpha 
\beta} 2 \emat, 
\qquad
v =\mat 1 + \frac {\gamma\delta}2 & \gamma\\ \delta& 1 - \frac {\gamma
\delta} 2 \emat.
\ee
 The matrices $u$ and $v$ satisfy the property that
\be
 u \sigma_3 u = \sigma_3, \qquad v\sigma_3 v = \sigma_3,
\ee
with $\sigma_3$ the third Pauli-matrix. Notice that we have not
imposed a conjugation on the Grassmann variables. This means that 
for the Grassmann variables we have relaxed the condition \cite{class} 
$U^\dagger L U = L$ that defines $U(1,1|2)$. 

To calculate the Berezinian of the polar variables we consider the variations
\be
u^{-1} d t_{12} v &=& u^{-1} d u \mu + d \mu - \mu v^{-1} 
d v,\nn\\
v^{-1} d t_{21} u &=& v^{-1} d v \mu^* + d \mu^* - 
\mu^* u^{-1} d u,\nn\\
\ee
where $\mu$ and $\mu^*$ are diagonal matrices with diagonal elements
$\mu_1$ and $i\mu_2$, and $\mu_1^*$ and $i\mu_2^*$, respectively.
The block of the Berezinian corresponding to differentiation of the 
boson-fermion blocks of $u^{-1} \delta t_{12} v$ and $v^{-1} \delta t_{21} u$ 
with respect to $\mu $ and $\mu^*$ is equal to zero. 
The determinant of the block
corresponding to differentiation of the boson-boson block with respect to
these variables is equal to one. The only nontrivial contribution to the
Berezinian then arises from the variation of the boson-fermion blocks
of $u^{-1} d t_{12}$ and $v^{-1} d t_{21} u$ with respect
to the off-diagonal elements of $u^{-1}du$ and $v^{-1}dv$. The Berezinian
from the $t_{12}$ and $t_{21}$ variables to the 
polar variables is thus given by the inverse determinant of
\be
\matf i \mu_2 & 0 & -\mu_1 & 0 \\
      0   & \mu_1  &0   & -i\mu_2 \\
        -\mu_1^* & 0 & i\mu_2^* & 0 \\
          0 & -i\mu_2^*& 0 & \mu_1^* \ematf.
\ee
The value of the Berezinian is thus equal to
\be
 B = \frac 1{(\mu_1^* \mu_1 + \mu_2^* \mu_2)^2}.
\ee
To derive the final result for the two-point function we also need
\be
{\rm Trg} k_2 Q_{11} &=& \lambda_1 + \lambda_2 + 
2\alpha\beta(\lambda_1 -\lambda_2),\nn\\
{\rm Trg} k_2 Q_{22} &=& -\lambda_1 - \lambda_2 - 2 \gamma\delta 
(\lambda_1 -\lambda_2),
\ee
where $\lambda_1$ and $ \lambda_2$ are the diagonal elements of $Q_{11}$
given by
\be
\lambda_1 = 1 + 2|\mu_1|^2,\nn\\
\lambda_2 = 1 - 2|\mu_2|^2.
\ee
For the exponent in the partition function we obtain
\be
e^{iN\omega \Sigma (\lambda_1 -\lambda_2)}.
\ee 
After performing the angular integrations, the final result for the two-point
correlation function is given by
\be
\langle G(x+i\epsilon) G(y-i \epsilon) \rangle &=& \frac {\Sigma^2}{4} 
\int_1^\infty d \lambda_1 \int_{-1}^1 d\lambda_2
\frac {d\alpha d \beta d \gamma d \delta}
{(\lambda_1 - \lambda_2)^2}e^{iN\omega \Sigma (\lambda_1 -\lambda_2)}
\nonumber \\ &&\times
(\lambda_1 + \lambda_2 + 2\alpha 
\beta(\lambda_1 -\lambda_2))(\lambda_1 + \lambda_2 + 2\gamma
\delta(\lambda_1 -\lambda_2)).\nn\\
\label{correl}
\ee
A factor of $1/(2\pi)^2$ in the integration measure has been canceled
by the angular integrations.
According to the Efetov-Zirnbauer theorem, the terms of second order in 
the Grassmann variables do not contribute to the final result. Because of
the extra factor $\lambda_1 - \lambda_2$ there are no problems with 
singularities. However, the term of zeroth order in the Grassmann variables
is the product of a divergent result, from the integration over the bosonic
variables, and zero from the Grassmann integrations. In this case, a careful
treatment of the singularity will result in
 a finite contribution known as an Efetov-Wegner term.

\subsection{Final Result and Efetov-Wegner Terms}
\label{twopo6}

We first consider the term of zeroth order in the Grassmann variables.
In section (\ref{integ4}) we discussed the appearance of Efetov-Wegner
terms for a Gaussian integral. Following the same procedure for the present
case, we will regularize the
integrals by excluding a small region around the singular
point $a = b = 0$. Using the procedure of ref. \cite{GPW} we replace
the integration measure by
\be
da da^* db db^* \rightarrow da da^* db db^* \, 
\theta(\sqrt{a^*a + b^* b} - \rho),
\ee
and take the limit $\rho \rightarrow 0$ at the end of the calculation.
We express this regularized measure in terms of the polar variables
$\mu_1$, $\mu_1^*$, $\mu_2$ and $\mu_2^*$. Some algebra using the 
explicit polar decomposition of $t_{12}$ and $t_{21}$ results in
\be
a^*a + b^* b = &&\mu_1^* \mu_1 + \mu_2^* \mu_2 +  
 (\mu_1^* \mu_1 - \mu_2^* \mu_2)(\alpha\beta + \gamma\delta)
+2i \mu_1 \mu_2^*\beta \gamma + 2i \mu_1^*\mu_2 \delta \alpha 
\nonumber \\ &+&
\frac 32 \alpha\beta \gamma \delta(\mu_1^* \mu_1 + \mu_2^* \mu_2),
\ee
For $ \sqrt{a^*a + b^* b}$ we obtain 
\be
\sqrt{a^*a + b^* b} = R + \frac 1{2R}\left [
 (\mu_1^* \mu_1 - \mu_2^* \mu_2)(\alpha\beta + \gamma\delta)
+2i \mu_1 \mu_2^*\beta \gamma + 2i \mu_1^*\mu_2 \delta \alpha\right ]
+\frac R2 \alpha\beta \gamma \delta, \nn \\
\ee
where
\be
R = \sqrt {\mu_1^*\mu_1 + \mu_2^* \mu_2} = \sqrt{\lambda_1 - \lambda_2}.
\ee
 
Keeping only the terms of zeroth and maximum order in the Grassmann variables
the $\theta$ function can be expanded as
\be
\theta(\sqrt{a^*a + b^* b} - \rho) = \theta(R-\rho) + \frac R2\delta(R-\rho)
\alpha\beta \gamma \delta + 
\frac{R^2}4 \del_R\delta(R-\rho)\alpha\beta \gamma \delta .
\ee
The derivative of the delta function can be converted into a delta function
by partial integration. Since our contribution comes from the singularity
we only have to take into account derivatives of the singular terms. Near 
$R\to 0$ the integrand containing the $\delta-$ functions is of the form
\be
\frac{dR}R  \left [
\frac R2\delta(R-\rho)+ 
\frac{R^2}4 \del_R\delta(R-\rho)\right]
\ee
times an angular factor. Partial integration of the second term results in
\be
\frac{dR}R  \left [
\frac R2\delta(R-\rho)- 
\frac{R}4 \delta(R-\rho) \right]
=
\frac{dR}R \frac R4\delta(R-\rho)
=
\frac{dR}R \frac {R^2}2\delta(R^2-\rho^2).
\label{EW}
\ee
 
Performing the Grassmann integration and including the Efetov-Wegner
terms (\ref{EW}) after transforming back to the $\lambda_1, \lambda_2$ 
variables in the correlation function (\ref{correl}  results in 
\be
\langle G(x+i\epsilon) G(y-i \epsilon) \rangle = \frac {\Sigma^2}{4} 
\int_1^\infty d \lambda_1 \int_{-1}^1 d\lambda_2
\left [\frac {\delta(\lambda_1 - \lambda_2 - \rho^2)}{2(\lambda_1 -\lambda_2)}
(\lambda_1 + \lambda_2)^2
+ 4\right ] e^{iN\omega \Sigma (\lambda_1 -\lambda_2)}.\nn \\
\label{correl2}
\ee
The first integral can be performed as follows
\be
\int_1^\infty d \lambda_1 \int_{-1}^1 d\lambda_2
\frac {2\delta(\lambda_1 - \lambda_2 - \rho^2)}{\lambda_1 -\lambda_2}
(\lambda_1 + \lambda_2)^2
= \int_{1-\rho^2}^1 d\lambda_2 \frac {2 (\rho^2 + 2\lambda_2)^2}{\rho^2} =8,
\nn \\
\ee
where we have taken the limit $\rho \rightarrow 0$.
The final result for the correlation function is given by
\be
\langle G(x+i\epsilon) G(y-i \epsilon) \rangle = \Sigma^2 +
\frac {\Sigma^2}{4} 
\int_1^\infty d \lambda_2 \int_{-1}^1 d\lambda_2
 e^{iN\omega \Sigma (\lambda_1 -\lambda_2)}=
\Sigma^2 +2i \frac {\sin(N\omega \Sigma) e^{iN\omega \Sigma}} {(\omega 
N)^2},\nn\\
\label{correl3}
\ee
which is the well-known result for the two-point correlation function
of the Gaussian Unitary Ensemble.
The first term is the disconnected piece of the correlation function.

\section{Integrability and the Replica Trick}
\label{repli}

Integrability and Random Matrix Theory are closely related. In this 
section we discuss the Toda lattice equation and show that the
two-point function of the GUE can be obtained from the replica
limit of the Toda lattice equation \cite{SplitVerb1}. A closely related
method which preceded this work is the replica limit of the Painlev\'e
equation. For a discussion of this approach we refer to the 
original literature \cite{kanzieper}.

\subsection{The Two-Point Function of the Gaussian Unitary Ensemble}
\label{repli1}

In this section we derive the two-point function of the Gaussian Unitary
Ensemble from the Replica limit of the Toda lattice equation. In section
\ref{twopo11} we have shown that the 
partition function with $n$ fermionic flavors
can written as \cite{kanzieper2001}
\be
Z_n(x) = C_n [\det (\del_{ix})^{i+j} Z_1]_{0 \le i,j \le n-1},
\ee
and $n$ bosonic flavors as \cite{SplitVerb1}
\be
Z_{-n}(x) = n! [\det (\del_{ix})^{i+j} Z_{-1}]_{0 \le i,j \le n-1},
\ee
where
\be
Z_1(x)= \int_{-1}^1 du e^{ixu}\qquad {\rm and}\qquad 
Z_{-1}(x)= \int_{1}^\infty due^{ixu}.
\ee

We will now show that the bosonic and fermionic partition functions
are related by a recursion relation. The main ingredient of this derivation
is the Sylvester identity \cite{Sylvester,ForresterBook}
which is valid for determinant of an arbitrary
matrix $A$. It is given by
\be
C_{ij} C_{pq} - C_{iq}C_{pj} = \det (A)C_{ij;pq},
\ee
where the $C_{ij}$ are cofactors of the matrix $A$ and
the $C_{ij;pq}$ are double cofactors. We apply this identity
for $i=j=n-1$ and $i=j=n$, and remind the reader that differentiating
a determinant is differentiating either with respect to all columns
or with respect to all rows. Because of the derivative structure of
the partition function, only differentiating the last 
column or the last row gives a nonzero result, but this is exactly
the desired cofactor of the matrix with $n$ increased by 1. 
One can easily show that the relevant cofactors are given 
by
\be
C_{nn} &=& Z_{n-1}(x),\nn \\
C_{n-1\, n} &=& -\del_{ix} Z_{n-1}(x),\nn \\
C_{n\, n-1} &=& -\del_{ix} Z_{n-1}(x),\nn \\
C_{n-1\, n-1} &=& (\del_{ix} )^2Z_{n-1}(x),\nn\\
C_{n-1\, n-1; n\,n} &=& Z_{n-2}(x).
\ee
By increasing the index by  one we obtain the Toda lattice equation
\be
\del_{ix}^2 \log Z_n(x) = n^2 \frac {Z_{n+1}(x) Z_{n-1}(x)}{[Z_n(x)]^2}
\ee
where the factor $n^2$ follows from the choice of the
normalization constants. We have made this choice because the left hand
side is proportional to $n^2$. The replica limit
of the two-point correlation is given by
\be 
R_2(x) &=& -\lim_{n\to 0}\frac 1{n^2} \del_x^2 \log Z_n(x)
= Z_1 (x) Z_{-1} (x)\nn \\  
&=&
\int_{-1}^1 du e^{iux} \int_1^\infty e^{iux}\nn \\
& = & 2i \frac{\sin x}x \frac{ e^{ix}}x,
\ee
which agrees with the exact analytical result for the two-point function.
The fermionic partition functions, the bosonic partition functions and 
the super-symmetric partition function form a single integrable
hierarchy which are related by the Toda lattice equation \cite{SplitVerb1}.
A closely related way to derive the two-point function of the GUE is
to take the replica limit and the corresponding Painlev\'e equation.
This has the advantage that the bosonic partition function does not
directly enter in the derivation.
For a discussion of this result we refer to \cite{kanzieper} which preceded
our work \cite{SplitVerb1} on the Toda lattice.   

\subsection{Toda Lattice Equation}
\label{repli2}

The Toda lattice equation describes a one dimensional lattice
in which neighboring atoms interact via a potential that depends
exponentially on the distance. The Hamiltonian of such system is given by
\be
H=\frac{1}{2}\sum_{k=-\infty}^\infty p_{k}^2 +
\sum_{k=-\infty}^\infty e^{-(q_{k}-q_{k-1})}.
\ee
This integrable system has infinitely many constants
of motion. For example, this can be concluded from the existence
of a Lax pair meaning that the Hamiltonian equations of motion
can be written in the form
\be
\frac {dK}{dt} = [K,L],
\ee
where $K$ and $L$ are two operators.
Then
\be
\frac {d {\rm Tr }[K^n]}{dt} = n{\rm Tr}[ K^{n-1}\frac {d K}{dt}]
=n{\rm Tr} [K^{n-1}[K,L]] = 0,
\ee 
by using the cyclic invariance of the trace.

The Hamilton equations of motion can be written as
\be
\frac {d^2}{dt^2} q_{k} = e^{q_{k-1}-q_{k}}-e^{q_{k}-q_{k+1}} .
\ee
If we define 
\be
z_{k+1} = z_k e^{-q_k},
\ee
we obtain the recursion relation 
\be
-\frac {d^2}{dt^2} \log z_{k+1} +\frac {d^2}{dt^2}\log z_k
= \frac{z_{k+1} z_{k-1}}{z_k^2} -\frac{z_{k+2} z_{k}}{z_{k+1}^2}.
\ee
We then conclude that
\be
\frac {d^2}{dt^2}\log z_k
= \frac{z_{k+1} z_{k-1}}{z_k^2} +  c_0.
\ee
The constant $c_0$ can be eliminated by a redefinition of
$z_k \to z_k \exp(c_0 t^2/2)$.

This equation has been studied extensively in the literature and has appeared
in different areas of physics. As an example, we mention applications to
Chern-Simons theory \cite{dunne} and other  applications
to Random Matrix Theory  \cite{tracy,ForresterBook}. The application of
the Toda lattice equation discussed in the previous section 
\cite{SplitVerb1} illustrates the power of integrability 
which makes it possible  to obtain exact nonperturbative by means of the
replica trick. More applications along these lines will be discussed
later in these lectures. Other applications are parametric correlations
and the QCD$_3$ spectral density \cite{pak} which both have been
analyzed by the supersymmetric method \cite{simons,szabo}.

\section{Symmetries of the QCD Partition Function}
\label{symme}

It is well-known that the QCD action is greatly 
constrained by gauge symmetry, Poincar\'e invariance 
and renormalizability. These symmetries determine the structure
of the Dirac operator. In this section we will discuss
the global symmetries of the Euclidean Dirac operator. 
They play an essential role
for its spectral properties in the deepest infrared. In particular,  
the chiral symmetry, the flavor symmetry and
the anti-unitary symmetry of the continuum Dirac operator are discussed.

\subsection{The QCD partition function}
\label{symme1}

The QCD partition function in a
box of volume $V_3=L^3$ can be expressed in terms of the eigenvalues of the
QCD Hamiltonian $E_k$ as
\be
Z^{QCD} = \sum_k e^{-\beta E_k},
\ee
where $\beta$ is the inverse temperature. At low temperatures $(\beta 
\rightarrow \infty)$ the partition function is dominated by the lightest
states of the theory, namely the vacuum state, with an energy density of 
$E_0/V_3$, and massless excitations thereof. 
The partition function $Z_{QCD}$
can be rewritten as a
Euclidean functional integral over the nonabelian gauge fields $A_\mu$,
\be
Z^{QCD}(M) = \int d A_\mu \prod_{f=1}^{N_f} \det (D + m_f) e^{-S_{YM} /g^2},
\label{ZQCD}
\ee
where $S_{YM}$ is the Yang-Mills action given by
\be
S_{YM} = \int d^4 [-\frac 14{ F_{\mu\nu}^a}^2 + i\frac {\theta}{32\pi^2}
F_{\mu\nu}^a \tilde F_{\mu\nu}^a].
\label{SYM}
\ee
The field strength is given by
\be
F_{\mu\nu}^a = \partial_\mu A_{\nu}^a -\partial_\nu A_\mu^a + f_{abc}
A_\mu^b A_\nu^c.
\ee
The $f_{abc} $ are the structure constants of the gauge group $SU(N_c)$.
We also denote the gauge fields by
\be
A_\mu = A_{\mu a} \frac{T^a}{2},
\ee
where $T^a$ are the generators of  the gauge group. The dual of the field
strength is defined by
\be
\tilde F_{\mu\nu}= \frac 14 \epsilon_{\mu\nu\alpha\beta} F^{\alpha \beta}.
\ee
The integral $\nu \equiv 
\frac 1{32\pi^2}\int d^4x F_{\mu\nu}^a \tilde F_{\mu\nu}^a$
is a topological invariant, i.e. it does not change under continuous 
transformations of the gauge field. An important class of field configurations
are instantons. These are topological nontrivial field configurations that
minimize the classical action which are classified according to their
topological charge $\nu$. The parameter $\theta$ in (\ref{SYM}) is known
as the $\theta$-angle. Experimentally, its value is consistent with zero.
In (\ref{ZQCD}) the mass matrix is diagonal, $M = {\rm diag}
(m_1,\cdots, m_{N_f})$, but below we will also consider a general mass
matrix. 

The anti-Hermitian Dirac operator in (\ref{ZQCD}) is defined by
\be
D = \gamma_\mu(\partial_\mu +i A_\mu),
\ee 
where the $\gamma_\mu$ are the Euclidean Dirac matrices with
anti-commutation relation
\be
\{ \gamma_\mu, \gamma_\nu \} = 2 \delta_{\mu\nu}.
\ee
In the chiral representation the $\gamma$-matrices are given by
\be
\gamma_k = \mat 0 & i\sigma_k \\ -i\sigma_k & 0  \emat \quad
\gamma_4 = \mat 0 & 1 \\ 1 & 0 \emat \quad 
\gamma_5 = \mat 1 & 0 \\ 0 & -1 \emat.
\ee
In this representation the Dirac operator has the structure
\be
D = \mat 0 & id \\ id^\dagger & 0 \emat.
\label{diracstructure}
\ee

The integration measure can be defined by discretizing space time
\be
d A_\mu^a = \prod_x dA_\mu^a(x).
\ee 
A particular popular discretization is the lattice discretization where
the QCD action is discretized on a hyper-cubic lattice with spacing
$a$. The discussion of lattice QCD would be a lecture by itself. For
the interested reader we recommended 
several excellent textbooks on the subject \cite{creutz,rothe,montvay}.

Alternatively, we may define the measure in terms of a complete
set of orthonormal functions $\phi_{n\, \mu}^a$. In terms of the expansion
\be
A_\mu^a = \sum_{n=1}^N a_n \phi_{n\, \mu}^a
\ee
the measure is defined as
\be
\prod_n d a_n.
\ee

A field theory is obtained by taking the continuum limit, i.e. the 
limit $a\rightarrow 0$ or $ N\rightarrow \infty$ for the integration
measures discussed above. This limit only exists if we simultaneously
adjust the coupling constant, i.e. $ g \rightarrow g(a)$ or $g(N)$.
If such limit exists the field theory is called renormalizable.
For QCD it turns out that this adjusted coupling constant approaches
zero in the continuum limit, a property known as asymptotic freedom.

We will be mainly interested in the eigenvalues of the Euclidean Dirac operator
and how they fluctuate for gauge fields $A_\mu$ distributed according
to the QCD action. We will show that below a well-defined
scale the fluctuations
of the Dirac eigenvalues are given by the corresponding chiral RMT
discussed in the first chapter. It turns out that QCD and chiral RMT
have the same low energy limit which we will analyze by means of the
supersymmetric method.

\subsection{Axial Symmetry}
\label{symme2}

The eigenvalues of the QCD Dirac operator are given by
\be
D \phi_k = i\lambda_k \phi_k.
\ee
The eigenvalues are purely imaginary because the Dirac operator is 
anti-Hermitian (notice that $\partial_\mu^\dagger = -\partial_\mu$
and $A_\mu^\dagger = A_\mu$). 

The axial symmetry, or the $U_A(1)$ symmetry, can be expressed as
the anti-commutation relation
\be
\{\gamma_5, D\} =0.
\label{ua1}
\ee
This implies that all nonzero eigenvalues occur in pairs $\pm i\lambda_k$ with
eigenfunctions given  by $\phi_k$ and $\gamma_5 \phi_k$. If $\lambda_k = 0$ the
possibility exists that $\gamma_5 \phi_k \sim \phi_k$, so that $\lambda_k=0$ 
is an unpaired eigenvalue. According to the Atiyah-Singer theorem, 
the total number of such zero eigenvalues
is a topological invariant, i.e., it does not change under
continuous transformations of the gauge field configuration.
Indeed, this possibility is realized by  the field of an instanton 
which is a solution
of the classical equations of motion. On the other hand, it cannot
be excluded that $\lambda_k = 0$ while $\phi_k$ and $\gamma_5\phi_k$ are
linearly independent. However, this imposes additional constraints on
the gauge fields that will be violated by infinitesimal deformations.
Generically, such situation does not occur.

In a decomposition according to the total number of topological zero modes,
the QCD partition function can be written as
\be
Z^{QCD}(M) = \sum_\nu e^{i\nu\theta} Z_\nu^{QCD}(M),
\label{ZQCDdet}
\ee
where
\be
Z^{QCD}_\nu(M) = \langle\prod_f m^{\nu}_f \prod_k 
(\lambda_k^2 + m^2_f)
\rangle_\nu.
\ee
Here, $\langle \cdots \rangle_\nu$ denotes the average over gauge-field
configurations with topological charge $\nu$ weighted by the Yang-Mills
action. If we introduce  right-handed and left-handed masses as complex
conjugated masses we find that the $\theta$ dependence of the QCD
partition function is only through the combination $m e^{i\theta/N_f}$.
This property can be used to obtain the $\theta$-dependence of the
low energy effective partition function.
 
\vspace*{0.2cm}\noindent{\it Exercise.} Show that the $\theta$ dependence
of the QCD partition function is only through the combination
$ e^{i\theta/N_f}$.
\vspace*{0.2cm}

\subsection{Flavor Symmetry}
\label{symme3}

A second important global symmetry is the flavor symmetry. This symmetry
can be best explained by writing the fermion determinant in the QCD partition
function as a functional  integral over Grassmann variables,
\be
\prod_f \det(D+m_f) = \int d \psi d \bar \psi e^{\int d^4x
\sum_{f=1}^{N_f} \bar \psi^f (D+ m_f) \psi^f}.
\ee
In a chiral basis with $\psi_R = \gamma_5 \psi_R$
and $\psi_L = -\gamma_5 \psi_L$, the exponent can be rewritten as
\be
\sum_{f=1}^{N_f} \bar \psi^f (D+ m_f) \psi^f=
\bar \psi^f_R D \psi^f_R + \bar \psi^f_L D \psi^f_L +
\bar \psi^f_R  M_{RL} \psi^f_L+ \bar \psi^f_L  M_{LR}\psi^f_R,\nonumber \\
\label{mlr}
\ee
where we have used that
\be
\bar \psi^f_R  D \psi^f_L= \bar \psi^f_R  \gamma_5 D \gamma_5 \psi^f_L 
=  -\bar \psi^f_R  D \psi^f_L = 0,
\ee
and similarly for the matrix elements between a left-handed
and a right-handed state. To better illuminate the transformation
properties of the partition function we have replaced the diagonal
mass matrix by $M_{RL}$ and $M_{LR}$.

\vspace*{0.2cm}\noindent{\it Exercise.} Construct $\psi_R$, $\psi_L$,
$\bar\psi_R$ and $\bar \psi_L$  from the original basis.
\vspace*{0.2cm}

For $m_f = 0$ we have the symmetry
\be
  \psi_L &\rightarrow& U \psi_L, \quad
\bar  \psi_L \rightarrow \bar \psi_LU^{-1}, \nn \\
\psi_R &\rightarrow& V \psi_R, \quad
\bar \psi_R \rightarrow \bar\psi_R V^{-1}.
\ee
The only condition to be imposed on $U$ and $V$ is that their inverse exists.
If the number of left-handed modes is equal to the number of right-handed modes
we thus have an invariance under $Gl_R(N_f) \times Gl_L(N_f)$, where 
$Gl(N_f)$ is the group of complex $N_f\times N_f$ matrices with nonzero
determinant.  
However, if
the number of left-handed modes is not equal to the number of right-handed
modes, the axial-symmetry group
is broken to an $Sl(N_f)$ subgroup whereas the vector
symmetry with $U=V$ remains unbroken.
For $m=0$ the flavor symmetry is thus broken explicitly to 
$Gl_V(N_f)\times Sl_A(N_f)$ by instantons or the anomaly.
A $Gl_V(1)$ subgroup of $Gl_V(N_f)$ corresponds to baryon number
conservation and is usually not considered when flavor symmetries
are discussed. 

What is much more important, though, is the spontaneous breaking of
the axial flavor symmetry. From lattice QCD simulations and phenomenological
arguments we know that the expectation value
\be
\langle \bar \psi \psi \rangle =
\langle \bar \psi_R \psi_R \rangle +
\langle \bar \psi_L \psi_L \rangle\approx -(240\, MeV)^3
\ee
in the vacuum state of QCD instead of the symmetric possibility
$\langle \bar \psi \psi \rangle = 0$.
Phenomenologically, this is known 
because of the presence of Goldstone modes. The pions are much lighter than
the $\sigma$ mesons. The spontaneous breaking of the axial symmetry  
also follows from the absence of
parity doublets. For example, the pion mass and the $\delta$ (or $a_0$) mass
are very different ($m_\pi = 135 MeV$ and $m_\delta = 980 MeV$).

For fermionic quarks there is no need to extend the
symmetry group to   $Gl_R(N_f) \times Gl_L(N_f)$. In that case we
will only consider the usual $SU_R(N_f) \times SU_L(N_f)$ flavor symmetry
and it spontaneous breaking to $SU_V(N_f)$. 
In the case of bosonic quarks we will see
in section \ref{spect3} that it is essential to consider the complex
extension of $SU(N_f)$.
Notice that by extending
the symmetry group to complex parameters the number of conserved 
currents remains the same.

On easily verifies  
that $\langle \bar \psi \psi \rangle$ is only invariant for $U=V$.
The vacuum state thus breaks the chiral symmetry down to $Gl_V(N_f)$.
Notice that only the axial symmetries are broken. This is in agreement
with the Vafa-Witten theorem \cite{Vafa}
which states that vector symmetries cannot
be broken spontaneously in vector-like theories such as QCD. 
We also observe that the $complete$
axial group is broken. The reasons behind this maximum breaking 
\cite{Shifman-three} of chiral symmetry are less well understood.

\subsection{ Anti-Unitary Symmetries and the Three-fold Way}
\label{symme4}

The QCD partition function with three or more colors in the fundamental
representations has no anti-unitary symmetries. For two colors and
fundamental fermions and for adjoint fermions, the Dirac operator
has an anti-unitary symmetry which will be discussed below. The classification
of QCD according to anti-unitary symmetries was introduced in \cite{V}.

\subsubsection{QCD with Two Colors}
\label{symme41}

For QCD with two colors in the fundamental representation the gauge
field is given by
\be A_\mu = \sum_k a_k \frac{\tau_k}2,
\ee
where the $\tau_k$ are the Pauli matrices acting in color space. 
Because of the pseudo-reality of $SU(2)$ we have that 
\be
A_\mu^* = - \tau_2 A_\mu \tau_2.
\ee
Using the explicit representation for the $\gamma$-matrices one can easily
show that
\be
\gamma_\mu^* = \gamma_2 \gamma_4 \gamma_\mu \gamma_2\gamma_4 .
\label{gammac}
\ee
For the Dirac operator $iD = i\gamma_\mu \partial_\mu + \gamma_\mu A_\mu$
we thus have
\be
\tau_2 \gamma_2 \gamma_4 (iD)^* \tau_2 \gamma_2 \gamma_4  = -iD.
\ee
This can also be written as
\be
[KC\gamma_5\tau_2, D] = 0,
\label{anti1}
\ee
where $K$ is the complex conjugation operator and $C= \gamma_2\gamma_4$ is the 
charge conjugation matrix.
Because
\be
(KC\gamma_5\tau_2)^2= 1
\label{sq1}
\ee
we can always find a basis such that the Dirac matrix is real for any $A_\mu$.
The proof goes along the same lines as the proof that time reversal symmetry
results in real matrix elements for the Hamiltonian in quantum mechanics
(see section \ref{rando4}).

\vspace*{0.2cm}\noindent{\it Exercise.} Construct a basis for which $D$
becomes real (without relying on the properties of $D$ other than
the anti-unitary symmetry (\ref{anti1}) and (\ref{sq1})).
\vspace*{0.2cm}

\subsubsection{QCD in the Adjoint Representation}
\label{symme42}

For QCD with gauge fields in the adjoint representation the 
Dirac operator is given by
\be
D = \gamma_\mu\partial_\mu\delta_{ab} + f^{abc} \gamma_\mu A_{a\mu},
\ee
where the $f^{abc}$ denote the structure constants of the gauge group. 
Notice that $f^{abc}$ is antisymmetric in its indices so that 
$D$ is anti-Hermitian.
Because of the complex conjugation property of the $\gamma$-matrices we 
have that
\be
[\gamma_2\gamma_4 \gamma_5 K,D] = 0.
\ee
One easily verifies that in this case
\be
(\gamma_2\gamma_4 \gamma_5K)^2 = -1
\ee
so that the eigenvalues of $D$ are doubly degenerate 
(see section \ref{rando4}).
This corresponds to the case $\beta_D = 4$, so that it is possible to construct
a basis in which the matrix elements of the Dirac operator can be 
organized into real quaternions.

\section{Low Energy Limit of QCD}
\label{lowen}
In this section we derive the chiral Lagrangian that  provides an exact 
description of QCD at low energies. In the domain where the kinetic
term of this chiral Lagrangian can be ignored this Lagrangian constrains
the Dirac spectrum to satisfy sum rules for its inverse eigenvalues.

    \subsection{The chiral Lagrangian}
\label{lowen1}

For light quarks the low energy limit of QCD is well understood. It is
given by a chiral
Lagrangian describing the interactions of the pseudo-scalar mesons. 
The reason is that pions are Goldstone bosons which
 are the only light degrees of freedom in a confining theory such as
QCD. To lowest order in the quark masses and the momenta, 
the chiral Lagrangian is completely determined by chiral symmetry and 
Lorentz invariance. In the
case of $N_f$ light quarks with chiral symmetry breaking according to 
$SU_L(N_f) \times SU_R(N_f) \rightarrow SU_V(N_f)$ the Goldstone 
fields are given by $U \in SU(N_f)$. Under an  $SU_L(N_f) \times SU_R(N_f)$
transformation of the quark fields, 
\be
\psi_R \to U_R \psi_R,\nn \\
\psi_L \to U_L \psi_L,
\label{uquarks}
\ee
the Goldstone fields  $U$ transform in the same way as
the chiral condensate
\be
U \to U_L U U_R^{-1} .
\ee
The symmetry (\ref{uquarks}) is broken the mass term. However, the full 
symmetry can be restored if we also transform the mass term 
in (\ref{mlr}) as
\be
M_{RL} \to U_R M_{RL} U_L^{-1},\qquad 
M_{LR} \to U_L M_{LR} U_R^{-1} .
\label{umass}
\ee
The low energy effective theory should have the same invariance 
properties. To second order in the momenta and first order in the
quark mass matrix we can write down the following invariant terms:
\be
{\rm Tr} (\partial_\mu U \; \partial_\mu
U^\dagger),\qquad  {\rm Tr} (M_{RL}  U), \qquad 
{\rm Tr} (M_{LR} U^{-1}).
\ee
Since the QCD partition function is invariant under $M_{RL} 
\leftrightarrow M_{LR}$, the effective partition function should
also have this symmetry. The action of the Goldstone fields is
therefore given by the so called 
Weinberg Lagrangian \cite{Weinberg,GL}
\be
{\cal L}_{\rm
eff}(U)=\frac{F^2}{4}  \;{\rm Tr} (\partial_\mu U  \partial_\mu
U^\dagger)-\frac{\Sigma}{2} \; {\rm Tr} (M_{RL} U+
M_{LR} U^{-1}),
\ee
where $F$ is the pion decay constant, and $\Sigma$ is the chiral condensate.
The second constant can be identified as the chiral
condensate because it is the mean field value of the mass derivative 
of the partition function.
The Goldstone fields can be parametrized as
\be
 U={\rm exp} (i\sqrt 2 \Pi_a t^a/F),
\ee
with the generators of $SU(N_f)$ normalized according to ${\rm Tr}\, t^a t^b
=\delta^{ab}$.
This chiral  Lagrangian has been used extensively
for  the description of pion-pion scattering amplitudes \cite{GL}.

To lowest order in the pion fields we find for equal quark masses
\be
{\cal L}_{\rm eff}(U) = \frac 12 \partial_\mu \Pi^a \partial^\mu \Pi^a
              + \frac{\Sigma m}{F^2} \Pi^a \Pi^a.
\ee
This results in the pion propagator $1/(p^2 +m_\pi^2)$ with pion mass
given by the Gellmann-Oakes-Renner relation
\be
m_\pi^2 =\frac{2m\Sigma}{F^2}.
\ee

In the long-wavelength limit the order of magnitude of the different
terms contributing
to the action of the chiral Lagrangian is given by \cite{GL}
\be
S = \int d^4 x {\cal L} (U) \sim L^{d-2} {\Pi^{a}_{NZM}}^2
      + L^d \frac {\Sigma m}{F^2} ({\Pi^{a}_{ZM}}^2+{\Pi^{a}_{NZM}}^2).
\ee
Here, $\Pi^{a}_{ZM}$ represents the zero momentum modes with no space time 
dependence, whereas the nonzero momentum modes are denoted  by
by $\Pi^{a}_{NZM}(x)$. This decomposition has two immediate 
consequences. First,
for ${\Sigma m}/{F^2} \gg  1/V$ the fluctuations of the pion fields
are small and it is justified to expand $U$ in powers of $\Pi^a$. Second,
for
\be
\frac {\Sigma m}{F^2} \ll\frac 1{\sqrt V}
\ee
the fluctuations of the zero modes dominate the fluctuations of the
nonzero modes, and only the contribution from the zero modes 
has to be taken into account for the calculation of an observable.
In this limit the so called finite volume partition function is given by
\cite{LS}
\be
Z^{\rm eff}_{N_f}({M},\theta)
\sim \int_{U\in SU(N_f)} dU e^{V\Sigma {\rm Re\,Tr}\,
{ M} Ue^{i\theta/N_f}},
\label{zeff}
\ee
where the $\theta$-dependence follows from the dependence of the QCD
partition function on the combination  $m e^{i\theta/N_f}$ only
(see section \ref{symme2}). 
We emphasize that any theory with the same pattern of chiral symmetry
breaking as QCD can be reduced to  the same extreme infrared limit.

The effective partition function at fixed $\nu$ follows by Fourier
inversion
\be
Z_\nu(M) = \frac 1{2\pi} \int_0^{2\pi} d \theta e^{-i\nu \theta} Z(M,\theta).
\ee
Combining the integral over $SU(N_f) $ and the integral over $U(1)$ and
using that
\be
\det Ue^{i\theta/N_f} = e^{i\theta},
\ee
we find that
\be
Z^{\rm eff}_\nu(M) = \int_{U(N_f)} {\det}^\nu (U) 
e^{V\Sigma {\rm Re\,Tr}\,
{M} U}.
\label{zeffnu}
\ee

We also could have derived this partition function directly from an
invariance argument. For field configurations with topological charge
$\nu$ the difference of the number of right-handed modes and left-handed
modes is also equal to $\nu$. Under the transformation
(\ref{uquarks}) and the transformation of the mass matrix (\ref{umass}),
the QCD partition in the sector of topological charge
$\nu$ thus transforms as
\be
Z_{\rm QCD}^\nu \to {\det}^{\nu}(U_L U_R^{-1}) Z_{\rm QCD}^\nu,
\label{chtrans}
\ee  
where $U_R$ and $U_L$ are both $U(N_f)$ transformations. 
To zeroth order in the momentum and first order in the quark mass,
this covariance property
immediately gives the partition function (\ref{zeffnu}).

\subsection{Leutwyler-Smilga Sum Rules}
\label{lowen2}

The Leutwyler-Smilga sum-rules \cite{LS} are obtained 
by expanding the partition
function  at fixed $\nu$ in powers of $m$ before and after
averaging over the gauge field configurations and equating the
coefficients. This corresponds to an expansion in powers of $m$ of
both the QCD partition function and the finite
volume partition function (\ref{zeffnu}) in the sector
of topological charge $\nu$.
As an example,
we consider the coefficients of $m^2$ in the
sector with $\nu = 0$. We expand the effective partition function to second
order in the mass and use the group integrals
\be
\int_{U \in U(n)} DU U_{ij}^p &=& 0,\nn \\
\int_{U \in U(n)} DU U_{ij} U^{-1}_{kl} &=& \frac 1n \delta_{il}\delta_{jk}.
\ee
One easily verifies that the second integral is consistent with the unitarity
relations for $U$. We then find
\be
Z^{\rm eff}_{0}(m) = 1 + \frac {\Sigma^2 V^2}{4} m^2.
\ee
On the other hand, the expansion of the QCD partition function 
in powers of $m$ is given by
\be
\frac{Z_{QCD}^{\nu=0}(m)}{Z_{QCD}^{\nu=0}(0)} = 
1 + m^2N_f \langle \sum_{\lambda_k >0} \frac 1{\lambda_k^2} \rangle.
\ee
Equating the coefficients of $m^2$ results in the 
Leutwyler-Smilga sum-rule \cite{LS}
\be
\langle {\sum}' \frac 1{\lambda_k^2}\rangle  = \frac {\Sigma^2 V^2}{4N_f},
\label{LS2}
\ee
where the prime indicates that the sum is restricted to nonzero positive
eigenvalues.
\vspace*{0.2cm}
\noindent{\it Exercise.} Show that in the sector of topological charge
$\nu$ the sum rule for $N_f$ massless flavors is given by
\be
\langle {\sum}' \frac 1{\lambda_k^2}\rangle  = \frac {\Sigma^2 V^2}
{4(N_f+\nu)}
\label{LS3}
\ee
\vspace*{0.2cm}

By equating higher powers of $m^2$ one can generate an infinite family
of sum-rules for the eigenvalues of the Dirac operator. However, they
are not sufficient to determine the Dirac spectrum. The reason is
that  the mass in the propagator also
occurs in the fermion-determinant of the QCD partition function. 
However, as will be shown
in the next chapter, the Dirac spectrum can be obtained from a chiral
Lagrangian corresponding to QCD with additional bosonic and fermionic 
quarks \cite{OTV}.
We conclude that chiral symmetry breaking leads to correlations
of the inverse eigenvalues which are determined by the underlying
global symmetries.

\section{The Dirac Spectrum in QCD}
\label{dirac}

In this section we give an interpretation of the Leutwyler-Smilga sum-rules
in terms of the smallest eigenvalues of the QCD Dirac operator. We show
that the smallest eigenvalues of the QCD Dirac operator are related to
the chiral condensate by means of the Banks-Casher relation.

The order parameter of the chiral phase transition,
$\langle \bar \psi \psi \rangle$,
is nonzero only below the critical temperature or a critical chemical
potential.
As was shown by Banks and Casher {\cite{BC}},
$\langle \bar \psi \psi \rangle$ is directly related to the eigenvalue density
of the QCD Dirac operator per unit four-volume
\be
\Sigma \equiv
|\langle \bar \psi \psi \rangle| =\lim\frac {\pi \langle {\rho(0)}\rangle}V,
\label{bankscasher}
\ee
where the spectral density of the Dirac operator with eigenvalues
$\{\lambda_k\}$ is defined by
\be
\rho(\lambda) = \langle \sum_k \delta(\lambda-\lambda_k) \rangle.
\ee

\begin{figure}[ht] 
\caption{ A typical Dirac spectrum. To derive the Banks-Casher
relation we integrate the resolvent over the rectangular contour
in this figure. (Figure taken from \cite{minn04}.)}
\centerline{\includegraphics[height=6cm]{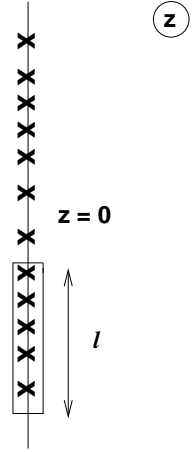}
}   
\label{fig1}
\end{figure}

Rather than the spectral density we often study the resolvent defined
by
\be
G(z) = \sum_k \frac 1{z+i\lambda_k},
\ee
which is related to the chiral condensate by 
\be
\Sigma
= - \lim_{m\to 0}\lim_{V\to \infty} G(m) / V.
\ee
The resolvent can be interpreted as the electric field at $z$ of charges
at $i\lambda_k$. Using this analogy it is clear that the resolvent 
changes sign if $z$ crosses the imaginary axis. Let us look at this
in more detail. A typical Dirac spectrum is shown in Fig. \ref{fig1}.
The average number of eigenvalues in the rectangular contour
in this figure is $\rho(\lambda) l$. Therefore, if
we integrate the resolvent along this contour  we obtain
\be
\oint G(z) = il(G(i\lambda+\epsilon) - G(i\lambda -\epsilon))
=2 \pi i \rho(\lambda) l,
\ee
where the second identity follows from Cauchy's theorem.
Using the symmetry of the spectrum we find
\be
{\rm Re } G(i\lambda+\epsilon) = \pi \rho(\lambda).
\ee
Near the center of the spectrum the imaginary part of the resolvent
in negligible which 
immediately results in the Banks-Casher relation (\ref{bankscasher}).

The order of the limits in (\ref{bankscasher}) is important. First we take the
thermodynamic limit, next the chiral limit and, finally, the field theory
limit. 

An important consequence of the Bank-Casher formula (\ref{bankscasher})
is that the eigenvalues near zero virtuality are spaced as
\be
\Delta \lambda = 1/{\rho(0)} = {\pi}/{\Sigma V}.
\label{spacing}
\ee
For the average position of the smallest nonzero eigenvalue 
we obtain the estimate
\be
\lambda_{\rm min} = {\pi}/{\Sigma V}.
\ee
This should be contrasted with the eigenvalue spectrum
of the non-interacting
Dirac operator. Then the eigenvalues are those of a free Dirac particle in a 
box with eigenvalue spacing equal to  $\Delta \lambda \sim 1/V^{1/4}$
for the eigenvalues near $\lambda = 0$.
Clearly, the presence of gauge fields leads to a strong modification of
the spectrum near zero virtuality. Strong interactions result in the
coupling of many degrees of freedom leading to extended states 
and correlated eigenvalues.
Because of asymptotic freedom, the spectral density of the Dirac operator for
large $\lambda$ behaves as $V\lambda^3$. In Fig. 2 we show a plot of
a typical average spectral density of the QCD Dirac operator for 
$\lambda \ge 0$. The spectral density for negative $\lambda$ is obtained 
by reflection with respect to the $y$-axis. More discussion of this figure
will be given in section \ref{parti5}.

\begin{figure}[ht] 
\caption{ Schematic picture of the average spectral density of 
QCD Dirac operator. (Taken from ref. \cite{minn00}.)}
\centerline{\includegraphics[height=6cm]{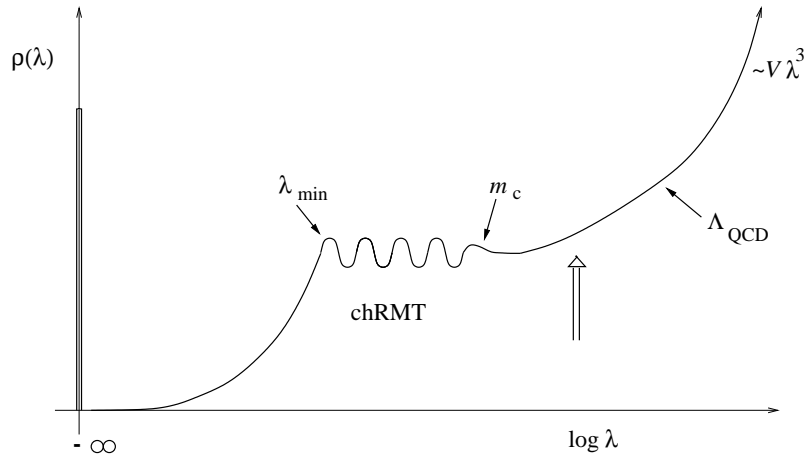}
}   
\label{fig1a}
\end{figure}

Let us study the Leutwyler-Smilga sum rule for equally spaced eigenvalues,
i.e.
\be
\lambda_n = \frac{\pi n}{\Sigma V}.
\ee
This results in the sum
\be
\sum_{n=1}^\infty \frac 1{\lambda_n^2} = \frac{\Sigma^2 V^2}{6}.
\ee
We obtain the right parametric dependence of the Leutwyler-Smilga sum rules.
The density of the smallest eigenvalues is suppressed by 
the number of massless quarks through the fermion determinant.
Therefore  it is not surprising that the exact result decreases with $N_f$.
 
Because the eigenvalues near zero are spaced as $\sim 1/\Sigma V$
it is natural to introduce the microscopic variable
\be
u = \lambda V \Sigma,
\ee 
and the microscopic spectral density \cite{SV}
\be
\rho_s(u) = \lim_{V\rightarrow \infty} \frac 1{V\Sigma} 
\rho(\frac u{V\Sigma}).
\label{rhosu}
\ee
We expect that this limit exists and converges to a universal function
which is determined by the global symmetries of the QCD Dirac operator.
The calculation of this universal function from QCD is the main 
objective of the next chapter. 
We will calculate $\rho_s(u) $ both from the simplest theory in this
universality class which is chiral Random Matrix Theory (chRMT) and 
from the partial quenched chiral Lagrangian which describes the low energy
limit of the QCD partition function. We find that the two results coincide
below the Thouless energy.
Universality can also directly be analyzed within the framework of 
of Random Matrix Theory. This can be done by showing that the microscopic
spectral density is insensitive to deformations of the probability
distribution. The interested reader is referred to the extensive 
literature on this subject 
\cite{sener,ADMN,Guhr-tilo,seneru,DN,FK,KV,A02}.

\section{Partial Quenching and the resolvent of the QCD Dirac Operator}
\label{parti}

In this chapter we study the spectrum of the QCD Dirac operator by means of
the resolvent. For simplicity we
will only consider the case with three or more colors and fundamental 
fermions.

\subsection{The resolvent in QCD}
\label{parti1}

In terms of the eigenvalues of the Dirac operator the resolvent is defined as
\cite{Vplb}
\be
G(z     ;m_1,\cdots, m_{N_f})& =&
\frac 1V \sum_k \left \langle \frac 1{i \lambda_k + z}
\right \rangle \nonumber \\
&=& 
\frac 1V
\int d\lambda \frac{\rho(\lambda; m_1,\cdots, m_{N_f})}
{i\lambda +z}.
\label{massdep}
\ee
Here,  $\langle \cdots \rangle$ denotes an average with respect to
the distribution of the eigenvalues. Sometimes the argument $z$ of
the resolvent is referred to as the valence quark 
mass \cite{Golleung}. It does not
enter in the fermion determinant as a regular quark and
is therefore a direct probe of the Dirac spectrum. 

The relation (\ref{massdep})
can then be inverted to give $\rho(\lambda;m_1,\ldots,m_{N_{f}})$.
As mentioned in \cite{OTV}, the spectral density
follows from the discontinuity across the imaginary axis,
\be
\left .{\rm Disc}\right |_{z = i\lambda}G(z)
= \lim_{\epsilon \rightarrow 0}
G(i\lambda+\epsilon) - G(i\lambda-\epsilon) = \frac{2\pi}V \sum_k
\langle \delta(\lambda +\lambda_k)\rangle
= \frac{2\pi}V \rho(\lambda),\nonumber\\
\label{spectdisc}
\ee
where we have suppressed the dependence on the sea-quark masses.
This relation follows by writing $G(z)$ as a sum over eigenvalues.

\subsection{Generating function}
\label{parti2}

In this section we introduce a generating function for the 
resolvent. Similarly to the supersymmetric method discussed before
this achieved by introducing the
ratio of two determinants in the partition function. 
This corresponds to a Euclidean partition function of the form
\be
Z^{\rm pq}_\nu(z,z',m_f)  ~=~ 
\int\! dA
~\frac{\det(D + z)}{\det(D + z')}\prod_{f=1}^{N_{f}}
\det(D + m_f) ~e^{-S_{YM}} ~,
\label{pqQCD}
\ee
which we will call the partially quenched or pq-QCD partition function.
For $z=z'$ this partition function  coincides with the original
QCD partition function. 
However, it is also the generator of the resolvent (see (\ref{massdep})).
In the sector of topological
charge $\nu$ we find
\be
G(z;m_1,\cdots, m_{N_f}) =
\frac 1V \left . \frac {\partial}{\partial z}
\right |_{z = z'} \log Z_\nu^{\rm pq}(z,z',m_f).
\ee

In addition to the regular quarks, the partition function (\ref{pqQCD})
has  additional bosonic and fermionic ghost quarks. 
Our aim is to find the chiral Lagrangian corresponding to (\ref{pqQCD}).
If we are successful, we have succeeded in deriving a partition function
for the extreme 
infrared limit of the spectrum of the QCD Dirac operator. The derivation
of this partition function and its domain of validity 
will be discussed in the next sections. We will first discuss the
subtleties that arise for the axial symmetry with bosonic quarks.

\subsection{Bosonic Quarks}
\label{parti3}

For bosonic quarks the Goldstone bosons cannot be parameterized by 
a unitary matrix. The reason is that symmetry transformations have
to be consistent with the convergence of the bosonic integrals.
Let us consider the case of one bosonic flavor. Then
\be
{\det}^{-1} \mat m & id\\ id^\dagger & m \emat
= \frac 1{\pi^2}\int d^2\phi_1 d^2\phi_2 \exp \left [-
\vect \phi_1^* \\ \phi_2^* \evect \mat m & id\\ id^\dagger & m \emat
\vect \phi_1 \\ \phi_2 \evect \right ],
\label{detinvbos}
\ee
so that the exponent is purely imaginary for $m=0$ and convergent for
${\rm Re}(m) > 0$.

The most general flavor symmetry
transformation of the action in (\ref{detinvbos}) 
is $Gl(2)$ which can be parameterized as 
\be
U = e^H V \quad {\rm with } \quad H^\dagger = H \quad {\rm and } \quad
V V^\dagger =1 .
\label{udecompos}
\ee
For $U$ to be a symmetry transformation for $m = 0$  we require that
\be
U^\dagger \mat 0 & id \\ id^\dagger & 0 \emat U =
 \mat 0 & id \\ id^\dagger & 0 \emat,
\ee
so that $H$ has to be a multiple of $\sigma_3$ and $V$ has to be a 
multiple of the identity. 
The $V$ transformations in (\ref{udecompos}) are not broken by  
the mass term and therefore represent the vector symmetry. Only the symmetry
transformation $\exp(s\sigma_3)$ is broken by the mass term
 so that the axial transformations are parameterized by
\be
U = \mat e^s & 0 \\ 0 & e^{-s} \emat \qquad {\rm with} 
\quad s \in \langle -\infty, \infty \rangle.
\ee
For $N_f$ bosonic flavors the axial transformations are parameterized by
\be
U = \mat  e^H & 0 \\ 0 & e^{-H} \emat \qquad{\rm with}\qquad H^\dagger  = H,
\ee
which is the coset $Gl(N_f)/U(N_f)$.

Also in the case of bosonic quarks the axial symmetry is
spontaneously broken:
\be
\langle \phi^*_1\phi_1\rangle + \langle \phi^*_2\phi_2\rangle =
\langle \bar\psi_R \psi_L\rangle +  \langle \bar\psi_L \psi_R \rangle.
\ee
The reason is that both          expectation values are equal to the
inverse trace of the Dirac operator.

\subsection{The infrared limit of QCD}
\label{parti4}

Ignoring for the moment convergence questions the global flavor symmetry
of (\ref{pqQCD}) is given by
\be
Gl_R(N_f+1|1) \times Gl_L(N_f+1|1).
\ee
We already have seen that the axial symmetry for bosonic quarks
is not $U(N_f)$ but rather $Gl(N_f)/U(N_f)$. Although the axial 
flavor symmetry
group of the fermionic quarks is not a priori determined by convergence
requirements
we will see in this section that supersymmetry necessarily imposes 
that this symmetry group is compact, i.e. equal to $U(N_f)$.

Under a $Gl_R(N_f+1|1) \times Gl_L(N_f+1|1)$ transformation
the quarks fields transforms as
\be
\psi_R \rightarrow U_R\psi_R, &&\qquad \psi_L \rightarrow U_L\psi_L\nn\\
\bar\psi_R \rightarrow \bar\psi_RU_R^{-1}, &&\qquad \bar\psi_L 
\rightarrow \psi_L U_L^{-1}.
\ee
The subscripts refer to the
right-handed (R) or left-handed (L) chirality of the quarks. 
The latter transformations transform the right-handed and the left-handed
fermion fields in the same way.

For $M= 0$ and $\nu =0$ this is a symmetry of the QCD action.
At nonzero mass this symmetry can be restored if we also transform
the mass term according to
\be
M_{RL} \to U_R M_{RL} U_L^{-1},\nn\\
M_{LR} \to U_L M_{LR} U_R^{-1}.
\ee
In the sector of topological charge $\nu$
the partially quenched partition function transforms as
\be
Z_\nu(M_{RL},M_{LR}) \rightarrow {\rm Sdet}^\nu U_R U_L^{-1} Z_\nu
(U_R M_{RL} U_L^{-1},U_L M_{LR} U_R^{-1})
\label{pqtrans}
\ee
The low energy effective partition function should have the
same covariance properties as the full QCD partition function.

Because  the expectation value of both the condensate of the
bosonic quarks and the condensate of the fermionic quarks is nonzero in
the chirally broken phase, the flavor symmetry of the partially quenched
QCD partition function (\ref{pqQCD}) is broken spontaneously
according to
\be
Gl_R(N_f+1|1) \times Gl_L(N_f+1|1) \rightarrow Gl_V(N_f+1|1).
\label{susybreaking}
\ee
Therefore the low energy limit of the 
partially quenched QCD partition function 
is a theory of weakly interacting Goldstone bosons parameterized by
\be
Gl_R(N_f+1|1) \times Gl_L(N_f+1|1) /Gl_V(N_f+1|1)\equiv Gl_A(N_f+1|1).
\ee
For a confining
theory such as QCD the only low lying modes are the Goldstone modes 
associated with the spontaneous breaking of chiral symmetry. 
However, the Goldstone manifold is not the full group $Gl_A(N_f+1|1)$. 
As we already discussed
for the case of bosonic quarks the  $U_A(1)$
transformation in which the bosonic fields with index 1 and 2
are transformed according
to an opposite phase factor is not a symmetry of the partition
function. What is a symmetry of the partition function is the 
axial transformation
\be
\phi_1 &\rightarrow& e^s \phi_1,\qquad \phi_1^* \rightarrow e^s \phi_1^*,\nn\\
\phi_2^* &\rightarrow& e^{-s} \phi_2^* \qquad
\phi_2 \rightarrow e^{-s} \phi_2.
\ee
Mathematically, this symmetry group is $Gl(1)/U(1)$. Had we restricted
ourselves to the unitary subgroup $U(N_f+1|1)$ of $Gl(N_f+1|1)$ 
from the start, we
would  have missed this class of symmetry
transformations.

Let us now consider the low energy limit of the partially quenched
QCD partition function in the sector of topological charge $\nu$.
Under a $Gl_R(N_f+1|1) \times Gl_L(N_f+1|1)$ transformation 
the Goldstone modes transform as
\be
Q\to U_R Q U_L^{-1}.
\ee
The low energy effective partition function should have the same
transformation properties a the partially quenched partition function
To lowest order in the mass and the momentum we can write down the
following invariants
\be
{\rm Str } \del_\mu Q^{-1} \del_\mu Q\quad {\rm  and} \quad 
{\rm Str}
(M_{RL} Q) + {\rm Str} (M_{LR} Q^{-1}).
\ee
If we factorize the Goldstone fields into the zero momentum modes
$Q_0$ and the nonzero momentum modes $Q(x)$ as
\be
Q = Q_0 Q(x),
\ee
the low energy effective
partition function that transforms in the same way as the partially
quenched QCD partition function (\ref{pqtrans}) is given by
\be
Z_\nu(M) = \int_{Q \in Gl(N_f+1|1)} dQ {\rm Sdet}^\nu (Q_0) 
e^{-\int d^4x {\cal L}^{\rm eff}(Q)},
\ee
where 
\be
{\cal L}^{\rm eff}(Q) = 
\frac {F^2}4{\rm Str } \del_\mu Q^{-1} \del_\mu Q + 
\frac 12\Sigma {\rm Str}(M_{RL} Q) + \frac 12 \Sigma{\rm Str} (M_{LR} Q^{-1}).
\ee
We already have seen that the boson-boson block of $Gl(N_f+1|1)$ is
 $Gl(1)/U(1)$. If we parameterize the field $Q$ as
\be
Q= e^{\sum_k T_k \pi_k/F},
\ee
with $T_k$ the generators of $G(N_f+1|1)$, to second order in the
Goldstone fields  the mass term is given by
\be
\frac {\Sigma}{2 F^2}{\rm Str} (M \sum_k T_k^2 \pi_k^2).
\ee
Let us take all (ghost-)quark masses positive. 
Because of the supertrace there is
a relative minus sign between boson-boson Goldstone modes and 
fermion-fermion Goldstone modes. The boson-boson Goldstone modes are noncompact
and require that the overall minus sign of the mass term is negative. 
In order to avoid tachyonic fermion-fermion Goldstone modes, we have
to compensate the minus sign of the supertrace. This can be done by
choosing the parameters multiplying the fermion-fermion generators
purely imaginary. This corresponds to 
a compact parametrization of the fermion-fermion Goldstone
manifold. This integration manifold is the maximum Riemannian
submanifold \cite{class,OTV,DOTV} of $Gl(N_f+1|1)$ and will be denoted as 
$\hat {Gl}(N_f+1|1)$.

\subsection{Domains in (Partially-Quenched) Chiral Perturbation Theory} 
\label{parti5}

 In chiral perturbation theory the different domains of validity where
analyzed in detail by Gasser and Leutwyler \cite{GaL}. A similar analysis
applies to partially quenched chiral perturbation theory \cite{Vplb}. 
The idea is
as follows. The $Q$ field can be decomposed as \cite{GaL}
\be
Q = Q_0 e^{i\psi(x)}.
\ee
where $Q_0$ is a constant (zero-momentum) field. The kinetic term for
the $\psi$ fields is given by
\be
\frac 12  \partial_\mu \psi^a(x) \partial_\mu \psi^a(x)  \sim L^{-2} \psi^2(x).
\ee
We observe that the magnitude of the fluctuations of the $\psi$ field
are of order  $1/L$ which justifies a perturbative expansion of
$\exp(i \psi(x))$. The fluctuations of the zero modes, on the other
hand, are only limited by the mass term 
\be
\frac 12  V \Sigma {\rm Str} {M}(Q_0+ Q_0^{-1}).
\ee
For quark masses 
$m \gg 1/V\Sigma$, the field $U_0$ fluctuates close to the identity and
the $U_0$ field can be expanded around the identity as well. If 
$m \ll \Lambda_{\rm QCD}$ we are in the  domain of chiral perturbation theory.

The same arguments apply to the partially quenched chiral Lagrangian.
However, there is an important difference. The mass of the ghost-quarks 
is an external parameter that can take on any value we wish.
The mass of the Goldstone modes containing
these quarks is given by
\be
M_{zz} = \frac {2 z \Sigma}{F^2}.
\ee 
Therefore, independent of the quark masses there is always a domain
where the fluctuations of the zero momentum modes dominate the 
fluctuations of the nonzero momentum modes. This domain is given by
\cite{Vplb}
\be
 z \ll \frac{F^2}{\Sigma L^2}
\label{domain}
\ee
and is known as the ergodic domain of QCD. Sometimes it is also referred to as
the epsilon regime\footnote{The epsilon regime is the regime \cite{GaL} where 
$m_\pi \sim O(\epsilon^2)$ and $1/L \sim O(\epsilon)$ in an expansion
in powers of $\epsilon$ of the chiral Lagrangian. These conditions
are more strict than the inequality (\ref{domain}). For example, we
are still in the ergodic domain if $1/L \sim O(\epsilon^{3/2})$}. 
In order that the non-Goldstone modes
do not contribute to the partition function we have to require that
\be
L \gg \frac 1{\Lambda_{\rm QCD}}.
\ee

In the Dirac spectrum we can distinguish three important scales. First,
the scale of the smallest eigenvalue,
\be
\lambda_{\rm min} = \frac \pi{\Sigma V}.
\ee
Second, the valence quark mass corresponding to a valence quark Goldstone
boson with Compton wavelength equal to the size of the box
\be
m_c = \frac {F^2}{\Sigma L^2}.
\ee
Third, the scale of $\Lambda_{QCD}$ which sets the mass scale of QCD.
Based on these scales we can distinguish four different domains.
In the domain where $z$ is of the order of $\lambda_{\rm min}$ or less
we can restrict ourselves to the zero momentum sector of the theory, but
we have to take into account quantum fluctuations to all orders. Also for 
$\lambda_{\rm min} \ll z \ll m_c$, we only have to include the zero momentum
modes but in this case 
the quantum fluctuations 
can be treated semi-classically, or in field theory language, by 
a loop expansion. 
The time scale conjugate to $m_c$ is of the order of the diffusion
time across the length of the box which explains the name 
the ergodic domain. For $m_c \ll z \ll \Lambda_{QCD}$, chiral perturbation
theory still applies, but the zero momentum modes no longer dominate the
partition function. 
For $z \gg \Lambda_{QCD}$, the masses of the Goldstone modes
and the other hadronic excitations are of the same order of magnitude. 
Chiral perturbation theory is no longer applicable to the spectrum of
the Dirac operator, and one has to take
into account the full QCD partition function.
A schematic drawing of the Dirac spectrum in the different domains
is given in Fig. \ref{fig2}.

In the theory of disordered mesoscopic systems 
the scale below which
random matrix theory is valid is known as the Thouless energy and
is given by \cite{Altshuler,Montambaux}
\be
E_c = \frac{\hbar D}{L^2},
\ee
where $D$ is the diffusion constant for the diffusive motion of electrons in
a disordered sample. The time conjugate to $E_c$ is the time scale over
which an electron diffuses across the sample. The
time scale in mesoscopic physics corresponding to
$\Lambda_{QCD}$ is the elastic scattering time. The domain in between
$E_c$ and $\hbar/\tau_e$, where $\tau_e$ is the elastic scattering time, 
is known as the diffusive domain. This domain is
characterized by diffusive motion of electrons in the disordered sample.

In the derivation of the
zero momentum limit of the partially quenched QCD partition function we
have only relied on symmetries and the spontaneous breaking of
symmetries. In particular, this means that any theory with the
same symmetry breaking pattern and only Goldstone modes as
light particles will lead to the same low energy effective
theory. The simplest theory in this class is chiral Random Matrix
Theory introduced in chapter \ref{rando2}. 
This theory has the flavor symmetries
of QCD and, in the limit of large matrices, the non-Goldstone modes 
decouple from the theory. Therefore, the zero momentum properties of
QCD can be derived from chiral Random Matrix Theory.
 
\subsection{Zero momentum limit of the effective partition function}
\label{parti6}
Since $z$ is a free parameter, independent of the quark masses, we can
always find a part of the Dirac spectrum inside the domain (\ref{domain}).
In this domain the zero momentum limit of the QCD partition
function in the sector of topological charge $\nu$ is given by \cite{OTV}
\be
Z^\nu_{\rm eff}( M) =
\int_{Q\in \hat{Gl}(N_f+1|1)} dQ \,{\rm Sdet}^\nu Q \, e^{
 V\frac{\Sigma}{2} \; {\rm Str} (M Q+M Q^{-1})}.
\label{superpart}
\ee
The number of QCD Dirac eigenvalues that is described
by this partition function is of the order (see (\ref{spacing}))
\be
\frac{F^2}{\Sigma L^2\Delta \lambda} = F^2 L^2.
\ee
This  number increases linearly in $N_c$ for $N_c\to \infty$ which
was studied recently by lattice simulations \cite{neuberger}.

In the next section we will study this partition function in the 
quenched limit ($N_f = 0$) and show that the resolvent
coincides with the result obtained
from chiral Random Matrix Theory \cite{OTV,DOTV}. The Random Matrix
result will be derived by means of the supersymmetric method in
section \ref{onepch}.

\subsection{Nonperturbative evaluation of $G(z)$ in the quenched limit} 
\label{parti7}

In this section we evaluate the resolvent of QCD for
the simplest case of $N_f = 0$ and $\nu =0$ in the
domain $z\ll F^2/\Sigma L^2$. In this domain
the partition function is given by
\be
Z(J) = \int_{Q\in \hat{Gl}(1|1)} dU \exp\left [ \frac {\Sigma V}2 {\rm Str}
\left ( \begin{array}{cc} z +J &0 \\ 0 & z \end{array} \right )
(Q + Q^{-1})\right ],
\ee
where the integration is over the maximum super-Riemannian sub-manifold
of $Gl(1|1)$. This manifold is parametrized by \cite{DOTV}
\be
Q =\exp {\left ( \begin{array}{cc} 0 &\alpha \\ \beta & 0 \end{array} \right )}
 \left ( \begin{array}{cc} e^{i\phi} &0 \\ 0 & e^s \end{array} \right ).
\ee
The integration measure is the Haar measure which in 
terms of this parameterization
it is given by
\be
{\rm Sdet } \frac {\delta Q_{kl}}{
\delta \phi \, \delta s\, \delta \alpha \, \delta \beta} 
\, \,d\alpha d\beta d\phi ds,
\ee
where $\delta Q \equiv Q^{-1} dQ$.

It is straightforward to calculate the Berezinian going from the
variables $\{\delta Q_{11},\,\delta Q_{22},\,
\delta Q_{12},\,\delta Q_{21}\}$ to the variables 
$\{\delta \phi, \, \delta s,\, \delta \alpha, \, \delta \beta\}$. The 
derivative matrix is given by
\be
B=\frac {\delta Q_{kl}}{
\delta \phi \, \delta s\, \delta \alpha \, \delta \beta} =
\left (\begin{array}{cccc} 
i & 0 & \frac \beta 2&
\frac \alpha 2\\
0  &1 & \frac \beta 2 &\frac \alpha 2\\
0 & 0& e^{s-i\phi}& 0 \\
0 & 0 & 0 &e^{-s+i\phi} 
\end{array}\right).
\ee
Using the definition of the graded determinant one simply finds that
${\rm Sdet} B = i$. Up to a constant, the integration measure is thus given by
$d\phi ds d\alpha d\beta$. In general, for $N_f \ne  0$, the Berezinian
is more complicated \cite{DOTV}.

We also need
\be
\frac 12(Q +Q^{-1} ) = \left ( \begin{array}{cc} 
\cos \phi (1 +\frac {\alpha \beta} 2 ) & \alpha (e^s - e^{-i\phi})\\
\beta(e^{i\phi} - e^{-s} ) & \cosh s ( 1- \frac{\alpha \beta }2) 
\end{array} \right ).
\ee
After differentiating with respect to the source term  
$(G(z) = \partial_J\log Z(J) |_{J=0}) $ this results in
\be
\frac{G (z)}{V\Sigma} =  \int \frac{d\phi ds d\alpha d\beta}{2\pi}
\cos \phi (1 +\frac {\alpha \beta} 2 ) 
e^{x \cos \phi (1 +\frac {\alpha \beta} 2 ) -
x\cosh s ( 1- \frac{\alpha \beta }2)}.\nonumber \\
\ee
With the Grassmann integral given by the coefficient of $\alpha\beta$ 
we  obtain
\be
\frac{G(z)}{V\Sigma} &=& \int \frac{ds d\phi}{4\pi} 
[\cos\phi  \nonumber +x(\cos\phi 
+ \cosh s)\cos\phi ]e^{x(\cos\phi - \cosh s )} .
\ee
All integrals can be expressed in terms of modified Bessel functions.
We find \cite{OTV,DOTV}
\be
\frac{G(z)}{V\Sigma} &=& I_1(x)K_0(x)
+\frac x2 ( I_2(x)K_0(x)+I_0(x)K_0(x)+ 2I_1(x)K_1(x)).
\ee
After using the recursion relation for  modified Bessel functions,
\be
I_2(x) = I_0(x) -\frac 2x I_1(x),
\ee
 we arrive at the final result \cite{Vplb,OTV,DOTV}
\be
\frac{G(z)}{V\Sigma} = x(I_0(x)K_0(x)+ I_1(x)K_1(x))
\label{resnf0}
\ee
where $x= zV\Sigma$.

This calculation can be generalized to arbitrary $N_f$ and arbitrary $\nu$. 
The calculation for arbitrary $N_f$ is much more complicated, but with a
natural generalization of the factorized parameterization, and using some known
integrals over the unitary group, 
one arrives at the following
expression in terms of modified Bessel functions
\be
\frac {G(z)}{\Sigma} = x(I_{a}(x)K_{a}(x)
+I_{a+1}(x)K_{a-1}(x)),
\label{val}
\ee
where $a = N_f+|\nu|$. This result is in 
complete agreement with the microscopic spectral density derived
from chRMT \cite{Vplb}.

\begin{figure}[ht] 
\caption{ The resolvent of quenched QCD. The points represent lattice
data obtained by the Columbia group, and the theoretical prediction
(\ref{resnf0}) is given by the solid curve.
(Taken from ref. \cite{minn04}.)}
\centerline{\includegraphics[height=14cm]{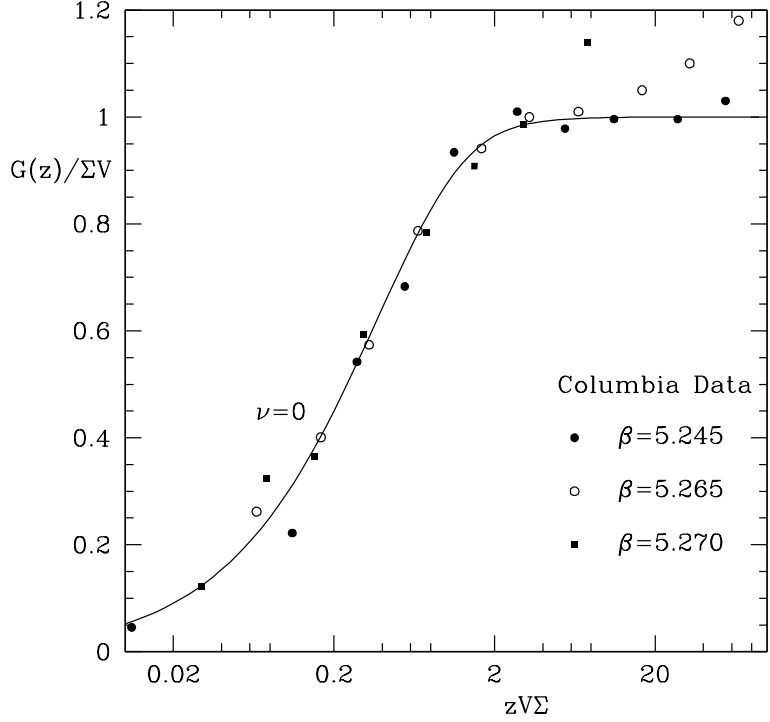}
}   
\label{fig3}
\end{figure}

For $a=0$ this result is plotted in Fig. \ref{fig3}. We observe that,
below some scale, lattice QCD data obtained by the Columbia group
\cite{Christ}
closely follow this curve.

\subsection{Random Matrix Calculation of $G(z)$ for the chGUE}
\label{onepch}

In this section we calculate the one-point function of the chiral Gaussian
Unitary Ensemble by means of the supersymmetric method 
\cite{simons-ch,brezin,sener,seneru}. 
The starting point
is the partition function
\be
Z(z,z+J) = \int {\cal D} W \ \frac{{\det}(z+J+D)}{\det(z+D)}
\ \exp[- n \Sigma^2 \,{\rm Tr}\, W W^\dagger], \ \nonumber\\
\label{zrandom}
\ee 
where $D$ is given by the anti-hermitian random matrix
\be
D = \left (
\begin{array}{cc} 0 & iW \\i W^\dagger  & 0
\end{array} \right ) \ \, .
\label{bigmatr}
\ee
In this section we only consider the case of zero topological charge
where $W$ is an $n\times n$ complex matrix.
The integral is over the real and imaginary parts of the matrix
elements of $W$. For definiteness we take the real part of $z$ to
be positive.
The resolvent is obtained by differentiating the partition function
\be
G(z) = \frac 1V {\rm Tr} \left \langle \frac 1{z +D}\right \rangle
\ = \frac 1V \left .\ \frac {\del \log Z(z,z+J)}{\del J}\right |_{J=0} \ \ ,
\label{resolvent} 
\ee

The determinant in the partition function is written as an integral over
Grassmann variables $\chi$ and $\chi^*$:
\be
\det(z +J+ D) & = & \int \prod_{i=1}^n d{\chi_1}_i^* d{\chi_1}_i
\prod_{i=1}^n d{\chi_2}_i^* d{\chi_2}_i
\nonumber \\
& {} & \qquad \times \exp -
\left ( \begin{array}{c} \chi_1^* \\ \chi_2^* \end{array} \right )
\left ( \begin{array}{cc} z+J & iW \\ iW^\dagger  &z+J
\end{array} \right )
\left ( \begin{array}{c} \chi_1 \\ \chi_2 \end{array} \right ) \ \ .
\label{zfermion}
\ee
Similarly, the inverse determinant can be written as an integral over the
commuting variables $\phi$ and $\phi^*$:
\be
{\det}^{-1}(z+ D) & = & \frac{1}{(- 2\pi i)^{2n}}\ \int
\prod_{i=1}^n d{\phi_1}_i d{\phi_1}_i^*
\prod_{i=1}^n d{\phi_2}_i d{\phi_2}_i^*
\nonumber \\
& {} & \qquad \times \exp -
\left ( \begin{array}{c} \phi_1^* \\ \phi_2^* \end{array} \right )
\left ( \begin{array}{cc} z & iW \\ iW^\dagger  &z
\end{array} \right )
\left ( \begin{array}{c} \phi_1 \\ \phi_2 \end{array} \right ) \ \ .
\label{zboson}
\ee
After performing the Gaussian integral over $W$ we obtain

\be
Z(z,z+J) &=& \int \prod_{i=1}^n d{\phi_1}_i d{\phi_1}_i^*d{\chi_1}_i^* d{\chi_1}_i
\prod_{i=1}^n d{\phi_2}_i d{\phi_2}_i^*d{\chi_2}_i^* d{\chi_2}_i \nonumber \\
&\times&\exp\left[ -\left (
\begin{array}{c} \phi_1^* \\ \phi_2^* \end{array} \right )
\left ( \begin{array}{cc}
           z & 0 \\
              0 & z\end{array} \right )
\left ( \begin{array}{c} \phi_1 \\ \phi_2 \end{array} \right )
-       \left (
\begin{array}{c} \chi_1^* \\ \chi_2^* \end{array} \right )
\left ( \begin{array}{cc}
           z+J &  0\\
              0  & z +J\end{array} \right )
\left ( \begin{array}{c} \chi_1 \\ \chi_2 \end{array} \right ) \right ]
\nonumber \\
&\times&
\exp -\frac 1{n\Sigma^2} (\chi_{2j}^* \chi_{1i} + \phi_{2j}^* \phi_{1i})
(\chi_{1i}^* \chi_{2j} + \phi_{1i}^* \phi_{2j}) \ \ .
\ee
The terms of fourth order in the fields are decoupled to second order
terms by means of the Hubbard-Stratonovitch transformation. There are
no convergence problems for the terms that contain Grassmann variables.
In these cases we use the identity
\be
e^{  \chi_1^*\cdot \chi_1 \, \chi_2^*\cdot \chi_2 /n\Sigma^2}
= c_1 \int d\rho_1 d\rho_2 \exp[ -n\Sigma^2(\rho_1^2 + \rho_2^2)
-(\rho_1-i\rho_2)\chi_1^*\cdot \chi_1
-(\rho_1+i\rho_2)\chi_2^*\cdot \chi_2 ]\  ,\nonumber \\
\label{rhosig}
\ee
for the term of fourth order in the Grassmann variables
and
\be
e^{-  \chi_1^*\cdot \phi_1 \, \chi_2\cdot \phi_2^*/n\Sigma^2}
&=& c_2 \int d\alpha^* d\beta \exp[ -n\Sigma^2\alpha^* \beta
+\alpha^*\chi_1^*\cdot \phi_1
-\beta\phi_2^*\cdot \chi_2 ] \ , \nonumber \\
e^{\chi_1\cdot \phi_1^* \, \chi_2^*\cdot \phi_2/n\Sigma^2}
&=& c_2 \int d\beta^* d\alpha  \exp[- n \Sigma^2 \beta^* \alpha
-\alpha \chi_1 \cdot \phi_1^*
+\beta^* \phi_2 \cdot \chi_2^* ] \ \ ,
\nn \\
\label{after-averaging}
\ee
for the mixed terms. In the above equations, $\rho_1$ and $\rho_2$ are
commuting variables and $\alpha$, $\alpha^*$, $\beta$ and $\beta^*$ are
Grassmann variables. The constants $c_1$ and $c_2$ are not necessary
for the calculation of the partition function. We can always fix the
overall normalization constant from the requirement that $Z(z,z) = 1$.
In the case of the the four boson term we have to
be more careful. This term can be rewritten as
\be
 e^{- \phi_1^*\cdot \phi_1 \, \phi_2\cdot \phi_2^*/n\Sigma^2}
=
 \exp- \frac 1{4n\Sigma^2} [
(\phi_1^*\cdot \phi_1 + \phi_2\cdot \phi_2^*)^2 
-(\phi_1^*\cdot \phi_1 - \phi_2\cdot \phi_2^*)^2].
\ee 
The uniform convergence of the $\phi_i$-integrations necessary to 
interchange the order of the integrations
can be achieved
if we perform a Hubbard-Stratonovitch transformation using the identity
\cite{seneru}
\be
e^{-a^2+b^2} = \frac {-n\Sigma^2}{\pi i} \int_{-\infty}^\infty
d s \int_{-\infty}^\infty
\sigma d\sigma e^{-n\Sigma^2\sigma^2 
-2i\sqrt n \Sigma a\sigma \cosh s - 2 i \sqrt n \Sigma b\sigma \sinh s} \ ,
\ee
where $a$ is real positive and $b^2 - a^2$
has a negative real part. Indeed, these conditions are fulfilled if we
choose
\be
a = \frac 1{2\Sigma \sqrt n}(
\phi_1^*\cdot \phi_1 + \phi_2\cdot \phi_2^*), \qquad
b = \frac 1{2 \Sigma \sqrt n}(\phi_1^*\cdot \phi_1 - \phi_2\cdot \phi_2^*). 
\ee
An alternative way to find correct transformation of the four-boson term
is to use the Ingham-Siegel integral instead of the Hubbard-Stratonovitch
transformation \cite{yan,spencer}. 

Combining the above results we obtain 
\be
Z(z,z+J) &=& c\int  d{\phi_1}d\phi_1^* d\phi_2 d\phi_2^*
d{\chi_1}d\chi_1^* d\chi_2d\chi_2^*  \ d [\sigma]
\exp \left[ -n\Sigma^2(\sigma^2
+\rho_1^2 + \rho_2 ^2 + \alpha^* \beta + \beta^* \alpha) \right]\nonumber\\
&\times&
\exp -\left (
\begin{array}{c} \phi_1^* \\ \phi_2^* \\ \chi_1^* \\ \chi_2^*
\end{array} \right )
\left ( \begin{array}{cccc}
           z+i\sigma e^s &  0 & \alpha & 0 \\
             0  & z+i\sigma e^{-s} & 0 & \beta \\
            \alpha^* & 0& z+J +\rho_1-i\rho_2 & 0\\
             0  & \beta^* & 0 & z  +\rho_1 +\rho_2
\end{array} \right )
\left ( \begin{array}{c} \phi_1 \\ \phi_2 \\ \chi_1 \\ \chi_2
\end{array} \right ),\nonumber \\
\label{voorlaatst}
\ee
where
\be 
d[\sigma] =
 \ \sigma d\sigma ds  \ d\rho_1 \ d\rho_2 \
 d\alpha \ d\alpha^* \ d\beta \ d\beta^* \ .
\ee
The  Gaussian integrals over the $\phi$ and $\chi$ variables are equal to 
the product of  inverse graded determinants. Writing out the $2\times 2$
graded determinants according to (\ref{detg}) we obtain
\be
Z(z,z+J) = &&\int d[\sigma]
\exp \left [ -n\Sigma^2(\sigma^2
+\rho_1^2 + \rho_2 ^2 + \alpha^* \beta + \beta^* \alpha) \right ]\,\,
\nn \\ &&\times
(z+J+\rho_1-i\rho_2)^n(z+J+\rho_1 +i\rho_2)^n
\nn \\ &&\times
(z +i\sigma e^s +\frac{\alpha \alpha^*}{z+J +\rho_1 -i\rho_2})^{-n}
(z+i\sigma e^{-s} +\frac{\beta \beta^*}{z+J +\rho_1+i\rho_2})^{-n}.\nn \\
\label{finalz}
\ee
The Grassmann integrals can be performed trivially by expanding the
exponent and the pre-exponent. The expression for the partition
function can be further simplified by shifting $\sigma \cosh s$ by $iz$ and
$\rho_1$ by $-z-J$. Using the polar coordinates
\be
\rho_1 = -z -J + \rho \cos \phi /\Sigma, \nn\\
\rho_2 = \rho \sin\phi/\Sigma,
\ee
and rescaling $\sigma$ by $1/\Sigma$ we find
\be
Z(z,z+J) &=& c
\int_{-\infty}^{\infty} \sigma d\sigma \int_{-\infty}^{\infty} ds
\int_0^{\infty}\rho d\rho \int_0^{2\pi} d\varphi
\left
[1+ \frac 1{\sigma^2\rho^2} \right ]
\nonumber \\
&\times& 
\left ( \frac {\rho^2}{-\sigma^2}
 \right )^n e^{-n(\sigma^2
+ \rho^2 +2iz\Sigma\sigma \cosh s - 2(z+J)\Sigma\rho \cos \varphi
+\Sigma^2((z+J)^2 -z^2))} \ .
\label{lastreal}
\ee
Because of the shift of the $\sigma$-variable its integration path
lies below the real axis.

The partition function (\ref{lastreal}) is valid for any $n$. In the
large $n$ limit the $\rho$ and the $\sigma$ integrations can be
performed by a saddle point approximation. The saddle point
equations are given by
\be
\rho^2 = 1, \qquad \sigma^2 =-1,
\ee
leading to the saddle points
\be
\bar \rho = \pm 1, \qquad \bar \sigma = \pm i.
\ee
The saddle point $\bar \rho =-1$ and and $\bar \sigma =-i$ are not accessible 
(remember that $\sigma$ contains an infinitesimal increment). 
By expanding $\rho= 1+\delta \rho$ and 
$\sigma=1+\delta \sigma$ to second order about the
saddle point we obtain
\be
Z(z,z+J) &=& c
\int_{-\infty}^{\infty}  d\delta \sigma \int_{-\infty}^{\infty} ds
\int_{-\infty}^{\infty} d\delta \rho \int_0^{2\pi} d\varphi
\nonumber \\&\times& 
\left[2z\Sigma \cosh s (\delta \sigma)^2 -2(z+J)\Sigma \cos \varphi
(\delta\rho)^2 -i(\delta \rho^2) + i(\delta\sigma)^2
\right ]
\nonumber \\&\times& 
e^{-2n((\delta\sigma)^2+(\delta\rho)^2) +z\Sigma \cosh s - (z+J)\Sigma \cos \varphi
+\Sigma^2((z+J)^2 -z^2))} \ ,
\label{largen1}
\ee
The terms that are odd in either $\delta \rho$ or 
$\delta \sigma$ integrate to zero and have not been written down.
 The Gaussian integrals over $\delta \rho$ and $\delta \phi$ are 
easily calculated. As final result for the partition function we find
\be
Z(z,z+J) &=& \frac {n}{\pi}
\int_{-\infty}^{\infty} ds
\int_0^{2\pi} d\varphi\nonumber \\
&\times& \left
[(z+J)\Sigma \cosh s  -z\Sigma \cos \varphi
\right ]
e^{-2n(z\Sigma \cosh s -z\Sigma \cos \varphi
+\Sigma^2((z+J)^2 -z^2))} \ ,\nn\\
\label{largen2}
\ee
The normalization constant in (\ref{largen1}) has been chosen such that
$Z(z,z) =1$.
The integrals can be expressed in terms of modified Bessel functions. This
results in 
\be
Z(z,z+J) = 2n\Sigma(z+J)K_0(2n\Sigma z )I_1(2n\Sigma(z+J)) +
2n\Sigma z K_1(2n\Sigma z)I_0(2n\Sigma(z+J)).
\ee
The normalized resolvent follows after differentiation with respect to the
source term. It is given by
\be
\frac{G(z)}{\Sigma}= x (K_0(x)I_0(x) + K_1(x)I_1(x))\qquad{\rm with}
\qquad x = 2n\Sigma z,
\ee
which we have obtained previously from the low energy effective
partition function (see eq. (\ref{val})).

    \section{Integrability and the Replica Limit of Low Energy QCD Partition 
Function}
\label{intqc}

In Chapter \ref{integ}
 we have shown that the GUE two point correlation function
can be obtained from the replica limit of the Toda lattice equation.
In this section we will show that a similar integrable structure exists
for the partially quenched QCD partition function. In particular, partition
functions with bosonic quarks, fermionic quarks and the supersymmetric
partition function form a single integrable hierarchy related
by the Toda lattice equation \cite{SplitVerb1,SplitVerb2,SplitVerb3}.

\subsection{Partition Function for $N_f$ Fermionic Flavors}
\label{intqc1}

We consider the zero momentum limit of the QCD partition function with
$n$ fermionic flavors with mass $m$. The effective partition function
for this case was already derived in section \ref{lowen1} and is given by
\be
Z^{\rm eff}_\nu(M) = \int_{U(n)} {\det}^\nu (U) 
e^{V\Sigma {\rm Re\,Tr}\,
{M} U}.
\label{zeffnu3}
\ee
In a diagonal representation of $U$ this partition function can be
rewritten as
\be
\int \prod d\theta_k \prod_{k<l} |e^{i\theta_k} - e^{i\theta_l}|^2 
e^{x\sum_k \cos \theta_k},
\ee
where $x = mV \Sigma$.
By writing the Vandermonde determinant as
\be
\prod_{k<l} (e^{i\theta_k} - e^{i\theta_l}) =\det [e^{ip\theta_q}]_{ 
0\le p\le n-1,\, 1\le q\le n}
\ee
and expanding the determinant, the angular integrals can be written as
modified Bessel functions. They can be recombined into a determinant as
follows: \cite{tan,KSS}
\be
Z_n^\nu(x) = c_n\det[I_{\nu+k-l}(x)]_{0\le k,l \le n-1}.
\label{zsm}
\ee
Using recursion relations for Bessel functions this determinant 
can be rewritten as a $\tau-$function
\be 
Z_n^\nu(x) = \frac{c_n}{x^{n(n-1)}}
\det[(x\del_x)^{k+1}Z_1^\nu(x)]_{0\le k,l \le n-1}.
\label{zfereig}
\ee
with
\be
Z_1^\nu(x) = I_\nu(x).
\ee
The simplest way to prove this is to start from (\ref{zfereig}). To
the columns of the matrix we apply the relation
\be
x\del_x [x^p I_{\nu+p}(x)] = (\nu+2p)I_{\nu +p}(x) + x^{p+1} I_{\nu + p +1}(x)
\ee 
starting with $p=0$ and increasing $p$ successively by 1 until all
derivatives in the first row are gone. The first term in the recursion
relation is canceled by the addition of a multiple of the previous column.
We arrive at
\be
Z_n = c_n \det [(x\del_x)^k I_{\nu+l}(x)]_{0\le k,l \le n-1}.
\ee
Next we apply the recursion relation
\be
x\del_x [x^p I_{\nu+p}(x)] = x^{p+1} I_{\nu+ p-1}(x) - \nu x^pI_{\nu+p}(x)
\ee
starting with the last row. The second term in the recursion cancels
by the addition of a multiple of the previous row. We continue until
the second row, and after that we start again in the same way with the last
row and continue until the third row. After repeating this procedure until all
derivatives are gone we end up with the representation (\ref{zsm}) which
completes the proof.

\vspace*{0.2cm}
\noindent{\it Exercise.} Work out the proof for $n =2$.
\vspace*{0.2cm}

\subsection{The Bosonic Partition Function}
\label{intqc2}

We already have seen in section \ref{symme3} that the Goldstone manifold for
$n$ bosonic quarks is given by $Gl(n) /U(n)$. Using the same invariance
arguments as before we obtain the low energy effective partition function
\be 
Z^\nu_{-n} = \int_{Q \in Gl(n)/U(n)} {\det}^\nu (Q) 
e^{\frac 12 V\Sigma {\rm Tr}\,
M (Q + Q^{-1})}.
\label{zeffnu2}
\ee
In this case $Q$ can be diagonalized as 
\be
Q = U {\rm diag}(e^{s_k}) U^{-1}
\ee
so that an eigenvalue representation of this partition function is
given by
\be
\int \prod_k ds_k \prod_k e^{\nu s_k} 
\prod_{k<l} (e^{s_k} - e^{s_l})(e^{-s_k} - e^{-s_l})
e^{x\sum_k \cosh s_k}.
\ee
The Vandermonde determinants can again be written as
\be
\prod_{k<l} (e^{s_k} - e^{s_l}) = \det 
[ e^{p s_q}]_{0\le p\le n-1, 1\le q\le n}
\ee
and a similar expression for $s_k \to -s_k$.
By expanding the two determinants, the integrals combine into modified
Bessel functions which can be combined into a determinant as follows
\cite{DV1}
\be
Z_{-n}(x) =c_{-n}\det[K_{\nu+k+l}(x)]_{0\le k,l \le n-1}.
\ee
Using recursion relations for Bessel functions this determinant 
can be rewritten as a $\tau-$function
\be 
Z_{-n}(x) = \frac{c_{-n}}{x^{n(n-1)}}\det[(x\del_x)^{k+l}Z_{-1}(x)]_{0\le k,l \le n-1}.
\label{bostau}
\ee
with
\be
Z_{-1}(x) = K_\nu(x).
\ee
This simply follows from the observation that $(-1)^\nu K_\nu(x)$ and
$I_\nu(x)$ satisfy the same recursion relations.

\subsection{Replica Limit and Toda Lattice Equation}
\label{intqc3}

Instead of the supersymmetric generating function
 one could have used the replica trick to calculate the resolvent.
The fermionic replica trick is defined by
\be
G(z) = \lim_{n \to 0} \frac 1{n} \log Z_{n}^\nu(z).
\ee
and the bosonic replica trick by
\be
G(z) = \lim_{n \to 0} \frac 1{-n} \log Z_{-n}^\nu(z).
\ee
If we take the replica limit of the fermionic or bosonic
partition functions directly, we will obtain a result that differs from the
supersymmetric calculation. We will see below
that these problems can be avoided if the take the replica
limit of the Toda lattice equation.

We now consider bosonic and fermionic partition functions with all
masses equal to $z$ which only depend on the combination
\be
x = z \Sigma V.
\ee

Next we use
the Sylvester identity \cite{Sylvester,ForresterBook}
which is valid for the determinant of an arbitrary
matrix $A$. It is given by
\be 
C_{ij} C_{pq} - C_{iq}C_{pj} = \det (A)C_{ij;pq},
\ee
where the $C_{ij}$ are cofactors of the matrix $A$ and
the $C_{ij;pq}$ are double cofactors. By applying this identity
to the determinant in (\ref{zfereig})
for $i=j=n-1$ and $p=q=n$,  and writing the cofactors as derivatives of
the partition function,
we easily derive the Toda lattice 
equation \cite{Kharchev,mmm}
\be
(x\del_x)^2 \log Z_{n}^\nu(x) =  2n x^2 
\frac{Z_{n+1}^\nu(x) Z_{n-1}^\nu(x)}{[Z^\nu_{n}(x)]^2}.
\label{toda}
\ee
This equation has also been derived as a consistency condition
for QCD partition functions \cite{paulconsist}. 

\vspace*{0.2cm}
\noindent{\it Exercise.} Show that the cofactors that enter in the
derivation of (\ref{toda}) can be written as derivatives of the
partition function.
\vspace*{0.2cm}

In the derivation of (\ref{toda}) we have only used that the
fermionic partition function is a $\tau$-function. Since the
bosonic partition function is also a $\tau$-function (see (\ref{bostau}), 
it satisfies
the same Toda lattice equation.

The two semi-infinite hierarchies are connected by
\be 
\lim_{n \to 0}\frac 1n (x\del_x)^2 \log Z_{n}^\nu(x).
\ee
By extending to Toda lattice hierarchy to include an additional
spectator boson, it can be shown that \cite{SplitVerb3}
\be
\lim_{n  \to 0}\frac 1n (x\del_x)^2 \log Z_{n}^\nu(x)&=& 
\lim_{y\to x} x\del_x(x\del_x + y \del_y) \log Z_{1,-1}(x|y)\nonumber\\
&=&x\del_x \lim_{y\to x} x\del_x \log Z_{1,-1}(x,y)
\nonumber \\ &=& x\del_x x G(x).
\ee
For the resolvent we obtain the identity
\be
x\del_x x G(x) = 2 x^2 Z_1^\nu(x) Z_{-1}^\nu (x),
\ee
which explains      this factorization property
of the resolvent. In the same way it can be shown the factorization of the
susceptibility in a bosonic and a fermionic partition function
\cite{SplitVerb2}.

Inserting the expressions for $Z_1$ and $Z_{-1}$ we find
\cite{SplitVerb1}
\be
G(x) = x(K_\nu(x)I_\nu(x) + K_{\nu-1}(x) I_{\nu+1}(x)) + a\frac 1x,
\label{fintodz}
\ee
where the integration constant $a$ is fixed by the condition that in
the sector of topological charge $\nu$ we have $\nu$ zero eigenvalues
resulting in a contribution of $\nu/x$ to the resolvent. In the calculation
with  the supersymmetric method we found (\ref{fintodz}) without the
topological term because we started
from a generating function for the nonzero eigenvalues only. 
This result was first obtained from chiral Random Matrix Theory and
has also been derived from the replica limit  of
a Painlev\'e equation \cite{kanzieper}.

\section{Dirac Spectrum at Nonzero Chemical Potential}
\label{qcmu}
In this Chapter we study the quenched microscopic spectrum of the QCD Dirac
operator at nonzero chemical potential. Using the replica limit of the Toda
lattice equation we obtain the exact analytical 
result \cite{SplitVerb2}.

\subsection{General Remarks}
\label{qcmu1}

At nonzero baryon chemical potential the Dirac operator is
modified according to
\be
D \rightarrow D + \mu \gamma_0 .
\ee
This Dirac operator does not have any hermiticity properties
and its eigenvalues are scattered in the 
complex plane \cite{all,Markum,maria,hands,tilo}. 
For small $\mu$ we expect that the width of the cloud of
eigenvalues $\sim \mu^2$ \cite{all}. 
The average spectral density is given by
\be 
\rho(\lambda) = \langle \sum_k \delta^2(\lambda- \lambda_k) \rangle,
\ee
and the average resolvent is defined as usual by
\be
G(z) = \frac 1V\left \langle\sum_k \frac 1{i\lambda_k +z}\right \rangle.
\ee
Using that $\del_{z^*} 1/z = \pi \delta^2 (z)$ we easily derive
\be
\del_{z^*} G(z)|_{z=\lambda} = \frac{\pi}V \rho(\lambda).
\ee
In this case it is possible to write down a supersymmetric generating
function for the resolvent. However, we did not succeed in evaluating
this partition function and leave it as a challenge to the reader
to solve the problem by this method. 
We will calculate the resolvent
and the spectral density
from the replica limit of the Toda lattice equation
\cite{SplitVerb2}.

The replica limit of the spectral density is given
by \cite{Girko,Goksch,misha}
\be
\rho(z,z^*) = \lim_{n \to 0} \frac 1{\pi n}\del_z \del_{z^*} \log Z_n(z,z^*),
\ee
with generating function given by 
\be
Z_n(z,z^*) = \langle {\det}^n(D + \mu\gamma_0 +z)
{\det}^n(-D + \mu\gamma_0 +z^*)\rangle. 
\ee

The low energy limit of this generating function is a chiral
Lagrangian that is determined by its global symmetries and 
transformation properties. 
By writing the product of the two determinants as \cite{TV}
\be
{\det}(D + \mu\gamma_0 +z)
{\det}(-D + \mu\gamma_0 +z^*)
= 
\det \left( \begin{array}{cccc} id+\mu &0 &z& 0 \\
                            0& id-\mu & 0& z^*\\
                 z & 0 & id^\dagger+\mu  & 0 \\
                  0 & z^* & 0& id^\dagger -\mu \\ \end{array} \right ),
\nonumber \\
\ee
where we have used the decomposition of the Dirac operator given
in (\ref{diracstructure}),
we observe that the $U(2) \times U(2)$ flavor symmetry is broken
by the chemical potential term and the mass term. Invariance is recovered
by transforming the mass term as in the case of zero chemical potential
and the chemical potential term by a local gauge transformation. 

If we expand the chiral Lagrangian for the spectrum of quenched QCD at
nonzero chemical potential we obtain the following types of terms
\be
p^2 \pi_a^2,\quad\mu p_0 \pi_a^2, \quad \mu^2 \pi_a^2, \quad 
m\Sigma \pi_a^2.
\ee
The zero momentum modes factorize from the partition function if
\be
\mu \ll \frac 1{L}, \qquad {\rm and} \qquad \frac{z\Sigma}{F^2} \ll \frac 
1{L^2}.
\ee
In this domain the partition function is uniquely determinant by invariance
arguments as in the case of zero chemical potential. In the addition to
the mass term we obtain a term from the zero momentum part of the covariant
derivative resulting in the partition function
\be
Z_n(z,z^*) = \int_{U(2n)} dU e^{-\frac{F^2\mu^2 V}4 {\rm Tr}[U,B][U^{-1},B]
+\frac {\Sigma V}2{\rm Tr}M(U+U^{-1})},
\label{genmu}
\ee
where
\be
B= \mat {\bf 1}_n &0 \\ 0 & - {\bf 1}_n \emat, \qquad
M = \mat z{\bf 1}_n &0 \\ 0 & z^* {\bf 1}_n \emat.
\ee

\subsection{Integration Formula}
\label{qcmu2}

By decomposing a $U(2n)$ matrix as
\be
U = \mat u_1 & \\ & u_2 \emat \mat v_1 & \\ & v_2 \emat  
\mat 
\sqrt{1-b^2}& b  \\ 
b & -\sqrt{1-b^2} \emat
\mat v_1^\dagger & \\ & v_2^\dagger \emat, \label{paramu}
\ee
with $u_1,\, u_2, \, v_1 \in U(n)$, $ v_2 \in U(n)/U^n(1)$ 
 and $b$ a diagonal matrix, 
the following integration formula can be proved \cite{SplitVerb2}
\be
&&\int_{U(2n)} dU {\det}^\nu U e ^{\frac 12 {\rm Tr} [ M(U + U^{-1}] + 
\sum_p a_p {\rm Tr}[(UBU^{-1} B)^p] } \nonumber\\ 
&=& \frac {c_n}{(xy)^{n(n-1)}} {\det} 
[ (x\del_x)^k (y\del_y)^l Z_{1}^\nu(x,y) ]_{
0\le k,l \le n-1},
\label{int}
\ee
where
\be
Z^\nu_{1}(x,y) = \int_0^1 \lambda d\lambda I_\nu(\lambda x) I_\nu(-\lambda y)
e^{2 \sum_p a_p \cos (2p\cos^{-1} \lambda)},
\ee
and $c_n$ is an $n$-dependent constant. 

\vspace*{0.2cm}\noindent{\it Exercise.} 
Show that the Jacobian of
of the transformation (\ref{paramu}) from $U^{-1}dU$ to
$u_1^{-1}du_1, u_2^{-1}du_2, v_1^{-1}dv_1, v_2^{-1}dv_2$  and
$\prod_k d\lambda_k$ is given by
\be
J = \prod_{1\le k<l\le n} (\lambda_k^2 -\lambda_l^2)^2 
\prod_{k=1}^n (2\lambda_k),\label{jactotText}
\ee
where $\lambda_k = \sqrt{1-b_k^2}$.
\vspace*{0.2cm}

\vspace*{0.2cm}\noindent{\it Exercise.} 
Show that for $n=1$ and all $a_p =0$ the integral $Z_1^\nu(x,y)$
is given by
\be
\left .Z_{1}^\nu(x,y)\right |_{a_p =0} &=& \int_0^1 \lambda d\lambda 
I_\nu(\lambda x) I_\nu(-\lambda y)\nonumber \\
&=&\frac{yI_\nu(x) I_{\nu-1}(-y) + xI_{\nu-1}(x) I_{\nu}(-y)}
{x^2 -y^2}.
\ee
It can be interpreted as the QCD partition function with two different 
masses \cite{jsv}.
\vspace*{0.2cm}

\subsection{Toda Lattice Equation}
\label{qcmu3}

Using the integration formula (\ref{int}) for $ p=1$ we find that the
zero momentum partition function $Z_n^\nu(z,z^*)$ (see eq. (\ref{genmu}))
is given by
\be
Z_n^\nu(z,z^*) = \frac{c_n}{(zz^*)^{n(n-1)}} \det[(z\del_z)^k 
(z^*\del_{z^*})^l 
Z_1^\nu(z,z^*)]_{0\le k,l\le n-1},
\label{mutau}
\ee
where
\be
Z_1^\nu(z,z^*) = \int_0^1 \lambda d\lambda e^{-2VF^2\mu^2 (\lambda^2-1)}
|I_\nu(\lambda zV\Sigma)|^2.
\ee 
By applying the Sylvester identity to the determinant in (\ref{mutau})
for $i=j=n-1$ and $p=q=n$ and expressing the cofactors as derivatives,
we find a recursion relation that can
be written in the form of a Toda lattice equation
\be
z\del_z z^*\del_{z^*} \log Z_n^\nu(z,z^*) =
\frac{\pi n}2 (zz^*)^2 \frac{Z_{n+1}^\nu(z,z^*) Z_{n-1}^\nu(z,z^*)}
{[Z_n^\nu(z, z^*)]^2}.
\ee
For the spectral density we find the simple expression
($Z_0^\nu(z, z^*)=1$)
\be
\rho(z,z^*) = \lim_{n \to 0} \frac 1{\pi n} \del_z \del_{z^*} \log
Z_n^\nu(z, z^*) =
\frac {zz^*}2 Z_1^\nu(z,z^*) Z_{-1}^\nu(z,z^*).
\label{rhosimple}
\ee
What remains to be done is to calculate the partition function with 
one bosonic
and one conjugate bosonic quark which will be completed in the
next subsection.

\subsection{The Bosonic Partition Function}
\label{qcmu4}

In this subsection we evaluate the low energy limit of the QCD
partition function at nonzero chemical potential
for one bosonic quark and one conjugate bosonic quark. 
Because of convergence 
requirements, the inverse determinants of nonhermitian operators
have to regularized \cite{Feinnh,efetovdir}. 
This is achieved by expressing them as the determinant
of a larger Hermitian operator 
\be
&&{\det}^{-1} \mat z& id + \mu\\id^\dagger +\mu &z \emat   
{\det}^{-1} \mat z^*& -id + \mu\\-id^\dagger +\mu &z^* \emat  \\
\label{detzm1}
& = & \lim_{\epsilon \to 0}
{\det}^{-1} \matf \epsilon &       0  &               z &    id+\mu\\
      0                  & \epsilon & id^\dagger +\mu & z \\
      z^*                &-id + \mu & \epsilon        & 0 \\
      -id^\dagger+\mu        & z^*      & 0               & \epsilon \ematf.
\nn
\ee 
To better expose the symmetries of this problem we rewrite the inverse
determinant in the r.h.s. of this equation as
\be
 \int d\phi_kd\phi_k^*
\exp[ i  
\vect \phi_1^*\\\phi_2^*\\\phi_3^*\\\phi_4^*\evect^T 
\matf \epsilon & z        &  id+\mu & 0\\
      z^*   &\epsilon   & 0& id - \mu         \\
    -id^\dagger +\mu&  0               & \epsilon &-z^*\\
      0 &  -id^\dagger -\mu &-z& \epsilon  \ematf
\vect \phi_1\\ \phi_2\\\phi_3\\\phi_4\evect] \nn. \\
\label{detzm2}
\ee 
In this case the mass matrices are given by
\be
\zeta_1 = \mat \epsilon & z \\ z^* & \epsilon \emat 
\qquad{\rm and}\qquad
\zeta_2 = \mat \epsilon & -z^* \\ -z & \epsilon \emat .
\label{masszeta}
\ee
The two mass matrices are related by 
\be
\zeta_2 = - I \zeta_1 I \qquad {\rm with} \qquad 
I =  \mat 0 & 1 \\ -1 & 0 \emat.
\ee
For $\mu =0$ and $\zeta_1=\zeta_2=0$ the basic symmetry\footnote{This is
the symmetry when we disregard convergence. Taking convergence of the
bosonic integrals into account the vector symmetry is $U(2)$ and the
axial symmetry is $Gl(2)/U(2)$.} 
of the partition 
function is $Gl(2) \times Gl(2)$,
\be
\vect \phi_1 \\ \phi_2 \evect \to U_1  \vect \phi_1 \\ \phi_2 \evect \qquad
\vect \phi_1^* \\ \phi_2^* \evect \to 
\vect \phi_1^* \\ \phi_2^* \evect U_2^{-1}\nn \\
\vect \phi_3 \\ \phi_4 \evect \to U_2  \vect \phi_3 \\ \phi_4 \evect \qquad
\vect \phi_3^* \\ \phi_4^* \evect \to  
\vect \phi_3^* \\ \phi_4^* \evect U_1^{-1}.
\ee
 This symmetry can be extended to 
nonzero mass or chemical potential if we adopt the transformation rules
\be
\zeta_1 \to U_2 \zeta_1 U_1^{-1}, &\qquad& \zeta_2 \to U_1 \zeta_2 U_2^{-1},
\nn \\ 
\mu_1 \to U_2 \mu_1 U_2^{-1}, &\qquad& \mu_2 \to U_1 \mu_2 U_1^{-1},
\ee
where $\mu_1$ is the chemical potential matrix that is added $id$ and
$\mu_2$ is the chemical potential matrix that is added to $-id^\dagger$.
These matrices are introduced for the sake of  discussing 
the transformation properties
of the partition function (\ref{detzm1}) and will ultimately be 
replaced by their original values
\be
\mu_1 = \mu_2 = \mat \mu & 0 \\ 0 & -\mu \emat.
\ee
The chiral symmetry is broken spontaneously
by the chiral condensate to $Gl(2)$. Because of convergence of the 
bosonic integral, the Goldstone manifold is not $Gl(2)$ but rather
$Gl(2)/U(2)$, i.e. the coset of positive definite matrices
as in the case of zero chemical potential. Under 
a  $Gl(2)\times Gl(2)$ transformation the Goldstone fields transform
as
\be
Q \to U_2 Q U_1^{-1}.
\ee
The
low energy effective partition function should have the same 
transformation properties as the microscopic partition function (\ref{detzm1}).
To second order in $\mu$ and first order in the mass matrix we
can write down the following invariants
\be
{\rm Tr} \zeta_1 Q, \quad {\rm Tr} \zeta_2 Q^{-1}, \quad
{\rm Tr} Q\mu_2 Q^{-1} \mu_1 ,\quad {\rm Tr} \mu_1 \mu_2.  
\ee  
 We also have the discrete symmetry that the partition function is
invariant under the interchange of $\zeta_1$ and $\zeta_2$. This symmetry
implies that the coefficients of the two mass terms in the effective
partition function are the same. Using that the integration measure 
on positive definite Hermitian matrices is given by $dQ/{\det}^2 Q$,
we finally arrive at the effective
partition function\footnote{Notice that the $Q$ variables used in this lecture
are the transpose the $Q$ variables used in \cite{SplitVerb2}. This
simplifies the discussion of the transformation properties}
\be
Z_{-1}^{\nu}(z,z*) = \int_{Q \in Gl(2)/U(2)} \frac{dQ}{{\det}^2 (Q)}  
{\det}^\nu(Q)
e^{-\frac{F^2\mu^2 V}4 {\rm Tr}[Q,B][Q^{-1},B]
+\frac {i\Sigma V}2{\rm Tr}(\zeta_1 Q+\zeta_2 Q^{-1})} .                        \nonumber\\
\label{z-1} 
\ee

\vspace*{0.2cm}\noindent{\it Exercise.} Prove that for a
parameterization of positive 
definite $2\times 2$ matrices as $Q= U {\rm diag}(e^{s_1}, e^{s_2}) U^{-1}$
the invariant measure is given by
\be
\frac{dQ}{{\det}^2 Q} = (e^{s_1} - e^{s_2})(e^{-s_1} - e^{-s_2})ds_1 ds_2
dU,
\ee
where $dU$ is the invariant measure on $U(2)/U(1)\times U(1)$.
\vspace*{0.2cm}
  
To evaluate the integral (\ref{z-1}) we do not use the parameterization
of the above exercise but rather
\be
Q = e^t \mat e^r \cosh s  & e^{i\theta}\sinh s \\
          e^{-i\theta}\sinh s &   e^{-r} \cosh s \emat.
\ee 
where
\be
r \in \langle -\infty, \infty \rangle, \quad
s \in \langle -\infty, \infty \rangle, \quad
t \in \langle -\infty, \infty \rangle, \quad
\theta \in \langle 0, \pi \rangle \quad.
\ee
Using that the Jacobian from the measure $dQ/{\rm det}^2Q$ to the
measure $dr ds dt d\theta$ is given by
\be
J = 4 e^{4t}\cosh s \sinh s,
\ee
we find the partition function ($ z= x+iy$)
\be
Z_{-1}^\nu(z,z^*) &=& 
\lim_{\epsilon\to0}C_\epsilon 
\int dr ds dt d\theta
\cosh s |\sinh s| e^{2\nu t}\\ &&\times
e^{i\frac{V\Sigma}{2}(4x\sinh s \cosh t \cos \theta - 4 y \sinh s \sinh t
  \sin \theta +4\epsilon \cosh r \cosh s \cosh t) -\mu^2 F^2 V 
(1+2 \sinh^2 s)}.
\nn \\ \nn
\ee
The integral over $r$ is equal to  
$2K_0(2N\epsilon \cosh s \cosh t)$ with leading singularity given by 
$\sim -\log \epsilon$. This factor is absorbed in the normalization of 
the partition function. 
Then the integral over $\theta$ gives a
Bessel function. Introducing $u = \sinh s$ as new integration variable
we obtain \cite{SplitVerb2}
\be 
Z_{-1}^\nu(z,z^*) &=&
C_{-1}\int_{-\infty}^\infty dt  \int_0^\infty du e^{2\nu t}
J_0(2V \Sigma u(x^2 \cosh^2 t
+y^2 \sinh^2 t)^{1/2})
e^{-\mu^2 F^2 V (1+2 u^2)}.\nn \\
\label{zminj}
\ee
To perform the integral over $u$ we use the known integral
\be
\int_0^\infty dx x^{a+1} e^{-\alpha x^2} J_a(\beta x) = 
\frac{\beta^a}{(2\alpha)^{a+1}} e^{-\beta^2/4\alpha}.
\ee
This results in
\be
Z_{-1}^\nu(z,z^*) &=& \frac {C_{-1}e^{-V\mu^2 F^2} }{4\mu^2 F^2 V } 
\int_{-\infty}^\infty  dt   e^{2\nu t}
e^{-\frac{V (x^2 \cosh^2 t +y^2 \sinh^2 t)}{2\mu^2 F^2}}.\nn \\
\ee
Using that $\cosh^2 t = \frac 12 + \frac 12 \cosh 2t$ and
$\sinh^2 t = -\frac 12 + \frac 12 \cosh 2t$, the integral
over $t$ can be rewritten as a modified Bessel function
resulting in \cite{SplitVerb2}
\be
Z_{-1}^\nu(z=x+iy,z^*) = 
C_{-1} \,
e^{\frac{V\Sigma^2 (y^2-x^2)}{4 \mu^2F^2}}
K_\nu\left ( \frac {V\Sigma^2(x^2+y^2)}{4 \mu^2 F^2}\right ).
\label{z-1a}
\ee

\subsection{The Dirac Spectrum at Nonzero Chemical Potential}

The final result for the quenched spectral density is obtained 
by substituting the partition functions $Z_{1}^\nu(z,z^*)$
and $Z_{-1}^\nu(z=x+iy,z^*)$ in expression (\ref{rhosimple})
obtained from the replica limit of the Toda lattice equation.
We find,
\be
\rho(x,y) &=& \frac {V^3\Sigma^4}{2\pi F^2\mu^2}(x^2+y^2)
e^{\frac{V\Sigma^2 (y^2-x^2)}{4 \mu^2F^2}}
K_\nu\left ( \frac {V\Sigma^2(x^2+y^2)}{4 \mu^2 F^2}\right )
\nonumber \\ && \times
\int_0^1 \lambda d\lambda e^{-2VF^2\mu^2 \lambda^2}
|I_\nu(\lambda zV\Sigma)|^2.
\label{final}
\ee 
The normalization constant has been chosen such that    the $\mu \to 0$  limit
of $\rho(x,y)$ for large $y$ is given by $\Sigma V/\pi$ (see below).
 
\begin{figure}[ht] 
\caption{ The quenched spectral density at nonzero chemical potential in
the ergodic domain of QCD (full curve). The dotted curve
represents the result obtained
from a nonhermitian eigenvalue model.  (Taken from \cite{minn04}.)  }
\centerline{\hspace*{1cm}\includegraphics[height=10cm]{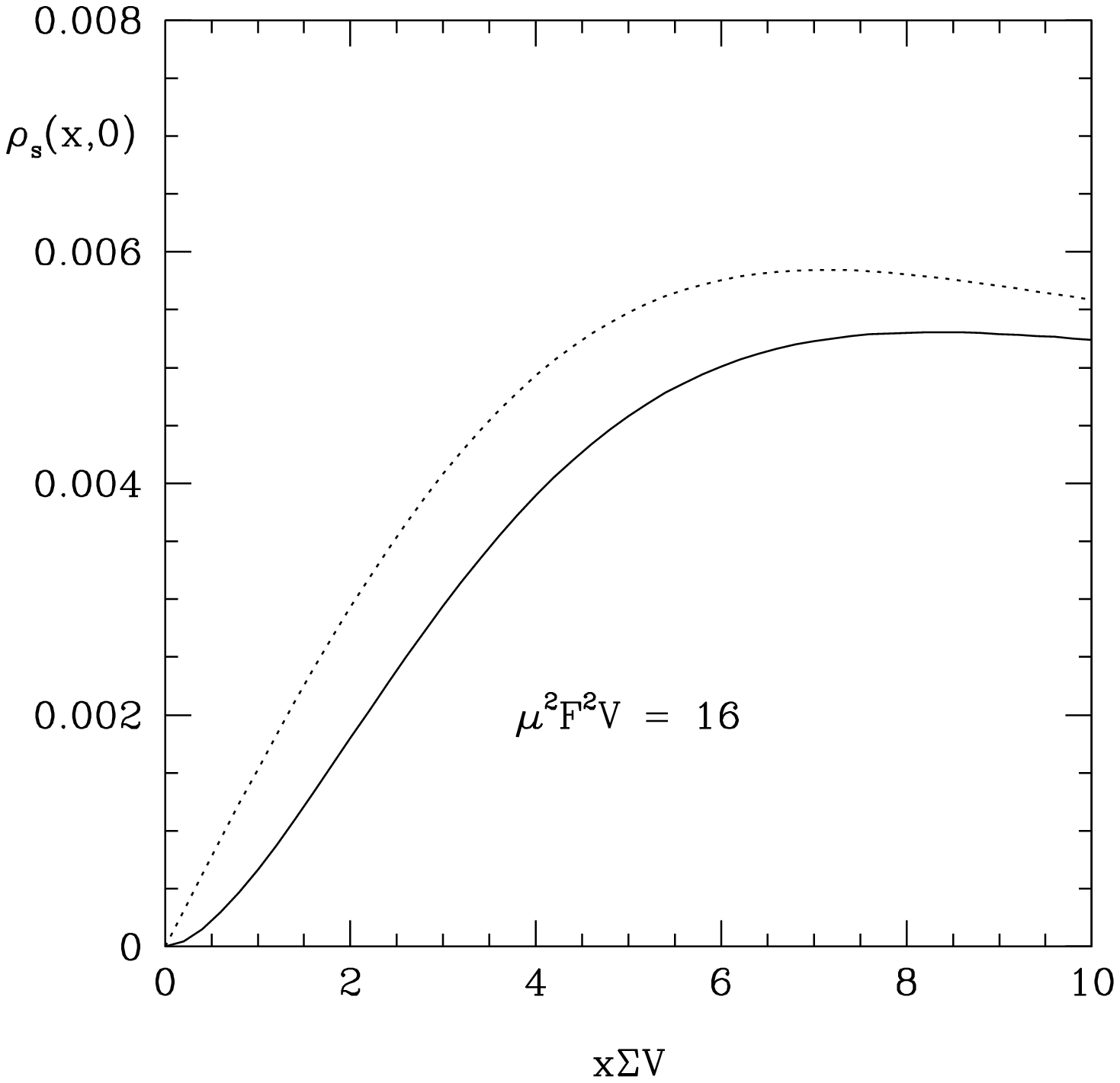}
\hspace*{-2.5cm}
\includegraphics[height=10cm]{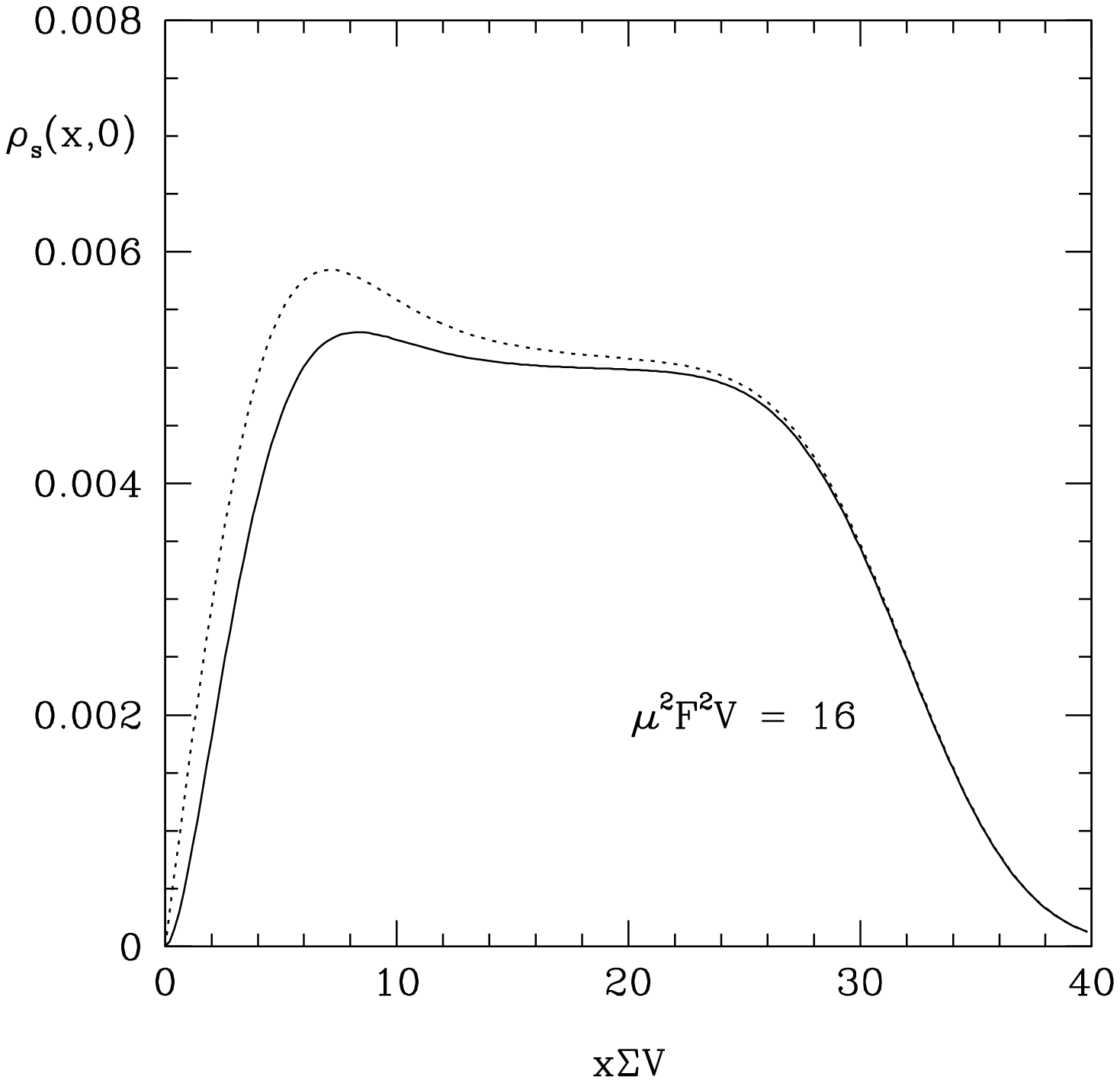} 
}   
\label{fig2}
\end{figure}

In Fig. \ref{fig2} we show a graph of the spectral density for $y=0$
and $\mu^2F^2 V =16$ (solid curve)
plotted as the dimensionless ratio
\be
\rho_s(x,y) = \frac {\rho(x,y)}{\Sigma^2 V^2}  .
\ee
Also shown is the result where
the Bessel function $K_\nu$ is replaced by
its asymptotic expansion (dotted curve). This result has been obtained
from a nonhermitian eigenvalue model \cite{akemann} that is not in
the universality class of QCD.
An important difference between the two results is that the spectral
density (\ref{final}) is quadratic in $x$ for $ x \to 0$, whereas 
the result given by the dotted curve is linear in $x$ for $x\to 0$.

In the limit ${\rm Re}(z)\Sigma/\mu^2 F^2 \ll 1$ the upper limit of the 
integral in (\ref{final}) can be extended to infinity. Using that
\be
\int_0^\infty \lambda d\lambda e^{-2VF^2\mu^2 \lambda^2} |I_\nu(\lambda 
zV\Sigma)|^2 = \frac 1{4\mu^2 F^2 V^2} e^{\frac{(z^2+z^{*\, 2}) \Sigma^2 V}
{8\mu^2 F^2}} I_\nu\left(\frac{zz^* V\Sigma^2}{4\mu^2F^2}  \right )
\ee
the spectral density can be expressed as
\be
\rho(x,y) = \frac {2}{\pi} u^2 zz^* K_\nu(zz^*u) I_\nu(zz^*u)
\qquad {\rm with} \qquad u = \frac{V \Sigma^2 }{4\mu^2 F^2}.
\ee
Therefore, the spectral density becomes a universal function that
only depends on a single parameter $u$. This parameter can be rewritten
in a more physical way as
\be
u =  \pi  \rho_{\rm asym}(x,y).
\ee
For the dimensionless ratio we obtain \footnote{This formula
arose as a result of a discussion with Tilo Wettig.}
\be
\frac{\rho(x,y)}{\rho_{\rm asym}(x,y)} = 
2 u zz^* K_\nu(zz^*u) I_\nu(zz^*u)
\ee
which is universal combination that depends only on  $zz^*u$.

In the thermodynamic limit the Bessel functions can be approximated
by their asymptotic limit,
\be
K_\nu(z) = \sqrt{\frac \pi{2z}} e^{-z},\quad 
I_\nu(z) = \frac 1{\sqrt{2\pi z}} ( e^z +i e^{-z}),\quad 
I_\nu(z^*) = \frac 1{\sqrt{2\pi z^*}} ( e^{z^*} -i e^{-z^*}).
\ee
This results in
\be
\rho(x,y) &=& \frac {V^2\Sigma^2}{2\pi F\mu \sqrt{2\pi V}}
\int_0^1 d\lambda e^{-2VF^2\mu^2(\lambda - \frac{|x|\Sigma}{2F^2\mu^2})^2}.
\ee
For $V\to \infty$ the integral over $\lambda$ can be performed by a
saddle point approximation. If the saddle point is outside the
range $[0,1]$ we find zero for $V\to \infty$. This results in 
the spectral density
\be
\rho(x,y) = \frac {V\Sigma^2}{4\pi \mu^2 F^2 } \quad {\rm for} 
\quad |x |< \frac {2F^2 \mu^2}\Sigma.
\ee
and $\rho(x,y) =0 $ outside this strip. This result is in agreement
with the mean field analysis of the effective partition function
\cite{TV}.
For the integrated eigenvalue density we find
\be
\int_{-\infty}^\infty dx \rho(x,u) = \frac{\Sigma V}\pi
\ee
in agreement with the eigenvalue density at $\mu = 0$.

\section{Conclusions} 

The supersymmetric method is a powerful method to analyze
random matrix ensembles. This approach makes it possible to formulate
the problem in terms of symmetries and the spontaneous breaking
of these symmetries. The final result for universal correlation functions
is uniquely given by a partition
function of Goldstone modes interacting according to an
effective Lagrangian with the symmetries of the microscopic
partition function. Convergence requirements necessary lead to
a Goldstone manifold that is  the product
of a compact and a noncompact submanifold. 
In the final answer for the correlation function this structure
is present in the form of compact and noncompact integrals and is
therefore the natural formulation of the problem.

In these lectures we have given an elementary introduction to
Random Matrix Theory and the supersymmetric method. We have 
applied this method to four problems which are technically
the least demanding: the one-point function of the Gaussian Unitary
Ensemble (GUE), the two-point function of the Gaussian Unitary Ensemble
and the one-point function of the chiral Gaussian Unitary Ensemble
both as zero and at nonzero chemical potential.
The one-point function of the GUE is not a universal quantity and was
only discussed to illustrate the method. The two-point function of
the GUE and the one-point function of the chGUE are universal
quantities. They are relevant for eigenvalue correlations of 
Hamiltonian systems with broken time reversal invariance and no
other symmetries  and for the distribution of the small eigenvalues 
of the QCD Dirac operator, respectively.

We have also shown that for $\beta =2$ the supersymmetric partition
function connects the semi-infinite bosonic and the semi-infinite
fermionic Toda lattice hierarchy of partition functions.
This makes it  possible to obtain the second
derivative of the partition function from the replica limit
of the Toda lattice equation. The final result is the product
of a bosonic (noncompact) and a fermionic (compact) integral. In 
this approach there is no need to analyze a supersymmetric non-linear
$\sigma$-model. The factorized structure of the final answer
is a direct consequence of the Toda lattice equation and make this
the minimal approach for the evaluation a supersymmetric partition function.
It is still an open problem whether integrable structures play a similar
role for other values of the Dyson index.  
 
The supersymmetric method is not the only method to derive nonperturbative
results for Random Matrix Theories. Perhaps the best known method is
the orthogonal polynomial method. This method starts from the
joint eigenvalue distribution and exploits orthogonality
relations to perform the integrals that lead to the correlation
functions we are interested in. Therefore, this method fails for
problems that can not be formulated in terms of a joint eigenvalue
density. Most notably, the statistical theory of $S$-matrix fluctuations
and the theory of parametric correlations have only been solved by
means of the supersymmetric method.  

A disadvantage of the supersymmetric method is that it 
requires a thorough knowledge of supersymmetry and superanalysis. 
A straightforward evaluation of the Grassmann integrals leads to
a factorial proliferation of terms and does not work in all but
the simplest cases. To find the right variables to solve the
supersymmetric integrals is often an arduous task and one wonders
why the symmetries of the problem do not provide us with a more 
directed path to the final result. One hint that more progress can
be made is that in some cases where the 
supersymmetric method fails because of technical problems 
the problem can be solved exactly by means of the replica limit of the Toda 
lattice equation. We have illustrated this for the quenched spectral
density of the QCD Dirac operator at nonzero chemical potential. Meanwhile,
this problem has also been solved by means of the orthogonal
polynomial method \cite{james,AOSV}, but the derivation of this solution from 
a supersymmetric nonlinear $\sigma$-model has remained elusive.

\begin{theacknowledgments}
G. Akemann,  M. Berbenni-Bitsch, D. Dalmazi, P. Damgaard, 
A. Garcia-Garcia, A. Halasz, 
A.D. Jackson, B. Klein, J. Kogut, S. Meyer, H. Nishioka, S. Nishigaki, 
Th. Seligman, A. Schaefer, Th. Schaefer, D. Son, M. Stephanov,   
J.C. Osborn, M. Sener, E. Shuryak, A. Smilga, K. Splittorff, 
M. Stephanov,
D. Toublan, H.A. Weidenm\"uller, T. Wettig, A. Zhitnitsky  
and M.R. Zirnbauer are
thanked for sharing their insights and collaboration on the joint publications
on which
these lectures are based. Kim Splittorff is acknowledged for a 
critical reading of the manuscript. 
Roelof Bijker is thanked for 
organizing this wonderful summer school, for the generous hospitality,
and for carefully editing the manuscript. 
 
\end{theacknowledgments}

\end{document}